# CYBERSECURITY THREAT ANALYSIS AND ATTACK SIMULATION FOR UNMANNED AERIAL VEHICLE NETWORKS (DRONES)

**Author:** Charles Abdulrazak

**Student ID:** UP2138844

**Date:** 15th September 2023

**Academic Supervisor:** Professor Simon Marsden

**This dissertation is submitted in part as a fulfilment of the**

**MSc Cyber Security and Forensic Information Technology**

**Ethics Review number for this submission is: TETHIC-2023-105910**

# Abstract

Drones, also known as unmanned air vehicles (UAVs), have revolutionised various industries, from farming to national security. (Wexler., Lesley. 2016) However, their broad use has revealed a severe weakness in cybersecurity. (Jean-Paul Yaacoub 2020) The urgent necessity to defend UAV networks from new cyber threats is explored in-depth in this research, making it a crucial subject for both technological development and national security. The two essential areas of our study are assault simulation and threat analysis in cybersecurity. This work demonstrates how easy it is to hack a drone mid-flight using only a Raspberry Pi3 and open-source online tools. This work illustrates the ability to penetrate a DJI drone currently used by the mercenary soldiers in the Ukraine war. (Greg Myre March, 2023)

This research examines strategies used to attack UAV networks, such as the de-authentic attack and the man-in-the-middle attack. This work investigates the weaknesses in these networks' sophisticated attack simulations with a Raspberry PI 3 and the Alpha network adaptor from Amazon, showing that basic tools are needed to perform cyberattacks on drones. This research proposes creative solutions and preventative methods for protecting UAV operations and highlights the seriousness of the problem. As drones become more prevalent daily, maintaining their security becomes crucial. This work provides a compelling perspective on protecting vital infrastructure and preserving our skies by bridging the gap between the latest technologies and cybersecurity.

# ACKNOWLEDGEMENT

I wish to convey my profound appreciation to Senior Lecturer Simon Marsden. Under his tutelage, I was first introduced to the world of ethical hacking, and his invaluable guidance was instrumental throughout my project's progression.

Additionally, I am deeply indebted to Dr. Peter Lee. Following his illuminating lecture on 'The Reaper Force,' he generously shared invaluable insights on drone warfare, which enriched my understanding further.

# TABLE OF CONTENTS

# LIST OF FIGURES

# Chapter 1: Introduction:

## 1.1 Introduction:

Drones are frequently referred to as Unmanned Aerial Vehicles (UAV), a name also employed by the Federal Aviation Administration (FAA) to define this category of aircraft, alternatively referred to as (ROA) remotely operated aircraft. (Bart Custers 2016).

Drones have a wide range of usage and have emerged as a transforming technology with various applications in several fields, such as surveillance, farming, legal work, and wars. More than ninety nations employ unmanned aerial vehicles (UAVs) in their armed forces. At least 23 of those drones are capable of being armed, and four are already using armed drones in operations. (Sayler K. 2015) . The rapid manufacturing of UAV networks become a primary target for hackers looking to find vulnerabilities for evil purposes. Hence, this research focuses on conducting a Cybersecurity Threat Analysis and Attack Simulation for Unmanned Aerial Vehicle (Marinenkov Egor., & Viksnin L. 2018). Since the start of the war in Ukraine, interest in drones has increased. Ukraine demonstrated just how quickly and simply drones may be activated and put into service. (Martin Broomhead, 2023)

## 1.2 Background of Study:

The utilisation of unmanned aerial vehicles (UAVs) equipped with advanced technology has brought about a significant transformation across diverse sectors. However, this advancement in technology presents new concerns, particularly in the field of cybersecurity. These drones' wireless communication protocols make the networks they depend on susceptible to cyber-attack vulnerabilities. (Faraji-Biregani., & Fotohi. 2021). This study extensively reviews the existing research about cybersecurity vulnerabilities encountered by unmanned aerial vehicles (UAVs). The primary objective of this investigation is to explain the principal concerns associated with current drone technology.

## 1.3 Research Question:

This study focuses on understanding and implementing cyber threats on unmanned vehicles. What are the primary cybersecurity threats faced by UAV networks, considering their functional attributes and communication framework along with the simulation and processing of the UAV systems to access the UAV navigation patterns and assess their resilience against cyber-attacks.

## 1.4 Research Aim:

There are below aims for this research to fulfil by the end of this research:

- **Aim 1: Literature Review and Analysis**
  Critically analyse the existing literature on threat inspection within the cybersecurity process of UAV networks, aiming to identify key trends, challenges, and gaps in the current understanding of cybersecurity threats in UAV networks.

- **Aim 2: Simulation of UAV Network Attacks**
  Design and conduct simulations of UAV network attacks that closely simulate real-life cybersecurity threats, providing insights into the vulnerabilities and potential countermeasures of UAV networks in the face of evolving security challenges.

- **Aim 3: Study of UAV Communication Methods**
  Comprehensively investigate the communication methods and data formats employed in UAV systems, to gain a deep understanding of the communication protocols and standards that reinforce UAV cybersecurity, and to provide recommendations for enhancing cybersecurity practices in UAV networks.

## 1.5 Research Objectives:

- Analyse the literature review for the existing threat in the cybersecurity process in UAV networks.
- Investigate and simulate UAV network attacks that show the processes of real-life cybersecurity threats encountered on UAV networks.
- Incorporate literature and practical experiments into cyber security recommendations.

## 1.6 Research Rationale:

The importance of this study is due to the vital necessity to address the new cybersecurity issues brought on by the growing use of UAV networks. The potential consequences of successful cyberattacks grow more troublesome, a hacked UAV network can also interrupt vital services and jeopardise public safety. (Garcia M., & Rishi G., 2020).

Since the start of the conflict in Ukraine, people all around the world have been more aware of the danger posed by rogue drone operators. The communication protocols utilised by drones must be studied. (Martin Broomhead, 2023) By giving a whole study of threats and vulnerabilities that are possible as particular to these systems, this research seeks to identify the current issue in UAV network cybersecurity. This study aims to increase knowledge of the vulnerabilities and the need for proactive safety precautions by simulating real-world attack scenarios. Additionally, the operational framework of the system is complex, requiring the involvement of a diverse range of individuals, including computer programmers, communication specialists and (UAV) operators, to ensure the reliability and security of UAV operations. (Dr. Peter Lee. 2018)

## 1.7 Research Significance:

Examining cybersecurity risk and developing attack models for UAV networks is vital as drones become integral in police, transport, and military surveillance sectors. This research identifies and simulates potential cyberattacks, aiming to develop robust countermeasures to safeguard sensitive data that drones handle daily.

The research examines the system's vulnerabilities and flaws by simulating actual cyberattacks through pen-testing attack simulations. A critical aspect is to enhance UAV operations' reliability and security. (Abdelouahid D. 2023) By identifying potential risks and vulnerabilities, we can enhance UAV networks against malicious attacks, fostering safer and more secure drone operations.

## 1.8 Research Framework:

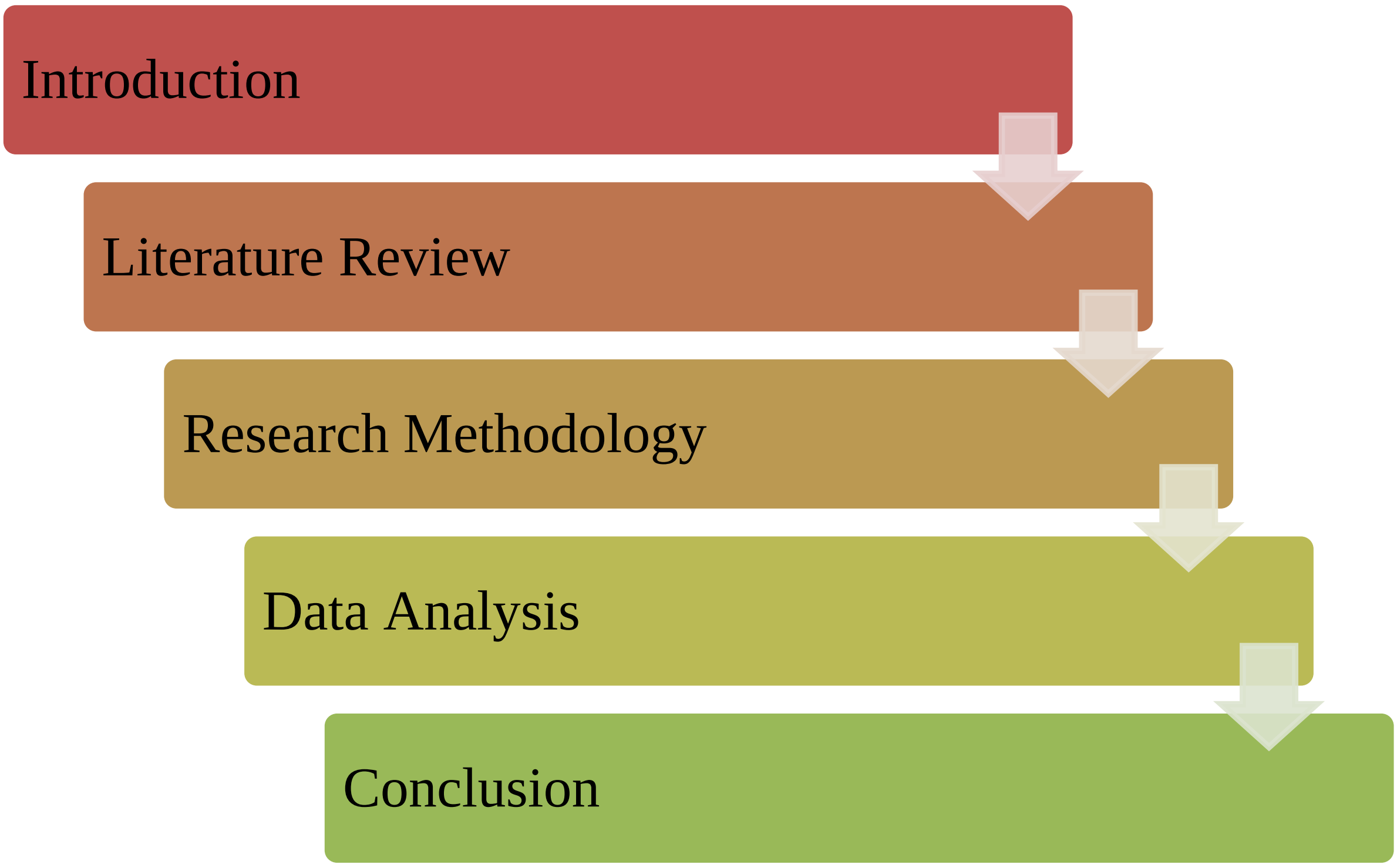

**Figure 1: Research Framework (Primary Source).**

## 1.9 Conclusion:

In summary, this study aims to improve the security and dependability of UAV networks through a thorough analysis of cybersecurity threats and attack simulation. This study intends to promote a more secure and safer environment for the ongoing integration and installation of drones in many sectors by addressing the particular problems given by drone networks and suggesting appropriate mitigation techniques (Pettit, 2020). The security of those networks is still a significant worry as UAVs proceed to transform numerous industries. This study contributes substantially to understanding the cybersecurity threats to UAV networks. This research seeks to enhance the security and confidentiality of unmanned aircraft networks by undertaking attack simulations and developing efficient mitigation measures, enabling the continuous development and responsible usage of drone technology within sectors. UAV network cybersecurity sets the road for the technology's continued growth and responsible adoption, providing various advantages across industries while reducing associated dangers.

# Chapter 2: Literature Review:

## 2.1 Introduction:

The security and accessibility of sensitive data, the protection of vital infrastructure and the avoidance of service interruptions depend on UAV networks' safe functioning. Multiple topics from various viewpoints are presented in journals and books of different types; consequently, the section on pragmatic studies is essential to this strategy. The comprehensive part on theories and models can then be used to carry on with the available literature gaps and produce critical views and interpretations appropriate for this assignment. This section can review specific deficiencies in the study summary crucial for achieving the goal in the following step.

## 2.2 Empirical Studies:

### Theme 1: UAV Network Architecture and Communication Protocols

Unmanned aerial vehicles are becoming more common in various businesses, with uses ranging from monitoring delivery. UAV operations' network structure and communication standards, which enable seamless data sharing and control, significantly impact their effectiveness (Zhang et al. 2019). The essential structure for how various UAVs, ground-based control stations (GCS), and other interconnected components interact inside the system is the UAV network architecture. An automatic node with sensors, computer power, and navigation formats is what a UAV represents. The GCS serves as the main command post and helps with real-time monitoring and control of the UAVs during operations. The communication links between the UAVs and the GCS are essential to the network architecture. (Asif Ali 2023)

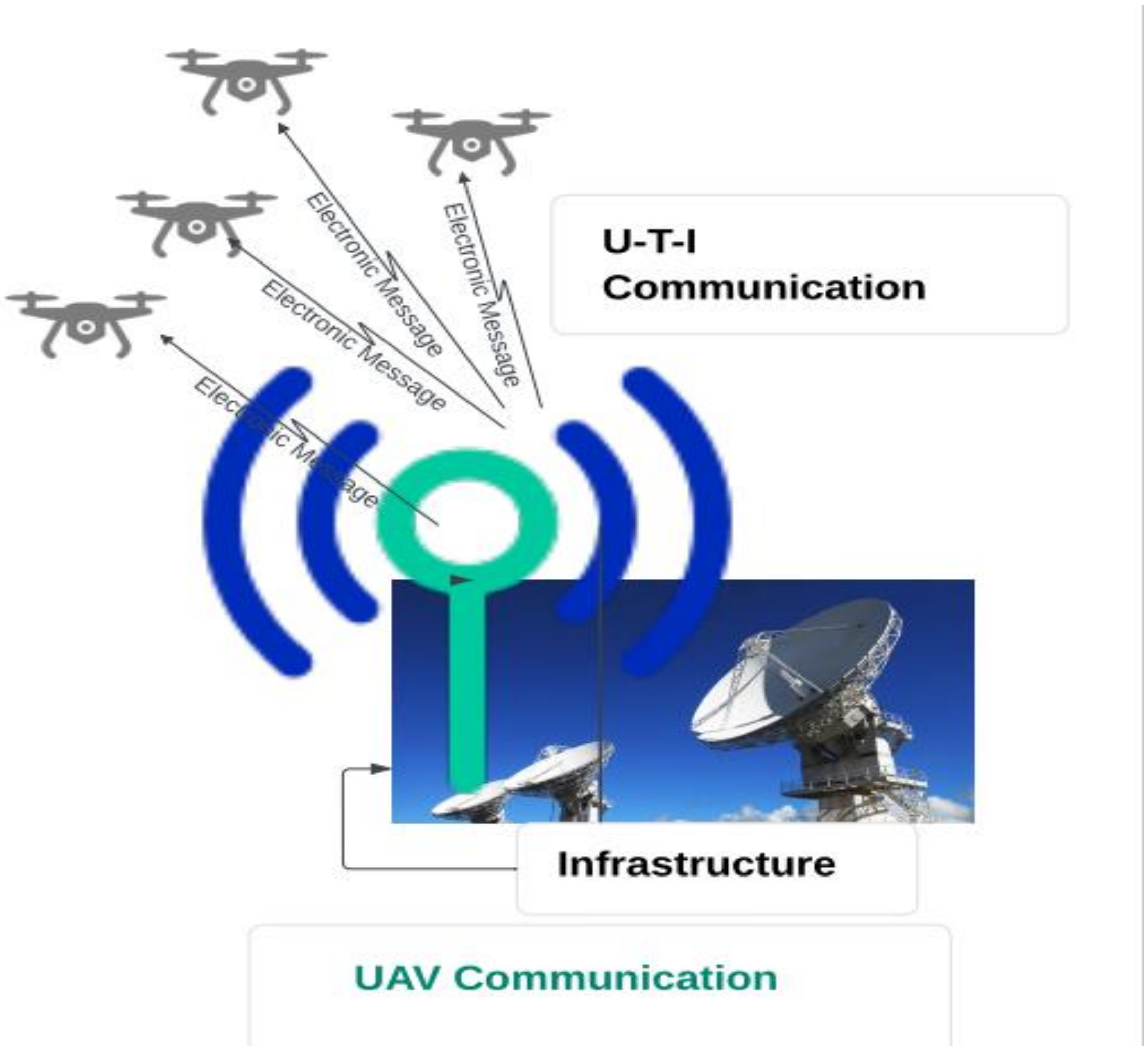

**Figure 2: UAV communication (Primary Source).**

In Figure 2, the communication method shows that these connections may use several mechanisms like mesh networks, Wi-Fi, cell phones, and satellite communication by using the tower infrastructure and drone controlling system. Low latency is vital for UAV networks, especially for tasks that must be completed rapidly, like screening and responding to emergencies. Another crucial component of UAV protocols for communication is reliability (Zhang et al. 2019). To effectively complete mission-critical jobs, it must be keeping a continuous connection with the GCS must be maintained. Data is transmitted safely without loss or damage because of communication protocols. Security is the top priority in UAV networks. Drones are vulnerable to cyber threats; thus, it is essential to safeguard their communication lines to stop unauthorised entry and data tampering. (Jean-Paul 2020)

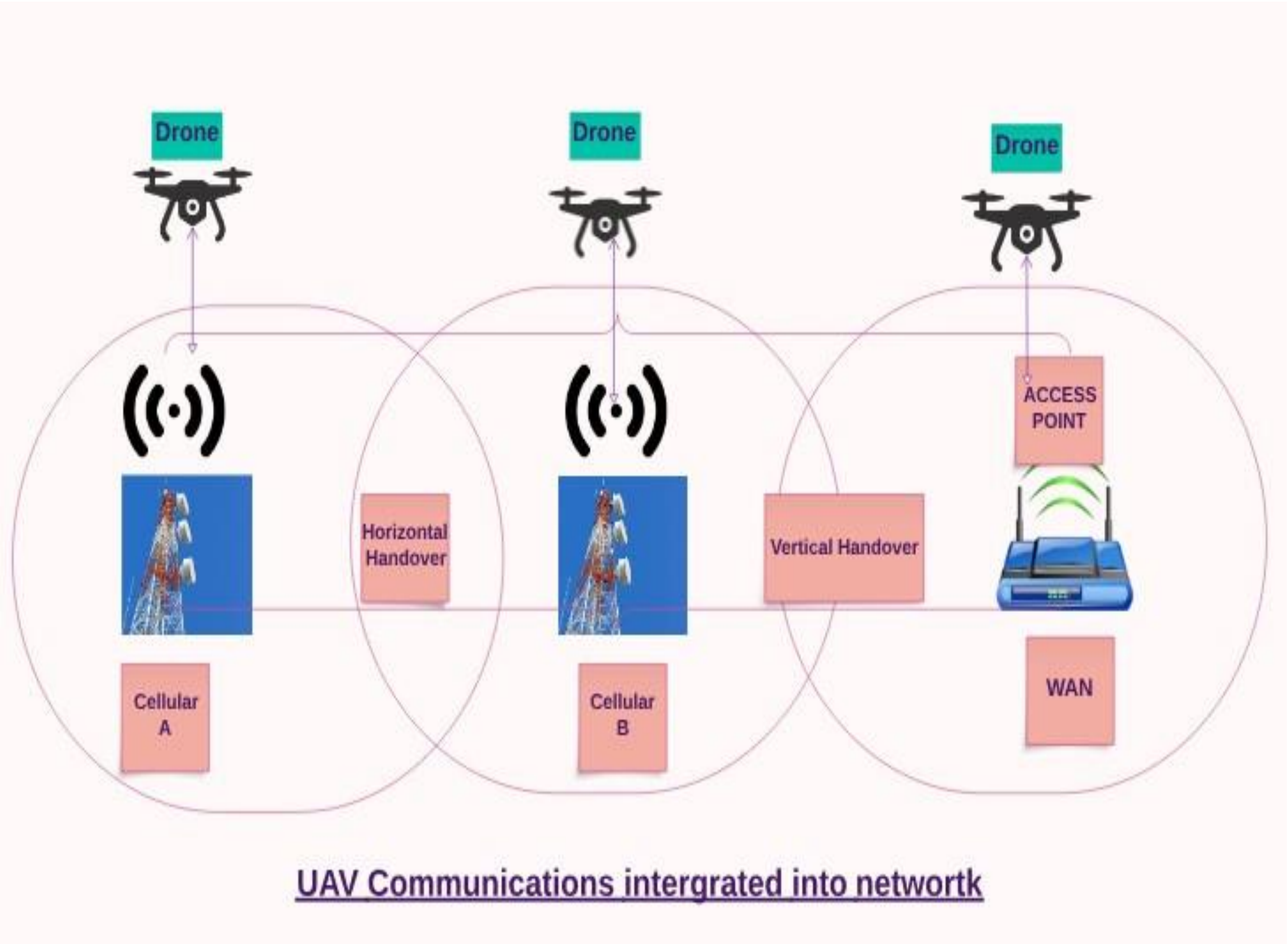

**Figure 3: UAV communication integrated into network traversal (Primary Source).**

Encryption and authentication procedures ensure the security and integrity of the data sent between the UAVs and the GCS. With expanding UAV fleets, scalability is a factor. In order to handle higher volumes of traffic and efficiently manage several UAVs without sacrificing performance, the network design and methods of communication need to be designed (Zhang et al. 2019). Figure 3. Shows an illustration as an example of conversations about using drones in the setting of security measures and UAV technologies. The UAV network structure also includes adequate data storage and processing components. The system is set up to handle and retain this information efficiently because UAVs produce significant sensor data, photos, and videos. UAV operations network architecture and communication strategies used by UAV operations are essential to their success. A well-designed architecture enables effective coordination and cooperation between UAVs and the GCS. (Nguyen MT. 2021)

The extent of the drone exploitation cases is mainly seen in cases of poor drone architecture. There are many cases or factors through which the extent of drone exploitation in UAV networks are assessed in the context of cybersecurity threat analysis, such as:

- **Vulnerabilities**: The drones can have an inherent vulnerability or weakness in the software or hardware parts and the connection process.
- **Attack vector** – Various types of attack vector cases, like the control signal interception or communication channel hijacking (Sergio Ramos, & Cruez, 2021).

**Theme 2: Cybersecurity Threats for UAV Networks:**

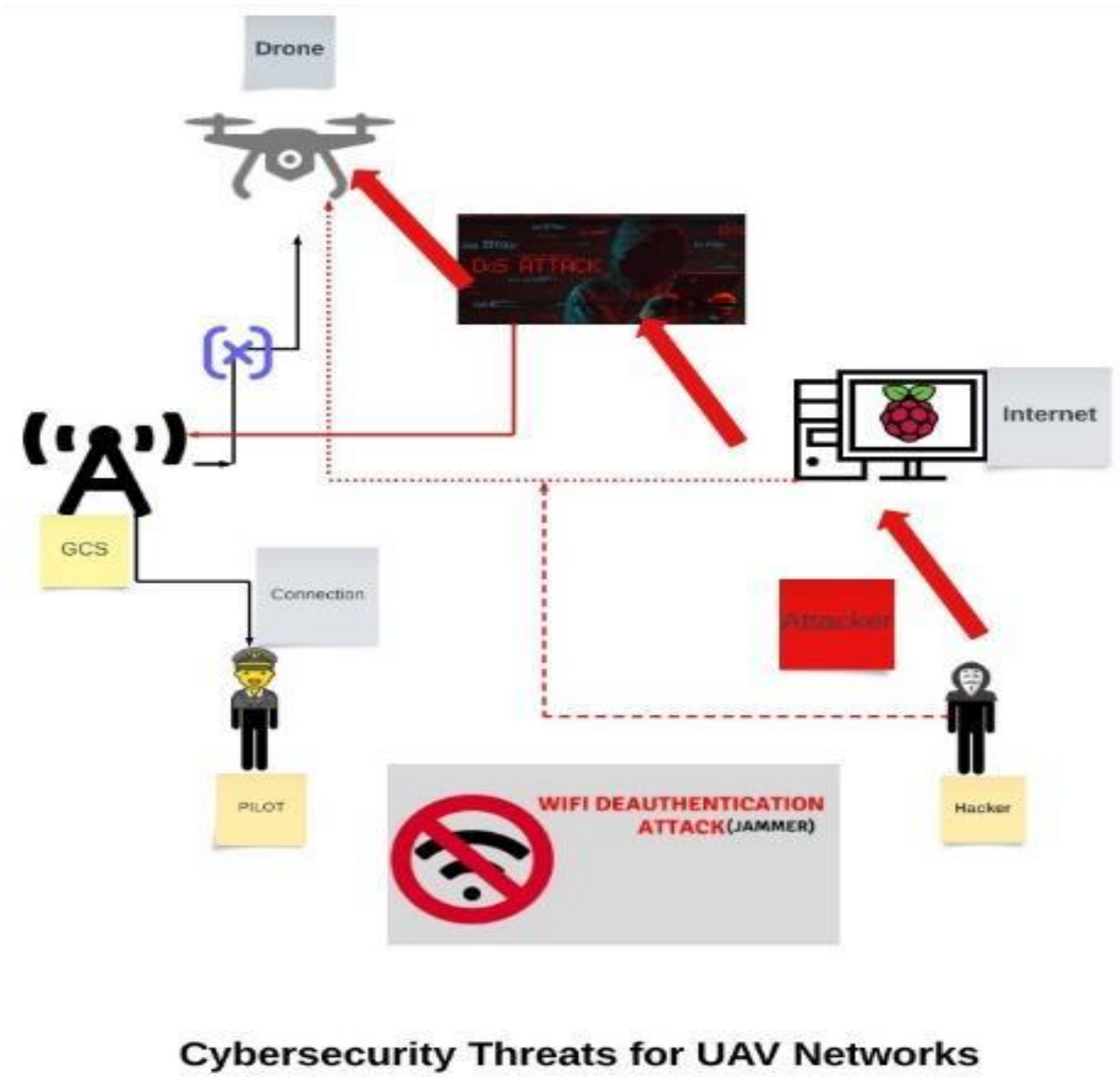

**Figure 4: UAV Deauthentication Attack (Primary Source)**

## De-Authentication Attack

A deauthentication attack is a form of cyber-attack involving removing an established authentication. The deauthentication attack is employed to make the previously enabled user incapable of further utilising the network. Once a station undergoes de-authentication, it loses the ability to access the WLAN unless it repeats the authentication process. (Frank Ohrtman Konrad Roeder. 2003)

Above Figure 4, shows how the UAV infrastructure is made and how this network works. The hacker, displayed in red, is trying to attack the user network and take control of the drone. Unmanned Aerial Vehicles (UAVs) rely on numerous essential communication components for effective operation. These are a few examples of data connectivity (wireless or satellite connections), internal cameras and sensors for data gathering, ground-based control focal points for remote piloting, and software for processing data, and signal transmission. This allows for immediate control and data transfer. UAV networks are susceptible to online dangers that could jeopardise their integrity, data privacy, and overall security (Haider. E. A., 2020). It is essential to comprehend these concerns to create effective mitigation plans and ensure the safe and ethical use of UAV technology. Unauthorised access is one of the leading cybersecurity risks for UAV networks. Malicious actors may try to take over ground-based control stations (GCS) or UAVs without authorization, which could result in drone hijacking. Threat actors might change the UAV's flight route, sensor data, or carrier after gaining unauthorised access, which can have severe repercussions, including unauthorised monitoring or malicious software exploits.

Another serious risk is the intercepting of data and listening in. Large volumes of data, particularly sensor readings, pictures, and video streams, are transmitted by UAVs. If this data is not sufficiently encrypted, cybercriminals may intercept it and gain access to confidential information, jeopardising its privacy and security. Jamming interferes with UAV communication, a cybersecurity risk (Haider et al. 2020). Threat actors can disable or deteriorate the communication channels between UAVs and the GCS by broadcasting interference signals. Operators may be unable to control drones during crucial missions due to this disturbance successfully. The precision and dependability of UAV operations are threatened by data tampering. (Sharma, & Mehra Pawan, 2023)

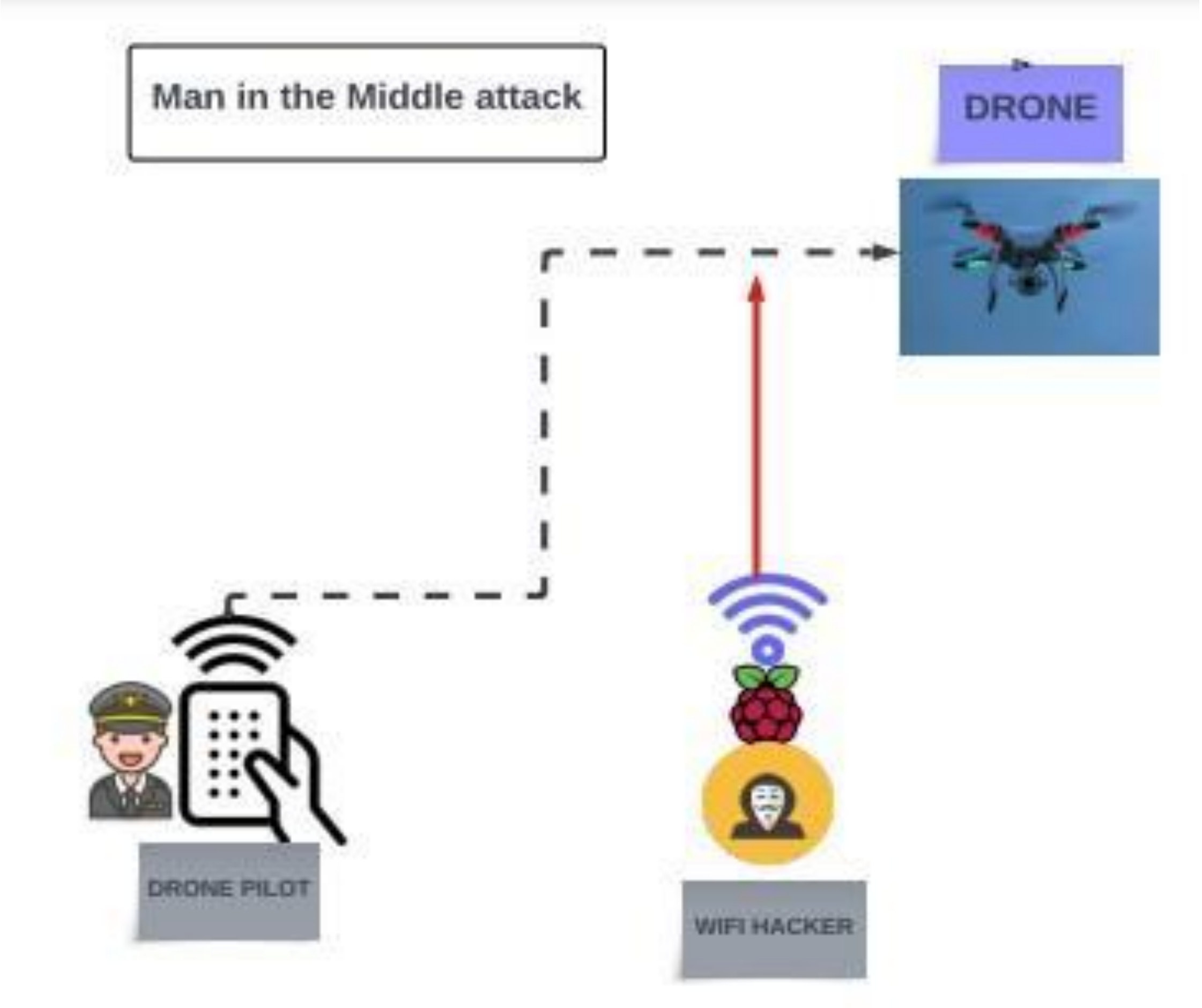

**Figure 5: Man in the middle attack (Primary Source).**

**Man In the Middle Attack**

Man-in-the-middle attacks require two hosts to be convinced that the computer in the middle is the other host to succeed. (Frank Ohrtman, Konrad Roeder 2003).

If attackers can alter the data that UAVs or the GCS collect, this could lead to misleading readings and incorrect judgments, jeopardising the mission's success or causing harm to people and property. Figure 5, demonstrates a malicious attacker attempts to infiltrate an unmanned aerial vehicle, or UAV, drone's operating system. This is by penetrating the drone's software to obtain unauthorised access and control. Denial-of-service (DoS) attacks can affect UAV networks as well. Threat actors who conduct DoS attacks overburden the system's capacity and cause disruptions by flooding communication channels with excessive traffic (Haider et al. 2020).

**Theme 3: Attack Simulation Techniques:**

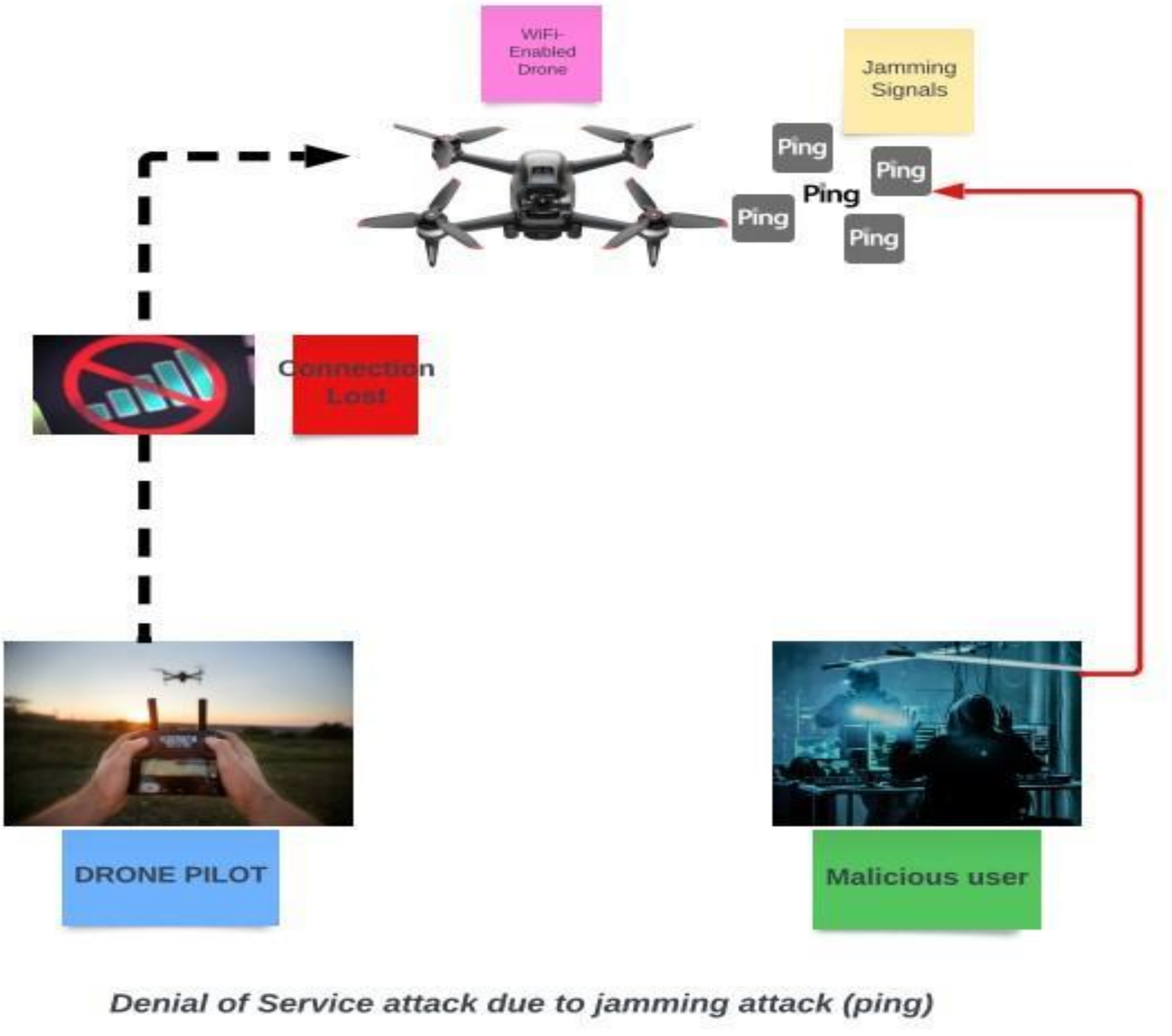

**Figure 6: Attack simulation Service Jamming Attack (Primary Source).**

The hacker might use ARP flooding, ping broadcasts, and Transmission Control Protocol (TCP) SYN flooding to try and use up all available network bandwidth. In this active attack on availability, the intrusive party prevents data from being transferred from the source to the target entities (Frank Ohrtman., & K. Roeder., 2003). (Shown in fig. 6). Utilising attack simulation techniques is essential for locating weaknesses, assessing defences, and enhancing UAV network cybersecurity. Using attack simulation techniques, evaluating a network's resistance to various threats is possible by simulating actual cyber-attack scenarios (De Melo et al. 2021). Researchers and security experts can create effective mitigation measures by simulating these attacks to gather insightful knowledge about the equipment's strengths and flaws. Penetration testing, or ethical hacking, is one of the most popular methods for simulating attacks against networks. This strategy aims to penetrate the network and exploit weaknesses by authorised people or teams. (Dazet E. Francis. 2016)

The attacker can identify network security gaps and suggest remedial actions by successfully breaking into the framework. Simulating particular assault situations is also essential. Researchers can examine the network's reaction and gauge the effect on UAV operations by simulating actual-life variabilities, including denial-of-service (DoS) assaults, data tampering, and cyber security threats. In assault simulation for UAV networks, threat modelling is a crucial technique. It entails detecting possible threats and weaknesses unique to the structure and features of the UAV system. (De Melo et al. 2021).

**Theme 4: Existing Security Measures for UAV Networks:**

The rapid adoption of drone (UAV) systems across various businesses proves that strong cybersecurity measures are essential to protect against potential attacks. To protect UAV networks and make sure they run securely and dependably, several existing security procedures have been created. For UAV networks, encryption is a crucial security component (Nguyen et al. 2021). The security of sensitive data is preserved by encrypting data exchanged between UAVs and ground-based control centres (GCS). By using encryption, it is possible to prevent unauthorised access and to ensure that malicious actors cannot read or use intercepted data. Verifying the authenticity of UAVs and GCS devices requires the use of authentication procedures. Digital certificates and two-factor authentication are frequently used to make sure that only authorised parties can access and manage the UAVs.

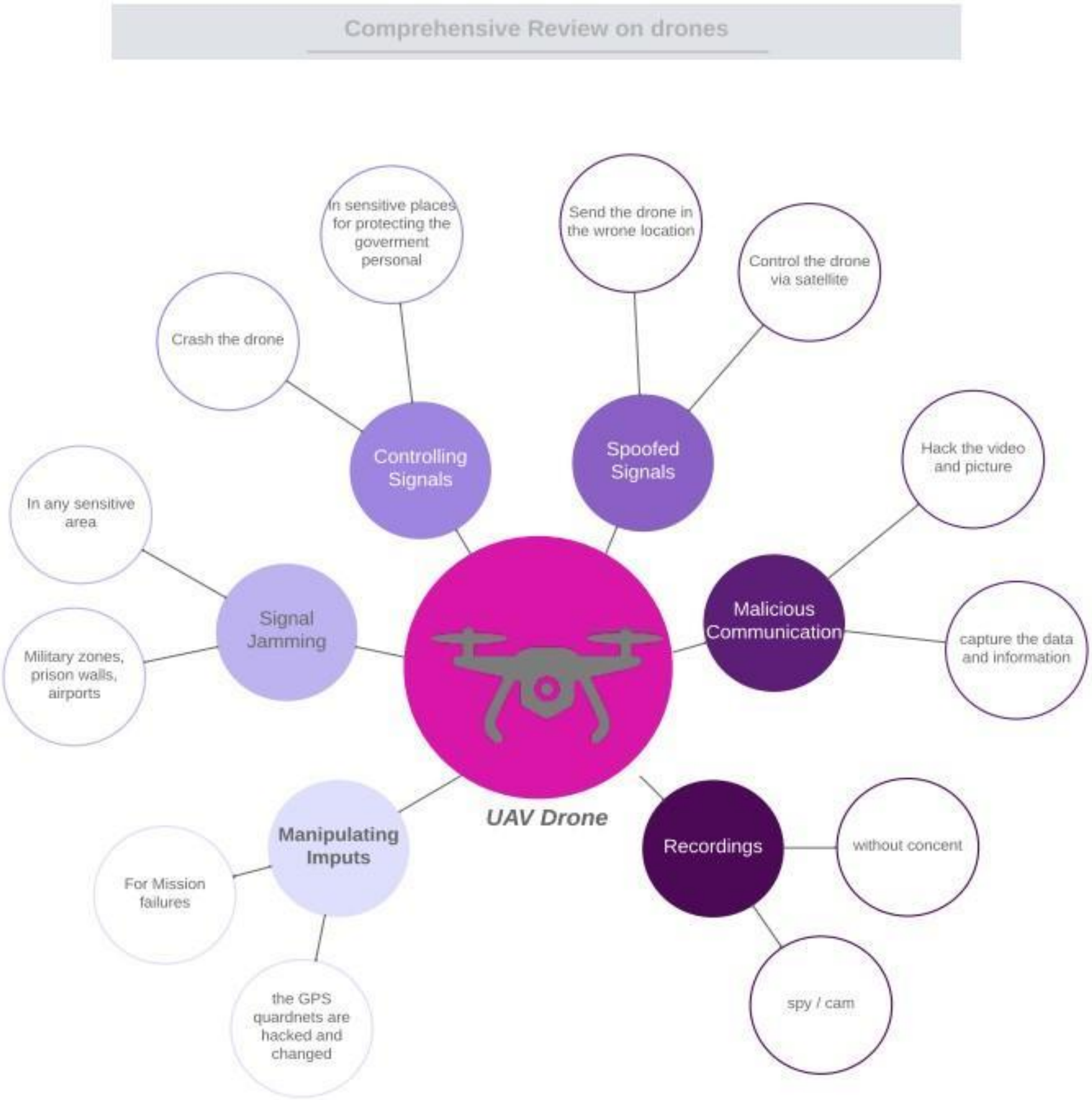

**Figure 7: Comprehensive Review of Drones.**

This precaution shields the network from rogue drones and helps avoid unauthorised manipulation. Segmenting the network is yet another essential security technique for UAV networks. It is possible to lessen the effect of a potential cyberattack by segmenting the network across smaller parts, each with regulated access (Nguyen et al. 2021). The likelihood of a broad breach is reduced by network segmentation, which helps isolate vulnerable components, as shown in Fig 8. In order to guarantee data integrity throughout transmission, secure communication techniques are essential. Secure communication channels are typically established using secure sockets and transport layer security. These protocols use encryption to protect data, check endpoints, and prevent data tampering while in transit. Interference detection and avoidance systems are employed to monitor the UAV system for any threats constantly. (Ben L. End-to-end encryption 2022).

The UAV techniques enhance the network's proactive defence capabilities by analysing network traffic, spotting suspicious activity, and reacting to possible vulnerabilities in real-time. Updates to software and firmware are essential UAV network security measures. Regular upgrades strengthen the system against new threats by fixing known faults and vulnerabilities (Nguyen et al. 2021). The tool used as an example to analyse the network traffic is Wireshark 4.08, which is discussed further with the practical experimentation in Chapter 4.

The steps for stopping the exploitation of drones and the probable vulnerabilities in the UAV networks are as follows:

- **Secured Communication Lines** – Add robust protocol systems for safe data transmission between drones and ground control traffic systems. This process stops unauthorised access of the data decryption and data interception.
- **Authenticated entry** implementation of a robust authentication system process so only authorised members can access the drone system. Multi-factor authentication, like a password or biometric system, can be used for verification. (Ko, Y. Kim J 2021)

## 2.3 Theories and Models:

### Cybersecurity Threat Landscape for UAV Networks:

It has become intensively difficult and concerning for UAV networks in recent years. As the network has advanced technologies and its applications expand, so are the potential vulnerabilities and risks that users can exploit. The excellent characteristics of UAVs have made them targets for numerous cyber threats. Initially, the networks faced the problem of unauthorised control and access to data. The hacker may penetrate the communicative signals or command the system, leading to a primary drone takeover and misusing it (Chaari L., Chahbani S., & Rezgui, 2020). Second, UAV networks have serious dangers from data interception and manipulation. These autonomous vehicles are vulnerable to data theft and espionage because they frequently send vital data, including real-time video updates, location data, and sensor data. The data can be utilised to compromise delicate missions or invade privacy if it is intercepted (Kim H. & Keir G., 2016).

Thirdly, denial-of-service attacks that overwhelm the network's resources or cause communication breakdowns can prevent UAV operations. A UAV's broken communication channels could cause it to lose control, land unexpectedly, or crash, endangering nearby persons and property. Additionally, malware and ransomware assaults may target UAV networks (Tunc, 2021). A hacked UAV could unintentionally introduce malware to crucial systems or join a botnet. Additionally, as UAVs are more thoroughly incorporated into IoT and innovative city ecosystems, they might be used as entry points for cyberattacks on more extensive networks. UAVs could act as entry points to other linked devices and systems if they are not encrypted (Jeaan Paul A. Yaacoub 2023).

**Cyber Threat Intelligence (CTI) for UAV Networks:**

Cyber Threat Intelligence is essential to protect Unmanned Aerial Vehicle systems from constantly changing cyber threats. CTI is the process of gathering, analysing, and disseminating data regarding prospective cyber threats and adversaries to support preventative defence actions. CTI (Cyber Threat Intelligence) in Unmanned Aerial Vehicle, or UAV, networks refer to the gathering, analysing, and using information about cybersecurity dangers and weaknesses that can harm these unmanned flight systems. It is obtaining data, including IP addresses, fields, hashes, and patterns about possible cyber threats, analysing that data to discover hazards and weaknesses, and delivering valuable intelligence to prevent or react to such attacks effectively. The ability to recognise, stop, and respond to cyberattacks can be greatly improved for UAV networks via CTI. This is by evaluating potential flaws in UAVs' software, hardware, and communication protocols. (Alzahrani Et., Al, 2023).

**Attack Surface Analysis for UAV Networks:**

A crucial cybersecurity procedure called an "attack surface analysis" evaluates the potential openings and weaknesses that could allow for the exploitation of a system or network by malevolent parties. Due to its direct ramifications on the privacy of unmanned aircraft networks, Attack Surface Analysis in UAV Systems is a significant part of this study. The attack surface in the setting of drone networks indicates all of the potential locations or entryways via which cyber attackers might exploit weaknesses or launch assaults. An attack surface analysis must be carried out in the unmanned aerial vehicle (UAV) systems framework to find and fix any vulnerabilities that can risk UAV operations' security, authenticity, and security. UAVs rely on various communication routes for remote piloting, data transfer, and command and control (Wang Z., Wu, & S., Zhou 2023).

These channels' security should be assessed using a surface for attack examination to ensure they are appropriately authenticated, encrypted, and guarded against interception, alteration, or unauthorised access. UAV management software and control technologies must thoroughly evaluate their security flaws by reviewing the source code, ensuring no known vulnerabilities, and using secure encryption techniques to prevent future exploits. (Javier S., & Ronny C., 2016). A crucial factor to consider is the actual safety of UAVs and the framework that supports them. An attacker who gains physical entry to a UAV may modify its parts, introduce malicious hardware, or take private information. (Yaacoub J. P, 2020).

**Threat Modelling for UAV Networks:**

Threat modelling, a systematic technique to detect, analyse, and mitigate potential cybersecurity threats in a network, is critical to improving the safety infrastructure of Unmanned Aerial Vehicle (UAV) networks. The initial phase needs a thorough inventory of the UAV network components, which includes UAVs, ground control stations, communication lines, data storage networks, and sensors (Ahmad Yazdan, W. Sun. 2012).

Following that, it is critical to categorise and prioritise threats based on their potential impact on network security, functionality, and data integrity, allowing for the development of tailored mitigation measures for each discovered threat (Javaid, Ahmad, 2022). This could include the use of strong authentication and encryption techniques, as well as regular software upgrades, network segmentation, and tight physical security standards.

## 2.4 Literature Gap:

The body of knowledge regarding Cybersecurity Threat Assessment and Attack Modelling for unmanned aerial vehicles (UAVs) is growing. However, there is still a substantial knowledge gap regarding cyber threats' thorough and practical effects of cyber threats on these systems. While some studies have examined the weaknesses and possible vectors of attack on UAVs, the extant literature frequently concentrates on certain attack types or isolated cybersecurity issues. The lack of thorough studies that cover a wide range of potential hazards and weaknesses that UAVs may encounter in various operational circumstances is the root cause of the literature gap. A holistic approach is essential as UAVs are rapidly incorporated into vital infrastructure, commercial uses, and military activities. Additionally, there is a need for additional empirical studies that use data from the real world to verify theoretical models and simulations.

Using UAVs in controlled situations for practical experimentation can offer essential insights into the possibility and efficacy of cyberattacks, assisting in creating effective defences and defensive tactics. Furthermore, the research frequently ignores the possible knock-on implications of UAV cyberattacks, such as the compromise of larger, interrelated systems or the effect on general security and confidentiality. Considering the more significant impacts of UAV cyber risks might help create more effective and flexible security systems. Future research should concentrate on undertaking thorough risk evaluations and assault scenarios that consider real-world circumstances and investigate the cascading impacts of UAV cyberattacks to close this gap.

## 2.5 Conceptual Framework:

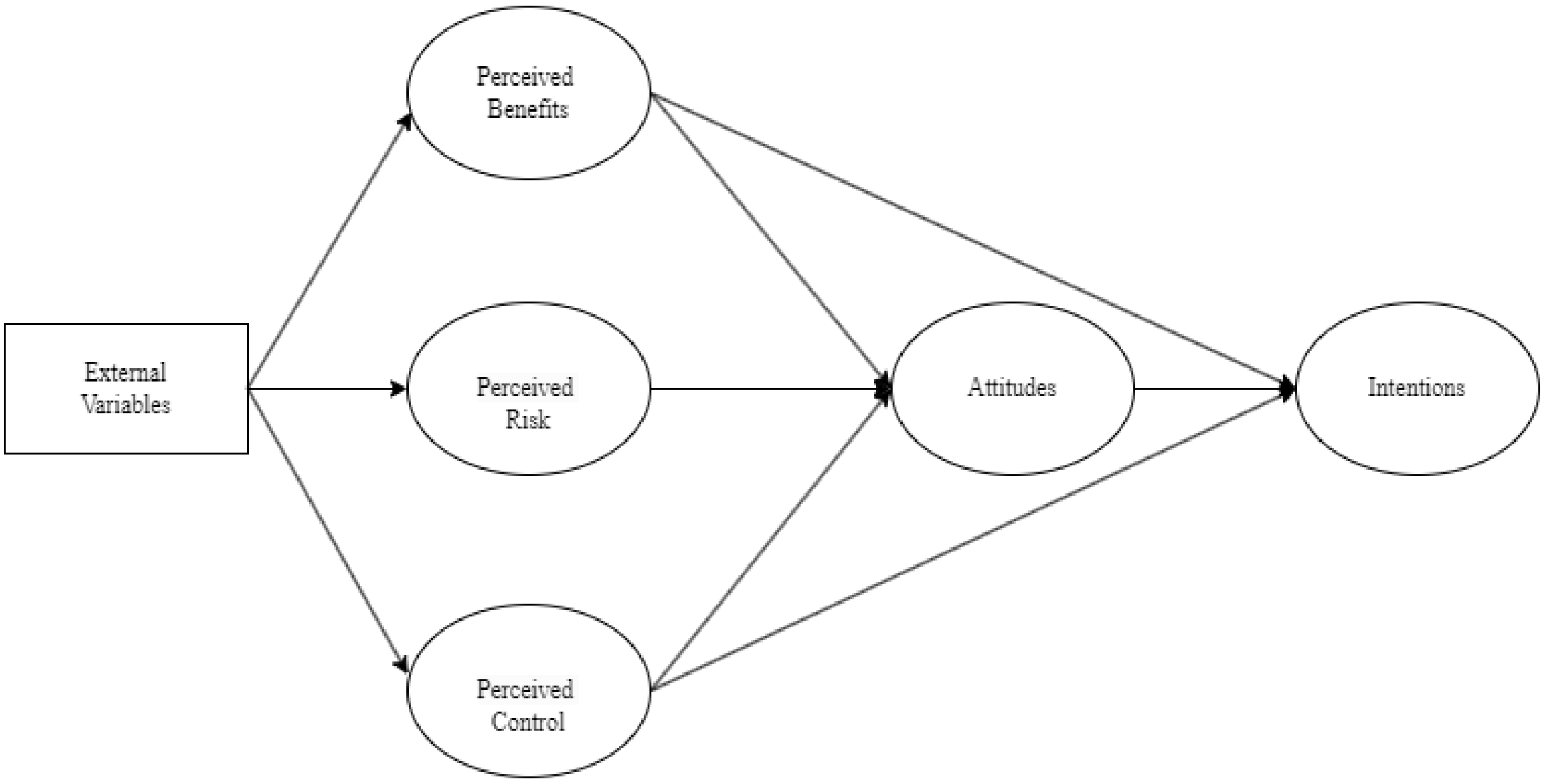

**Figure 8: Conceptual Framework (Primary Source).**

This conceptual framework provides a structured way to examine the factors that influence the decision-making and actions of individual organisations involved in UAV network operations in cybersecurity.

- External Variables: Benefits associated with using UAV networks
- Perceived Benefits: Surveillance, delivery, or data collection
- Perceived risk, the risk and threats concerns
- Perceived control: the ability of individuals or organisations to mitigate and control cybersecurity threats.
- Attitudes: The intention of responding to cybersecurity threats
- Intentions: the decision-making and action taken

## 2.6 Conclusion:

Real-time replies are ensured, and UAV operations are protected from cyber threats thanks to dependable and secure communication protocols. Cybersecurity concerns for UAV networks significantly hinder the legal and safe deployment of drones across a variety of businesses. Researchers can find vulnerabilities, hone defence mechanisms, and create efficient ways to protect UAV networks from new cyber threats by using hacking, red teaming, scenario-based tests, threat modelling, and CTF competitions. These methods, which range from network division and intrusion detection to encryption and authentication, work together to improve the resilience and safety of UAV operations.

# Chapter 3: Research Methodology:

## 3.1 Introduction:

Modern unmanned aerial devices deployed against unmanned aerial vehicles (UAVs) mainly depend on radio commonness jammers and refusal-of-assistance attacks against adversary, accomplishes this through RF jamming or radio connection resistance. (Martin Broomhead, 2023) However, this approach does not just go against well-conventional tactics, methods, and processes for possible cyber-attack. The research studies the new enhancement necessary to be utilised to protect drones against cybersecurity threats from malicious hackers.

## 3.2 Method Outline:

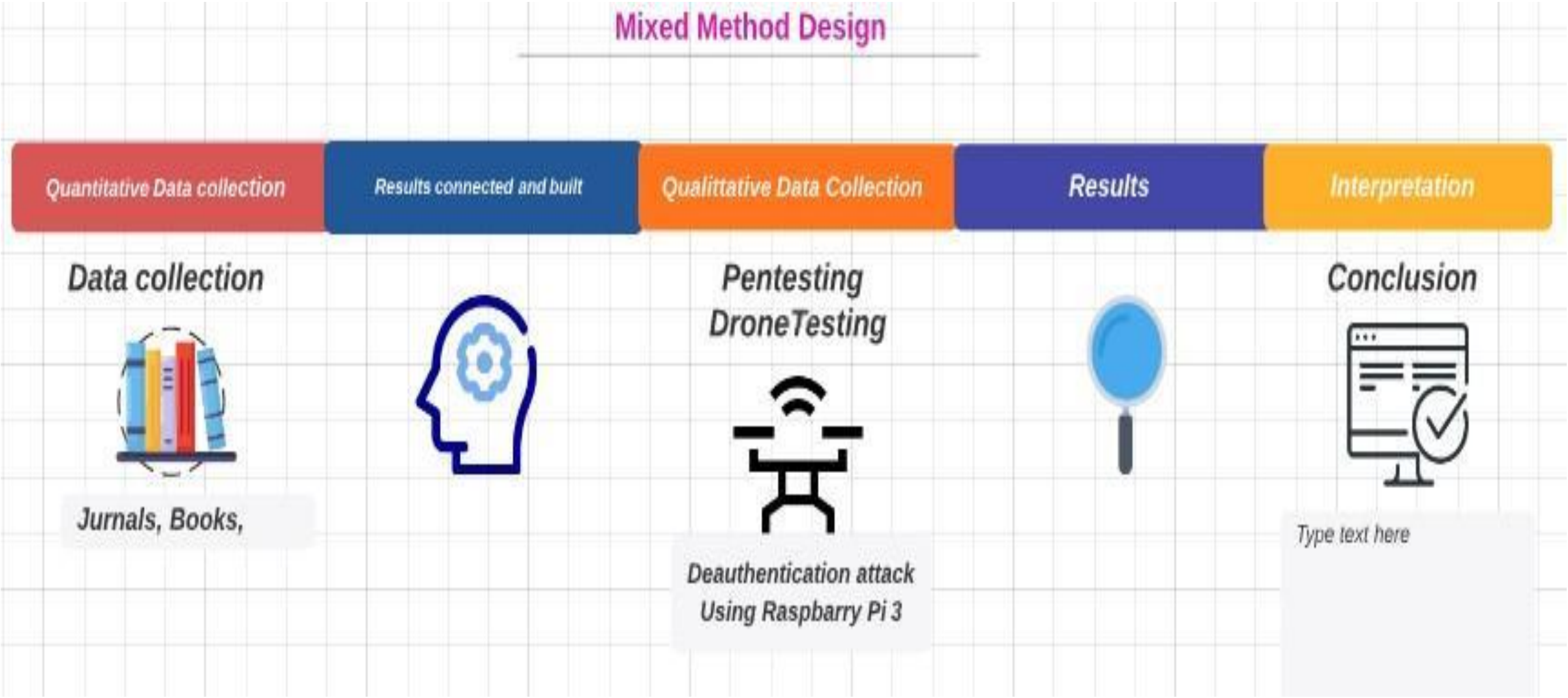

**Figure 9: Mixed method design for dissertation topic (Primary Source).**

The framework of this study will involve mixed-method research. Qualitative data by conducting a practical experiment focused on penetration testing of unmanned aerial vehicles (UAV) and secondary quantitative data research, as shown below in Fig. 1. Incorporating mixed techniques enhances the comprehensiveness of research by providing a more comprehensive description of the chosen study topic. (Clark, Tom, et al. 2021).

The paradigm approach is to study the essential and relevant data. (Kuhn’s. 1970) Here, the method outlined traces all the analysis section phases crucial to the methodology chapter. This chapter's experimental, qualitative strategy, design, and other sections are compiled adequately per the dissertation's main topic.

## 3.3 Research Philosophy:

The study's approach is based on the hypotheses of several academics and includes simulations of attacks and threat analysis for UAV objectives. The analysis of the examiners' philosophical approaches and the promotion of the research's goal. The idea is to render the principal philosophy case simpler to recognise the individual components of the various formats. The most important aspects of the entire research work depend on the research philosophy. The process alters the overall framework of the investigation by including trusted and reliable research approaches. In this study, two distinct research methodologies are adopted.

## 3.4 Research Approach:

- The first step is a thorough literature review to identify the threats and prospective attack vectors against UAVs.
- Vulnerability Assessment: The next stage is to use attack simulations to detect potential system weaknesses in UAVs.
- Following the simulations, the outcomes are critically evaluated to understand the potential implications of different attacks and recommend areas for improvement.
- Security Enhancement: Recommending and evaluating security measures and improvements based on the investigation findings to mitigate detected vulnerabilities and improve the overall cybersecurity posture of UAVs.

## 3.5 Research Design:

This is not a programmatic approach but it is important to also consider. The armed forces, governments of states, federal departments, and private businesses use existing C-UAS employing techniques that require significant amounts of energy to operate. In settings where operational machines use radio frequency communication for conversation, such as an army air base, a prominent firm event, or any place in a busy metropolitan area, some UAS approaches, such as frequency jamming, may not always be appropriate (Alkadi Et. Al. 2022). The study in this paper might be helpful to the government military air force, airports, UAV manufactures, universities computer research facility.

## 3.6 Data Collection Method:

This study examines UAV network cybersecurity infrastructure using the literature to identify known and unknown vulnerabilities. Our technique examines known concerns from relevant studies and exploratory analysis to find new vulnerabilities.

## 3.7 Research Method:

- **Stage 1: Issue Identification and Compilation**
    - Our list of OSI structure vulnerabilities in UAV networks will be based on the survey of cyber security threats for UAVs. (Kai Yun T., Vassilios G, 2022) & This stage will comprise a thorough literature analysis to identify security vulnerabilities and network communication interferences.

- **Stage 2: Penetration Testing**
    - Our approach involved the systematic execution of unauthorised root access using a Raspberry Pi 3 in conjunction with an Alpha Network adaptor. Employing penetration testing methodologies to conduct in-depth analysis and discover previously discussed vulnerabilities that have been addressed by theory only.
    - Installing the Alpha Network drivers is an important step in the pen-testing process. Refer to Appendix. This is how we enable 'Monitor mode'.
    - Open-source tools will also be used such as – Aircrack-ng and Wireshark.

- **Stage 3: Conclusion**
    - The method is to provide a comprehensive and up-to-date analysis of the UAV networks' cybersecurity environment by carefully addressing known and

potentially new vulnerabilities. The final objective is to develop suggestions to help UAV networks be defended against new cyber threats, thus improving overall reliability and security.

## 3.8 Research Ethics:

Drone technologies are rapidly in demand for a variety of uses, including both personal and professional ones. "Unmanned aerial vehicles" (UAVs) and drones are types of aircraft remotely operated without an in-flight pilot. It utilises the basic data link technique. Radio waves connect the drone to a controller on the ground and communicate (Vanitha and Padmavathi, 2021). The appraisal of the data-gathering procedure is the project's moral quandary. In order to manage the survey answers as needed, the requirement for information analysis may be supplied by several activities, data-collection publications, and genuine research. The implementation of regulations and the data gathering technique are essential elements in gaining approval of varied locations. The study may be easily investigated because non-classified information is employed to obtain and verify organisational information and strategies.

## 3.9 Professional Issues:

Important to note that the hacking experiment is only for educational purposes and must not be used for any malicious attacks. The drone vulnerabilities research and experiment will not be repeated and was intended for research purpose only. Confirming to the University ethics review form signed and completed.

The ethical dilemma of data collecting is an essential component of the research on drones. This project addresses a significant requirement for information analysis to be supplied by a number of activities, data-collection publications, and genuine research. This will ultimately promote ethical consistency in the procedures of acquiring data.

This research aims to adopt a systematic methodology, the study may be easily investigated because information is easily accessible to obtain and verify. By doing so, it aims to avoid any potential legal consequences and promote a secure, ethical, and robust framework for the implementation of drone technologies in diverse scenarios.

## 3.10 Research Limitations:

- **Research Capabilities**
  To fully evaluate the applicability of the results of this research, it is necessary to understand the following limitations of this study.

- **Sample Variety**
  The study was limited to a single drone model, which may have limited the results' applicability to other UAV systems currently on the market. As shown in Appendix 1 and 2. The other drones that were used malfunctioned because of the level of quality.
- **Resource Limitations:**
  The research could only cover a more limited set of potential vulnerabilities due to a lack of time and money, which prevented it from providing a comprehensive picture of the UAV cybersecurity landscape.

- **Access Restrictions for High-End UAV Systems:**
  The findings are limited to the characteristics of commercial drone systems due to the inability to access and test military or government-grade UAV systems, which excludes insights into the security standards and protocols enforced in high-security situations.

## 3.11 Project Planning and Management:

Comprehensive project planning and preparation is clearly outlined in a Gantt chart, found in Appendix 8, which lays out the overall dissertation plan and preparation along with the respective time scales.

## 3.12 Summary:

The initial method of comprehending these issues was structured. Starting with an in-depth look into other researchers' findings for variabilities. The challenge faced during this research was that only one drone model was tested, which limited the results. Considering ethical concerns, we conducted our research correctly and within bounds, and the experiment was conducted in a controlled environment.

# Chapter 4: Experiment of the UAV (Drones):

## 4.1 Analysis of Drone Security: An Exploration of Hacking a DJI Drone:

The rapid growth of the drone industry has underscored the immense potential drones offer for various military and civilian applications (Hayes et al., 2014). This study aims to meticulously analyse the contemporary landscape of drone security, focusing particularly on vulnerabilities associated with Wi-Fi-enabled drones. In this section, we delve into the communication links of these drones and reveal potential threats, notably highlighting the "De-authentication method." Our approach involved the systematic execution of unauthorised root access using a Raspberry Pi 3 in conjunction with an Alpha Network adaptor.

The De-authentication method functions by disabling the access privileges of previously authorised users, effectively preventing their continued utilisation of the network. This category of action involves the removal of the existing authentication process. Once a station experiences de-authentication, its ability to access the WLAN is suspended until the authentication process is successfully completed again (Ohrtman, 2018).

The application of drones in modern warfare has proven to be versatile and integral across a spectrum of combat and military operations. This is strikingly evident in the ongoing conflict in Ukraine, where drones have been employed in tandem with civilian populations. However, the accessibility of this technology also poses a potential threat for nefarious purposes. Notably, terrorist organisations find drones to be an attractive option, thereby raising significant security concerns (Clusters B. 2018).

The primary objective of the conducted experiment is to demonstrate the feasibility of hacking into drone Wi-Fi systems, thereby exposing the vulnerabilities inherent in such attacks. As aptly pointed out by Bart Custers, "Most of the security infrastructures that are aimed at limiting access to sensitive locations are ineffective in circumventing access by drones" (Custers B, 2018).

### Equipment Utilised for the Investigation

For the purposes of this research, the following hardware was employed:

- Tello Drone
- Raspberry Pi 3B
- Alpha Network AWUS036NHA Adapter
- 32GB MicroSD Card
- Power Supply: Either electric or a 1000-amp battery.

## 4.2 DJI-Powered Drone:

For this investigation, the DJI-powered drone was chosen over other alternatives due to its superior quality and stability, which are crucial for the experiment's accuracy and consistency. The controller facilitates a connection with the drone by leveraging its open Wi-Fi protocol. Subsequently, the user can establish a direct link with the drone via their smartphone.

The Ukrainian government denounced DJI last spring because Russian armed forces were utilising DJI drones for missile targeting and exploiting Ukraine's DJI drones' radio transmissions to pinpoint Ukrainian military personal. (Andy Greenberg., 2023,)
In addition to the DJI-powered drone, other models were evaluated for the experiment:

- Parrot Drone 2.0: This drone was found to be unstable during trials. Furthermore, its battery tended to overheat, emitting a concerning burning odour during operations. [Refer to Appendix 1]
- SIMREX X500: This more economically-priced toy drone demonstrated significant instability upon testing. A key limitation of the SIMREX X500 was its susceptibility to RC remote control interference. This posed a challenge in conclusively proving the execution of a de-authentication attack technique, given the drone's inherent vulnerability to such remote controls. [Refer to Appendix 4]

In light of these findings, the DJI-powered drone was deemed the most suitable for the objectives of this research.

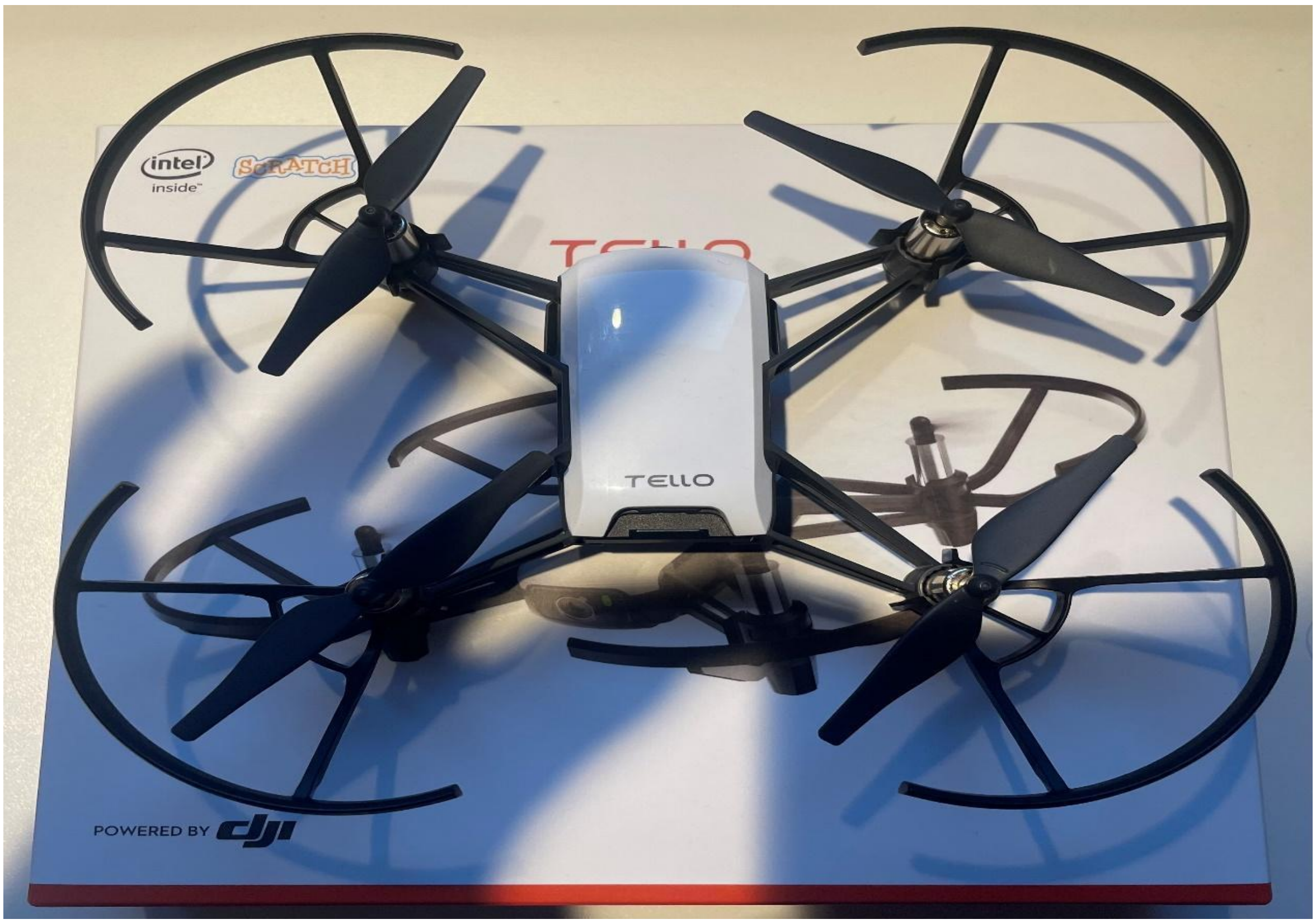

**Figure 10: Tello Drone by DJI (Primary Source).**

**Specifications**

| Feature | Details |
| --- | --- |
| Weight | Approx. 80 g (including propellers and battery) |
| Dimensions | 98×92.5×41 mm |
| Propeller | 3 inches |
| Built-in Functions | Range Finder, Barometer, LED, Vision System, 2.4 GHz 802.11n Wi-Fi, 720p Live View |
| Port | Micro USB Charging Port |

https://docs.alfa.com.tw

## 4.3 Alfa Network AWUS036NHA

The ALFA Network AWUS036NHA was used because it is a popular wireless adapter known for its capabilities in wireless penetration testing. Here are some of its advantages:

1. **High Sensitivity:** One of the most well-regarded features of the AWUS036NHA is its high reception sensitivity, which allows it to detect weaker signals that other adapters might miss.
2. **Compatibility with Aircrack-ng Suite:** The AWUS036NHA is known for its compatibility with open-source software i.e., Aircrack-ng and Github, making it a favourite for penetration testers.
3. Monitor Mode and Packet Injection: Crucial for wireless penetration testing, the AWUS036NHA can be easily set to monitor mode and supports packet injection, facilitating various network tests.

One of the primary misconceptions about 802.11b and other wireless protocols is that their range is confined to a mere 100 meters. It is a verifiable fact that through the implementation of appropriate engineering techniques, the 802.11b wireless communication standard has the capability to establish point-to-point connections reaching distances exceeding 20 miles. (Frank Ohrtman, 2018)

The only disadvantage with the Alpha network is that it is not plug-and-play with the Raspberry Pi-3. Open source software needs to be manually downloaded to configure the drivers onto the Raspbian operating system. Refer to Appendix: 5.

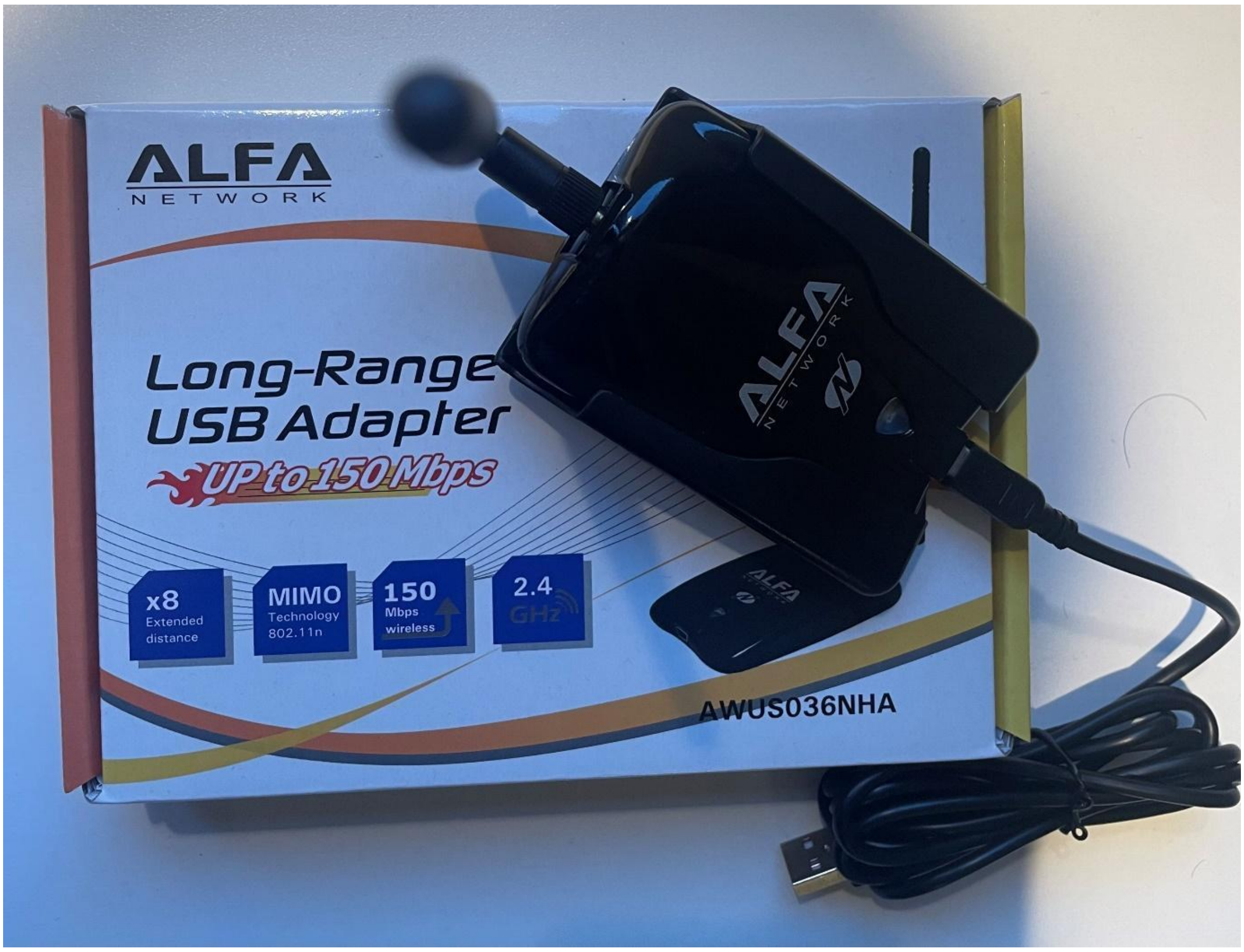

**Figure 11: Alfa Network AWUS036N (Primary Source).**

**Specifications Alfa AWUS036-NHA**

| Parameter | Specification |
|---|---|
| Chipset | Atheros AR9271 |
| Wireless Standards | IEEE 802.11b/g/n |
| Frequency | 2.4 GHz |
| Data Rate | - 802.11n: Up to 150 Mbps |
| Security | WEP 64/128, WPA, WPA2, TKIP, AES |
| Supported OS | Windows XP/Vista/7/8/8.1/10, Linux, macOS |
| Compatibility | Kali Linux, Backtrack, and other penetration |

Alfa Network long Range User manual https://www.manualslib.com/manual/2437578/Alfa-Network-Long-Range.html

### 4.4 Raspberry Pi 3 Model B

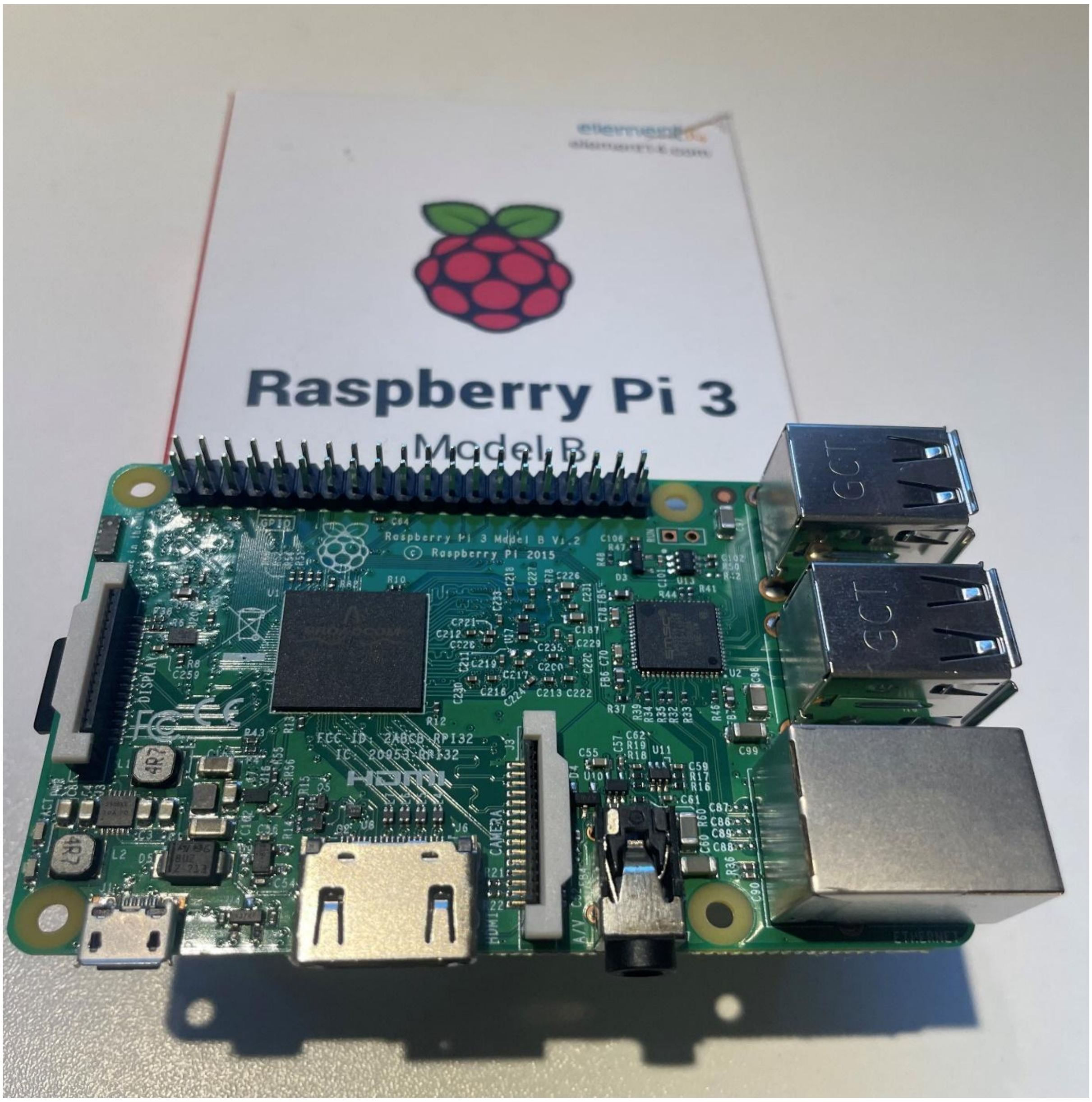

**Figure 12: Raspberry Pi 3 Model B (Primary Source).**

The Raspberry Pi 3 is a single-board computer with wireless LAN and Bluetooth functionalities. There are other intriguing applications for a Raspberry Pi; however, the primary motivation behind obtaining a Raspberry Pi for this project was to explore the potential for exploiting drone vulnerabilities (Pasolini, Bazzi & Zabini, 2017).

**Specifications**

| Feature | Specification |
|---|---|
| CPU | Quad-core 1.2GHz Broadcom BCM2837 64bit |
| GPU | Broadcom Video Core IV |
| RAM | 1GB LPDDR2 |
| Storage | microSD slot |
| Wireless | Wi-Fi 802.11n, Bluetooth 4.2 |
| Ethernet | 10/100 Ethernet |
| USB Ports | 4 x USB 2.0 |
| Video Outputs | HDMI, 3.5mm analogue audio-video jack |
| Audio Outputs | HDMI, 3.5mm analogue audio-video jack |
| GPIO | 40-pin |
| Power Source | 5V/2.5A DC power input |
| Dimensions | 85.6 x 56.5 x 21mm |

https://www.raspberrypi.com/documentation/

**The software utilised in this experiment.**

- **aircrack-ng**: I utilised the open-source software called aircrack-ng to find the drones BSSID and MAC address by switching the network adaptor into ‘monitor mode.’
- **aireplay-ng**: This is another open-source program, used to perform a de-authentication attack on the targeted drone.
- **Raspberry Pi OS**: Also known as Raspbian, is the operating system used.

## 4.5 Attack Brief

The selected attack strategy involves the utilisation of de-authentication techniques, as illustrated in Figure 1. This type of assault aims to disrupt the connection between a pilot and their drone (Pasolini, Bazzi, & Zabini, 2017). For instance, an attacker could exploit this method when attempting to wrest control of the drone from its rightful pilot. The consequence of executing such an attack is that the drone would come to an abrupt halt mid-flight (Pasolini et al., 2017). Subsequently, the attacker can swiftly establish a connection to the immobilised drone using a device, such as an iPhone running the drone control program. During this phase, even as the legitimate drone pilot maintains their connection, the attacker's system, based on a Raspberry Pi, can link up with the drone. It is noteworthy that identifying the Media Access Control (MAC) address of the drone's controller becomes relatively straightforward due to the attacker's presence on the shared network.

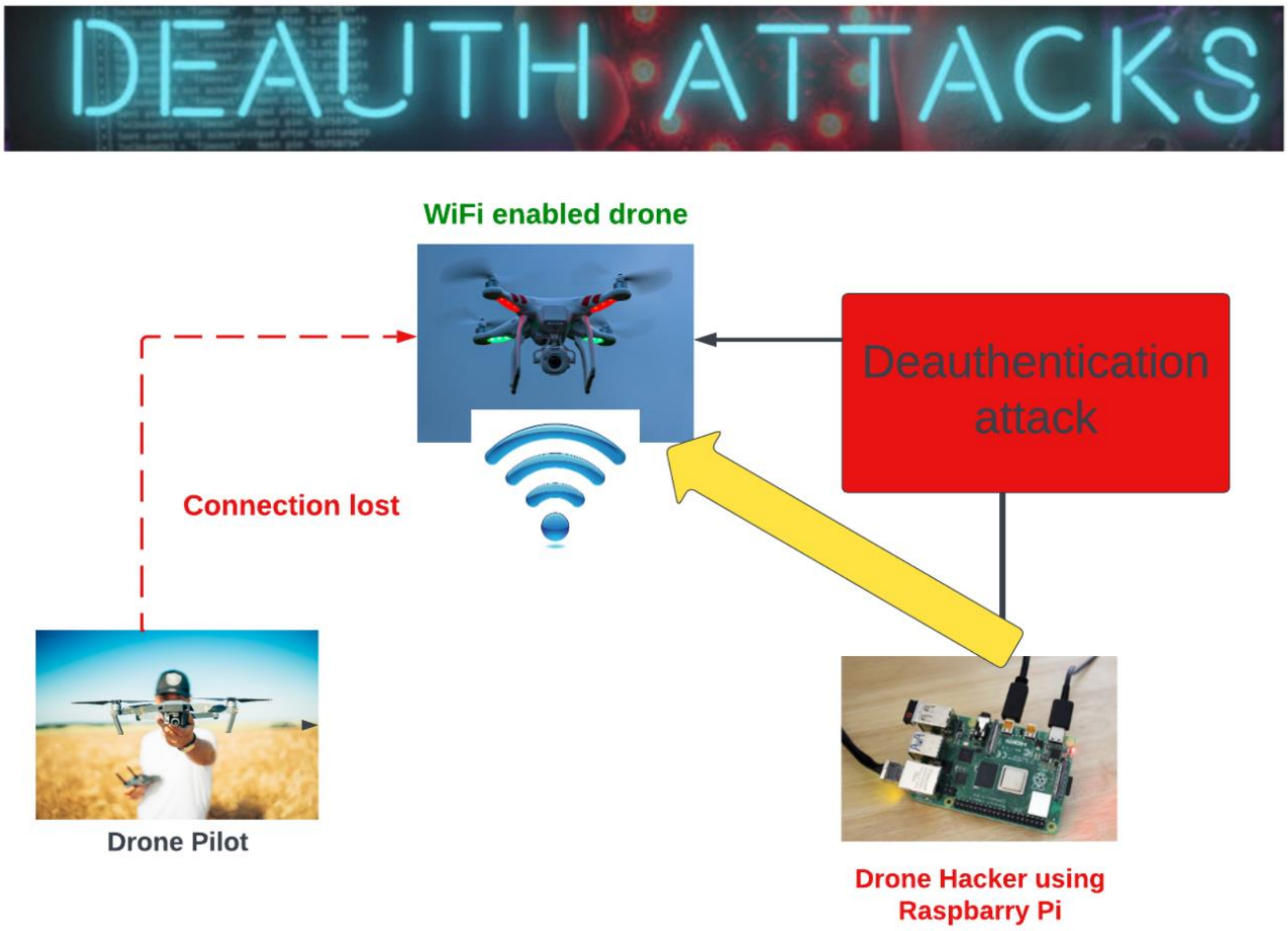

**Figure 13: Model of the DE authentication Attack (Primary Source).**

The execution of the deauthentication attack hinges on employing the Aireplay-ng module, a pivotal component within the Aircrack-ng suite. Aireplay-ng is distinguished for its proficiency in frame injections, a capability that proves critical for the success of this operation.

Throughout the execution of this operation, both the MAC address of the targeted drone and the MAC address of the designated client will be readily discernible (Pasolini et al., 2017). This level of precision underscores the effectiveness of this methodology, ensuring a focused and efficient de-authentication process. As a result, the risk of unintended disruptions is significantly minimised.

**Drone Penetration Testing Report**

This report delineates the systematic approach undertaken during our drone penetration testing procedures, leveraging a selection of tools to identify and potentially exploit vulnerabilities within drone communication systems.

**4.6 Testing Methodology:**

The testing was executed in a controlled environment, ensuring the non-interference of external networks. Below are the detailed steps that were followed:

1. **Environment Check:** Utilised `iwconfig` to assess the current wireless interfaces and their configurations.
2. **Tool Installation:** Installed the `Aircrack-ng` suite, a renowned toolkit for wireless network assessments, using the command `apt install aircrack-ng`.

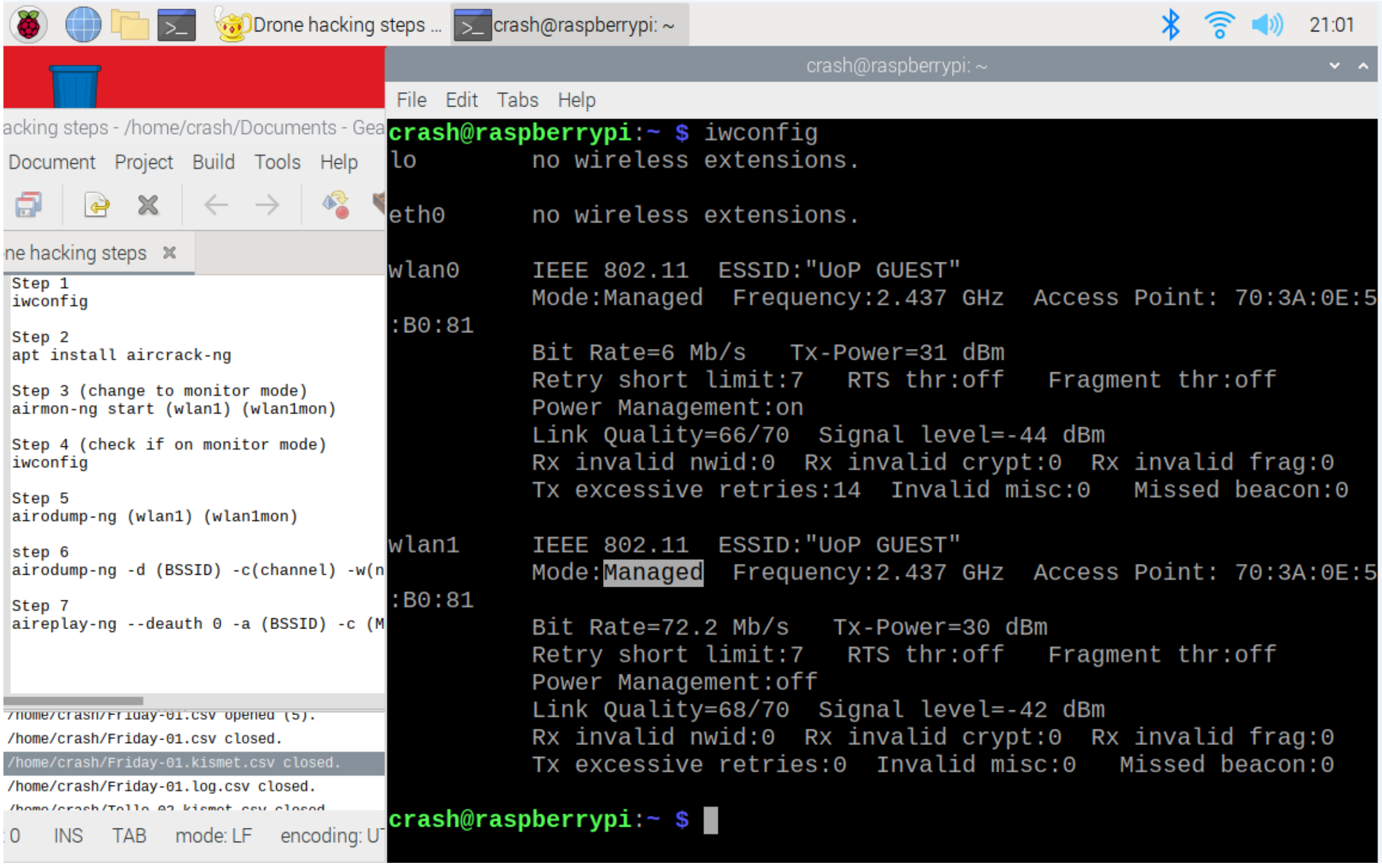

**Figure 14: Command `iwconfig` shows wlan1 and wlan0 in Managed mode (Primary Source).**

3. **Switching to Monitor Mode:** Transitioned the network interface to monitor mode using the command `airmon-ng start wlan1`, resulting in the new interface designation `wlan1mon`.

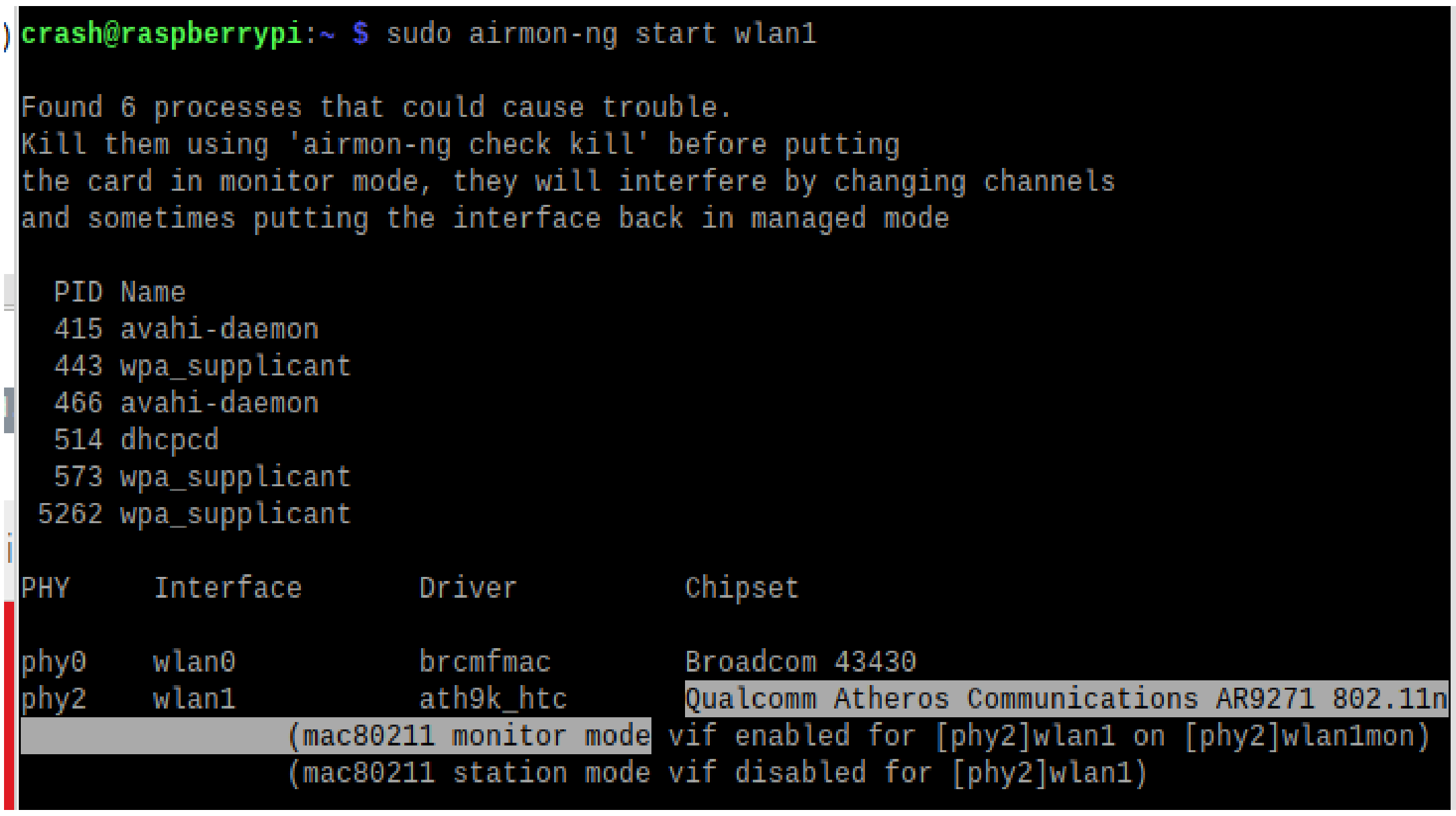

**Figure 15: The Alpha network had enabled into monitor mode (Primary Source).**

4. **Monitor Mode Verification**: Conducted a verification check to ensure the interface had successfully transitioned to monitor mode.

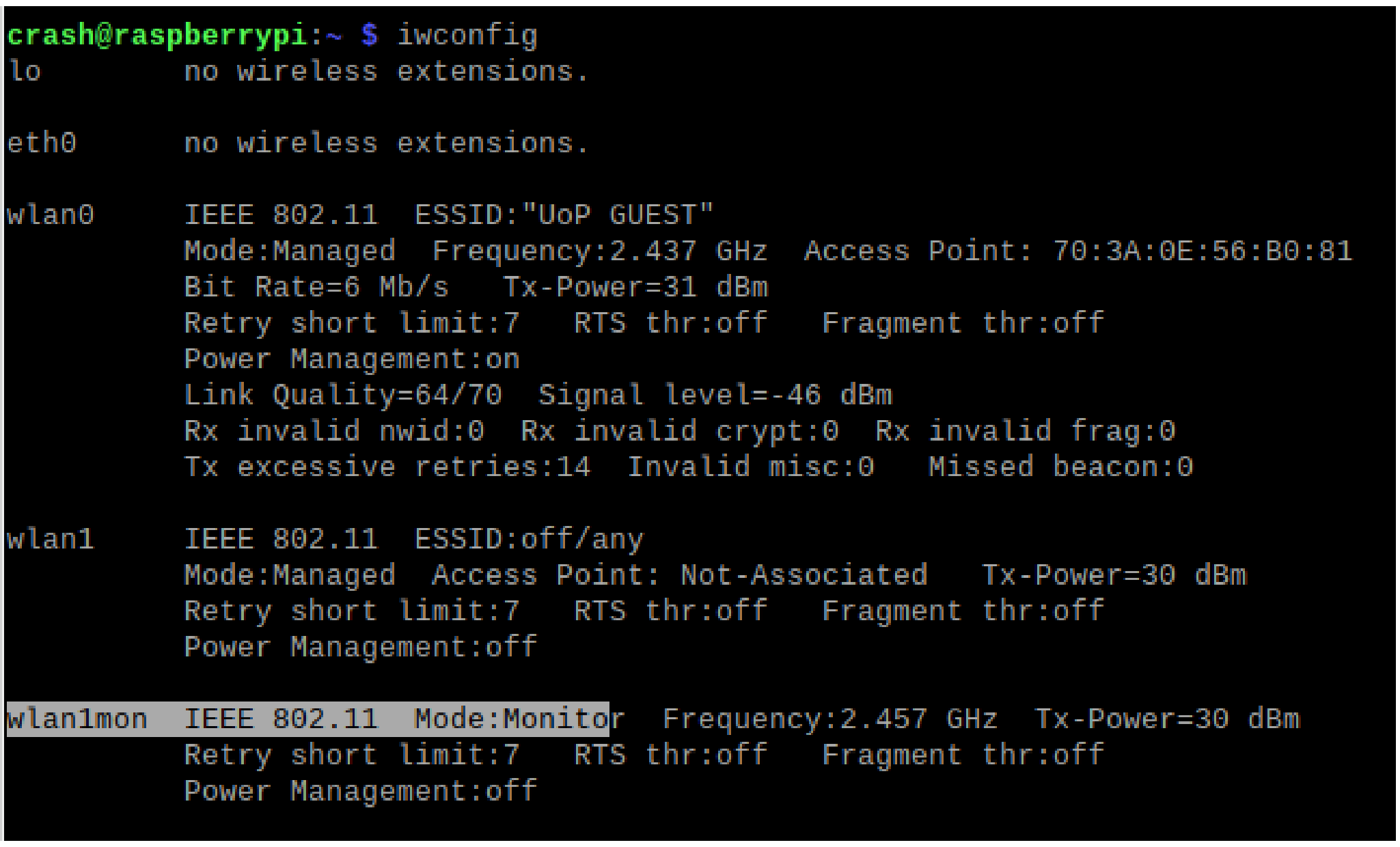

```
crash@raspberrypi:~ $ iwconfig
lo        no wireless extensions.

eth0      no wireless extensions.

wlan0     IEEE 802.11  ESSID:"UoP GUEST"
          Mode:Managed  Frequency:2.437 GHz  Access Point: 70:3A:0E:56:B0:81
          Bit Rate=6 Mb/s   Tx-Power=31 dBm
          Retry short limit:7   RTS thr:off   Fragment thr:off
          Power Management:on
          Link Quality=64/70  Signal level=-46 dBm
          Rx invalid nwid:0  Rx invalid crypt:0  Rx invalid frag:0
          Tx excessive retries:14  Invalid misc:0   Missed beacon:0

wlan1     IEEE 802.11  ESSID:off/any
          Mode:Managed  Access Point: Not-Associated   Tx-Power=30 dBm
          Retry short limit:7   RTS thr:off   Fragment thr:off
          Power Management:off

wlan1mon  IEEE 802.11  Mode:Monitor  Frequency:2.457 GHz  Tx-Power=30 dBm
          Retry short limit:7   RTS thr:off   Fragment thr:off
          Power Management:off
```

**Figure 16: Switching to monitor mode wlan1mon (Primary Source).**

It is important to use the Alpha adaptor, for the penetration to work we have to change the network adaptor into monitor mode.

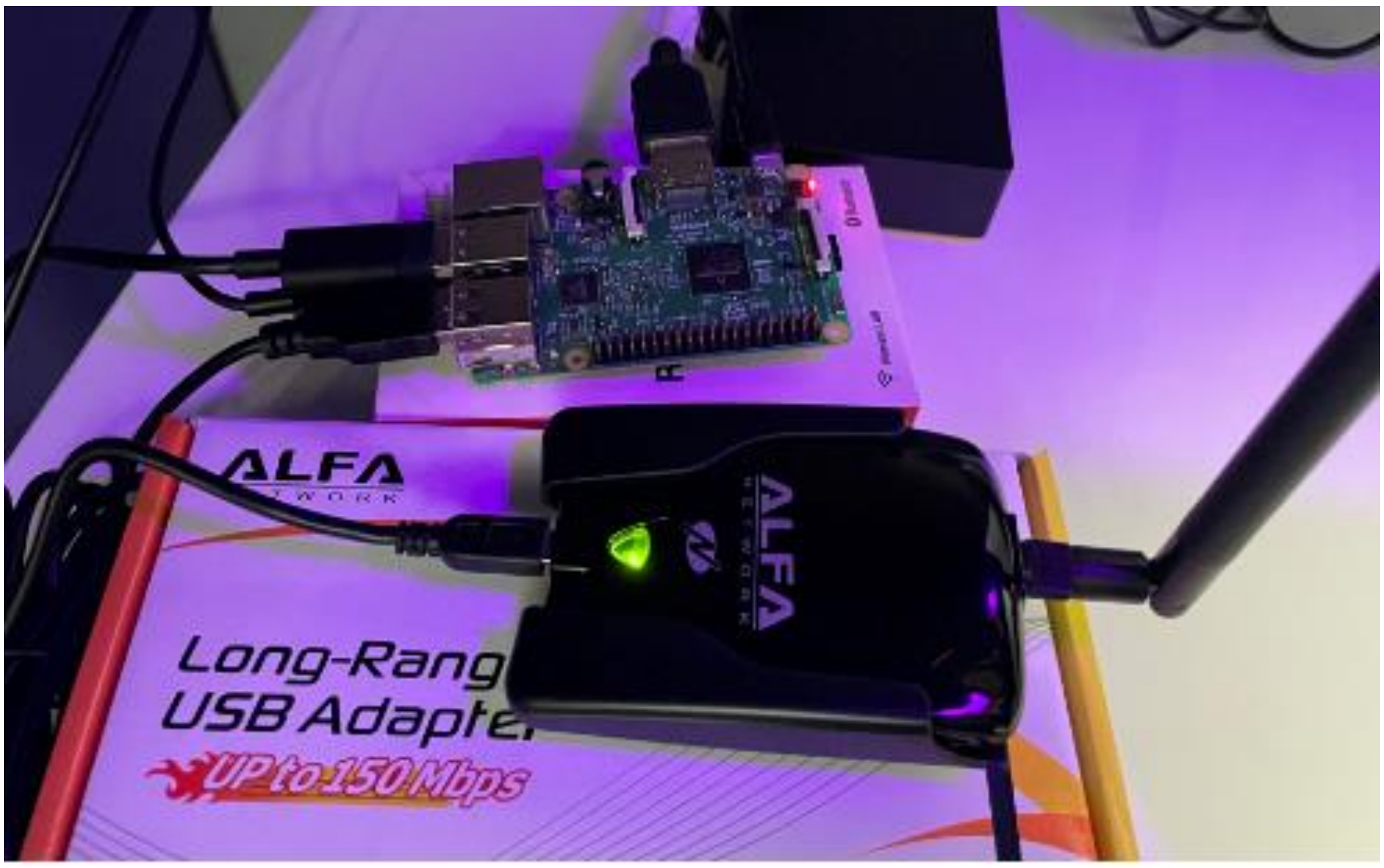

**Figure 17: Alpha Wi-Fi adaptor and Raspberry- Pi 3 (Primary Source).**

5. **Network Scan:** Initiated a network scan with `airodum-ng wlan1mon` to identify active wireless networks and connected clients. `-$ sudo airdump-ng wlan1mon`

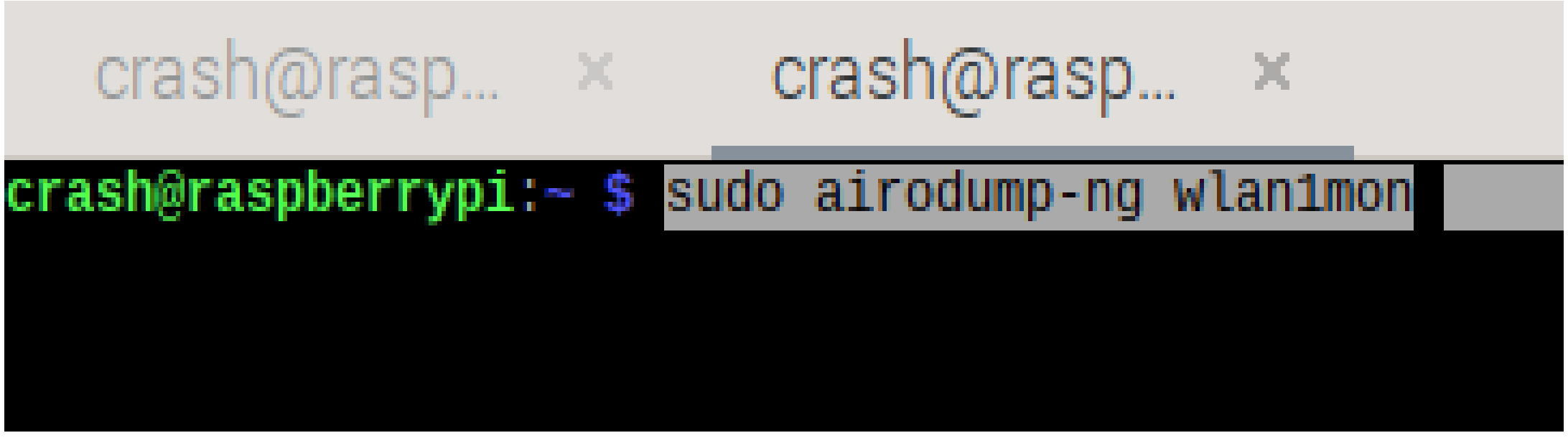

**Figure 18: A network scan (Primary Source).**

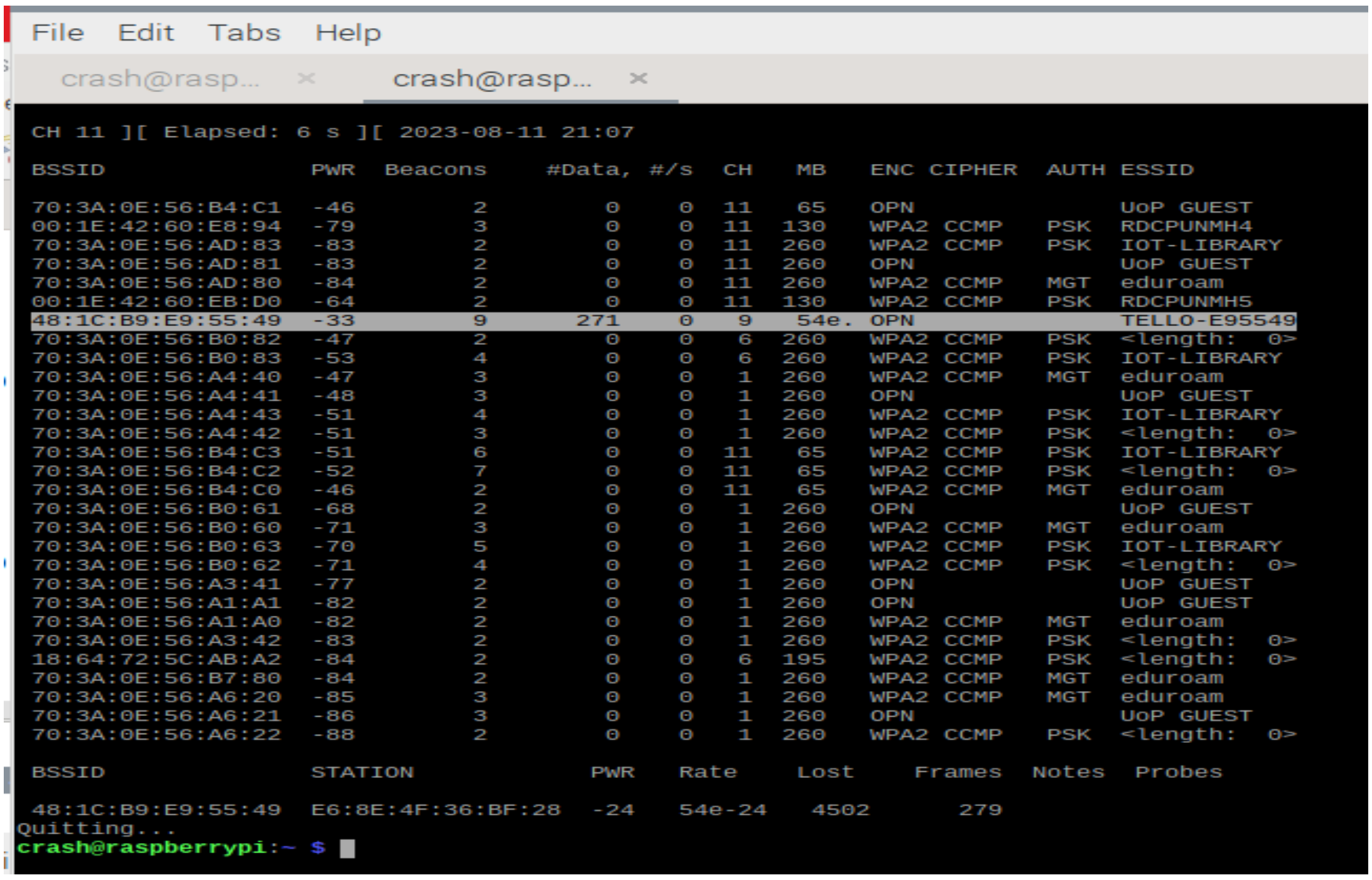

**Figure 19: Tello- Drone with the BSSID number and channel number 9 (Primary Source).**

Note: It is important to note down the channel number and the BSSID number, as we will be using this for the attack.

6. **Targeted Packet Capture:** The process began by identifying the target network's Basic Service Set Identifier (BSSID) and channel. Subsequently, the command "`airodump-ng -d BSSID -c channel -w` name of capture packet `wlan0mon`" was executed. The packets were acquired only from the identified drone's Basic

Service Set Identifier (BSSID) on the designated channel, to preserve them for subsequent analysis. The packet file capture was named hackTello.

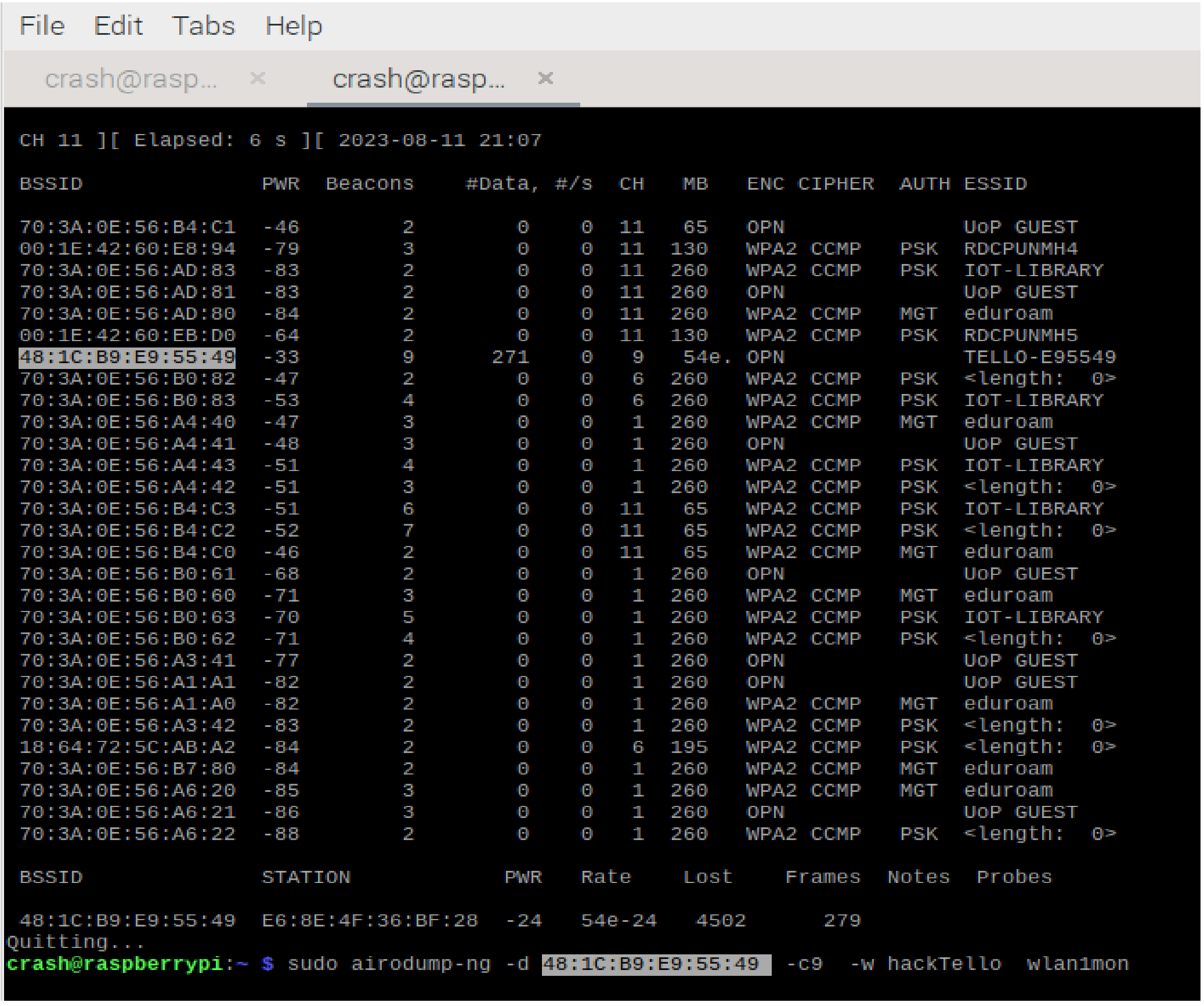

**Figure 20: The command used (Primary Source).**

`sudo airdump-ng -d (BSSID) -c (Channel) -w (packet name) Wlan1mon`

7. **Deauthentication Attack:** Initiated a deauthentication attack on the target drone using `airplay-ng --deauth 0 -a BSSID -c MAC wlan1 or wlan0mon`. This command forced the drone off its connected network, paving the way for potential unauthorised control access.

The security provided by MAC addresses in wireless networks is not strong, as they can be easily discovered using a sniffer. (Frank Ohrtman., 2018)

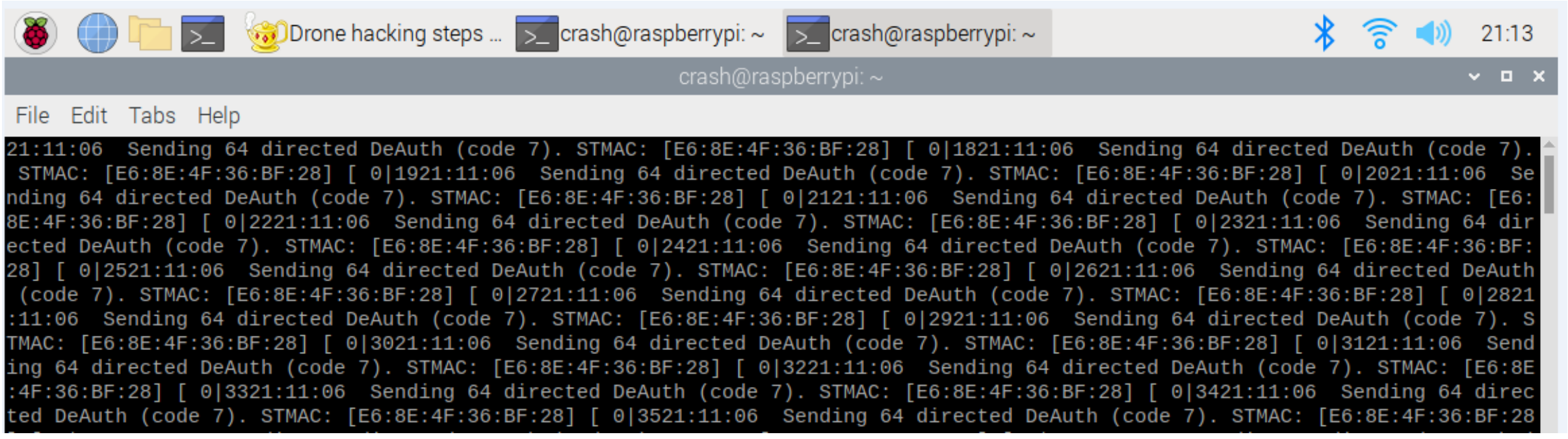

**Figure 21: Shows successfully Deauth (Primary Source).**

**Cybersecurity Analysis Report using Wireshark**

Subject: Deauthentication Attacks Targeting Tello Drones

Date: 08-08-2023

## 4.8 Executive Summary:

This report provides an extensive analysis of the network traffic associated with Tello drone operations, uncovering compelling evidence of a Deauthentication attack.

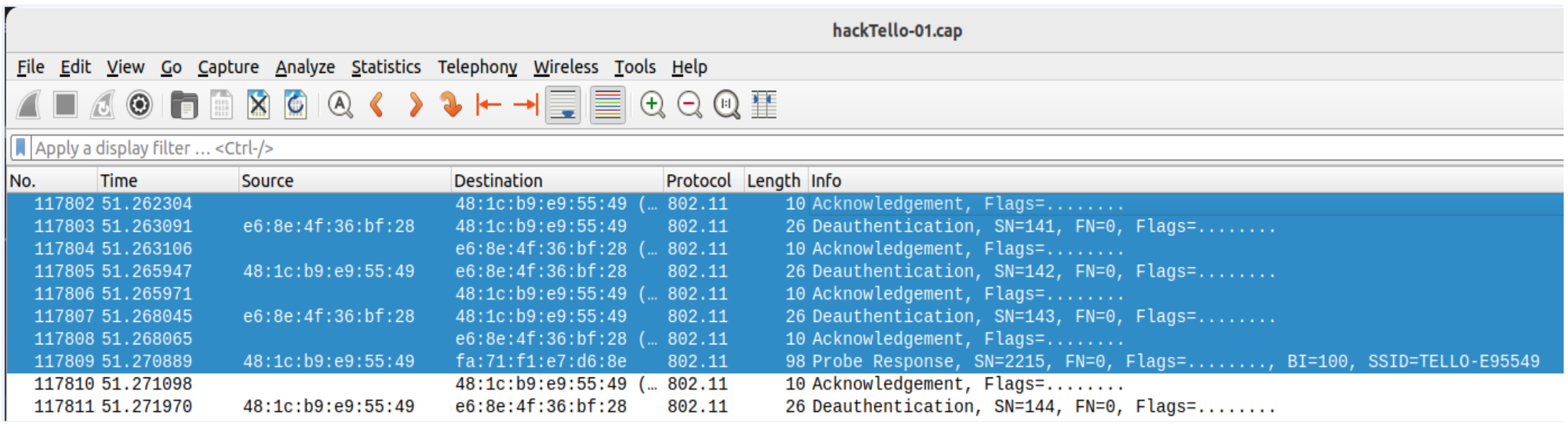

**Figure 22: Network traffic: SSID=Tello using Wireshark 5.0 (Primary Source).**

2. Findings of the Analysis:

The collected data reveals a series of Deauthentication frames that possess the capability to sever the connection between a drone and its controller (Pasolini, Bazzi, & Zabini, 2017).

Specifically, a Deauthentication frame, labelled with the sequence number (SN) 141, originates from the MAC address e6:8e:4f:36:bf:28 and is directed towards the MAC address 48:1c:b9:e9:55:49.

Subsequently, another packet is transmitted, originating from the MAC address 48:1c:b9:e9:55:49 and targeting the MAC address e6:8e:4f:36:bf:28.
Following this, a third transmission is observed, with the source MAC address (SN 143) traced back to e6:8e:4f:36:bf:28. The intended recipient of this frame is identified as the MAC address 48:1c:b9:e9:55:49. The rapid succession of these frames strongly indicates the involvement of a malicious actor executing a de-authentication assault.

**Detection of Probe Response:**

A noteworthy observation pertains to a Probe Request frame originating from the MAC address 48:1c:b9:e9:55:49, associated with the SSID TELLO-E95549. This observation signifies active communication by the drone, as it actively seeks controllers to establish connections. Notably, frames of this nature hold the potential to be exploited by malicious individuals for unauthorised access point establishment.

**Deauthentication Attack and Implications**

The successful execution of the deauthentication attack was achieved through the application of `Aircrack-ng`, a tool renowned for its frame injection capabilities (Pasolini, Bazzi, & Zabini, 2017). By transmitting '`deauth`' frames to the connected network, the attack effectively severed the communication between a drone and its controller. This necessitated the acquisition of MAC addresses for precise targeting. The Rasberian operating system's intricacies were navigated by disabling the Network Manager before initiating the assault, as Network Manager Incompatibility with `aireplay-ng` was observed.

**Experimental Process**

The experiment utilised the `airmon-ng` tool for directed deauthentication packet transmission, primarily targeting the control device. Each instance of de-authentication entailed dispatching 128 packets, with a total of five authentications issued. Successful de-authentication led to the interruption of communication between the controller and the drone (Pasolini et al., 2017).

**Replay Attack and Vulnerabilities**

The steps illustrate the potential vulnerability of drone communication systems. Successful execution of these tactics can result in the de-authentication of legitimate drone pilots, revealing critical weaknesses in communication protocols.

**Implications:**

The successful execution of these steps can lead to de-authentication of genuine pilots from their drones, revealing critical vulnerabilities in the drone's communication mechanisms.

## Conclusion

The DE authentication of the drone connection was achieved through the utilisation of Air crack-ng. Air crack-ng possesses the capability to do frame injections, enabling the transmission of a '`deauth`' frame to the network that is currently connected. In order to execute a focused attack, it was necessary to obtain the MAC addresses of the devices. Initially, it was necessary to ascertain the MAC address of the drone, as well as the MAC address of the target controller. There exist other methodologies for identifying the MAC address, however the approach employed in this particular scenario involved initiating a ping to each device and afterwards utilising the ARP -A command to retrieve the associated MAC address. Prior to conducting the experiments, it was hypothesised that in the event of a lost connection, the Tello drone would descend in a straight path and crash. However, contrary to this assumption, the drone's built-in sensors enabled it to activate safety mechanisms and maintain a hovering position above the ground, thus preventing the user from attaining complete control.

The `aireplay-ng` command included the specification of the targets, as well as the interface utilised on the attacker Raspberry Pi. Nevertheless, for the successful execution of the assault, it was necessary to deactivate and subsequently reactivate the network interface, as depicted in Figure 4. The occurrence of this event can be attributed to an anomaly within the operating system employed during the attack. In the Rasberian operating system, it has been observed that the Network Manager encounters difficulties with the `aireplay-ng` process. Consequently, it is necessary to disable the Network Manager prior to initiating the attack.

Figure 3 illustrates the process of initiating the assault by utilising the `airmon-ng` tool.

After being initialised, the tool begins to transmit directed deauthentication packets on the network, specifically aimed at the control device. In the context of a deliberate assault, a total of 128 packets are dispatched for every instance of de-authentication that is defined. A total of five authentications were transmitted in the command issued. Figure 8 illustrates the presence of a field indicating [23—64 ACKs] following the initial de-authentication sequence. This implies that the client, namely the controller, got a total of 64 packets. The controller ceases to receive packets subsequent to the initial set, as a result of the successful deauthentication. The outcome of the experiment was deemed successful, leading to the de-authentication and subsequent absence of the link between the controller and the drone.

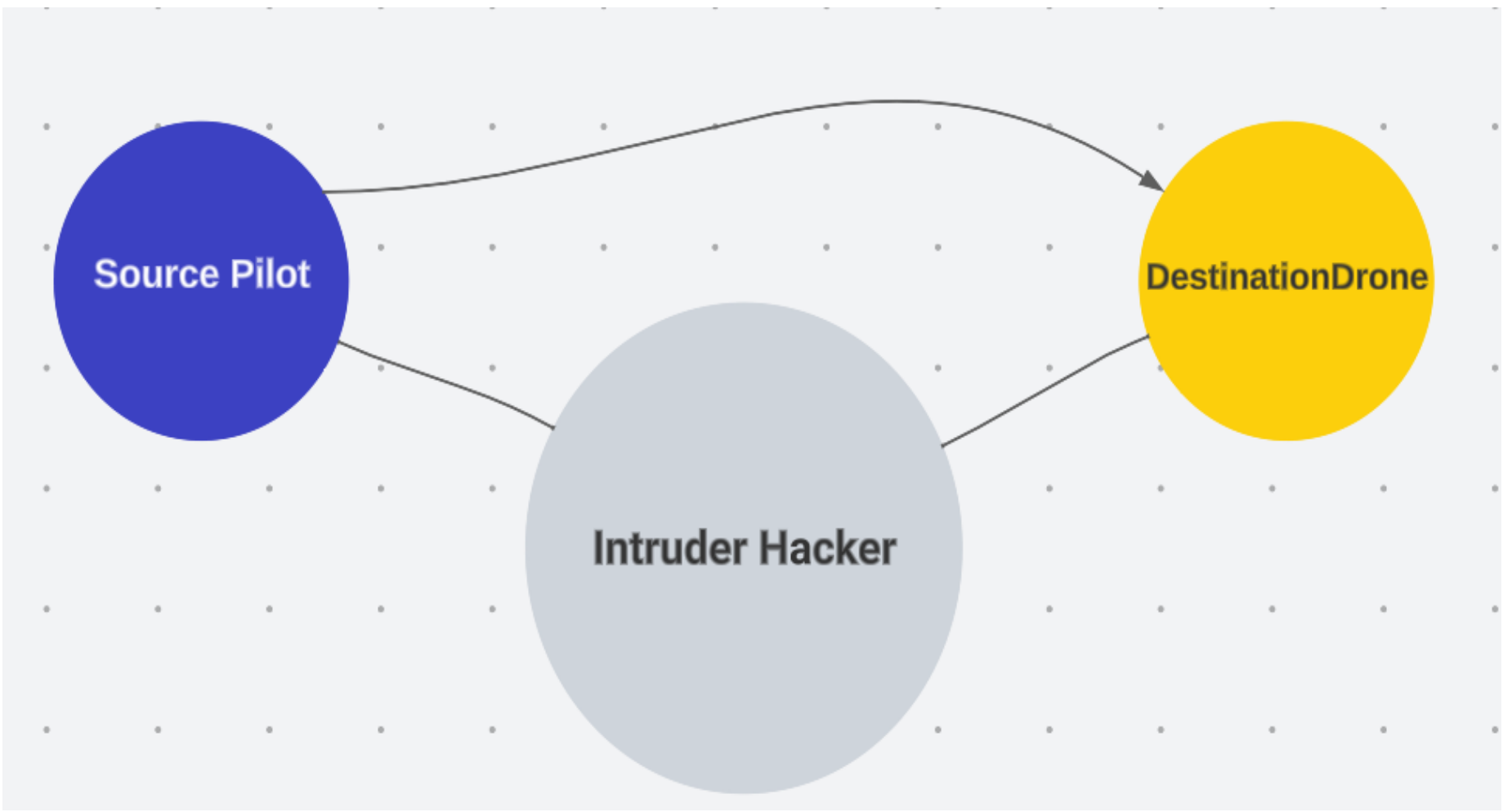

**Figure 23: Replay attack on a network (Primary Source).**

## 4.9 Recommendations:

### Monitor Deauthentication Frames :

Deploying network monitoring tools with real-time alert capabilities for heightened occurrences of DE authentication frames can greatly assist in promptly identifying and mitigating attacks (Pasolini et al., 2017).

**Implement Robust Encryption:**

Enhancing security measures through the adoption of robust encryption protocols, such as WPA3, can significantly reduce the effectiveness of DE authentication attacks. (Frank Ohrtman, 2018)

## 4.10 Extension of experiment

My DJI drone did not have a strong encryption method. I wanted to redo this experiment. However, I could not buy another higher spec drone. I used my personal Amazon Alexia which has WPA2 Wi-Fi encryption.
The identical methodology was employed in the case of the Tello AR drone. The device in question is the Amazon Alexa, which is connected to the same wireless network as previously mentioned.

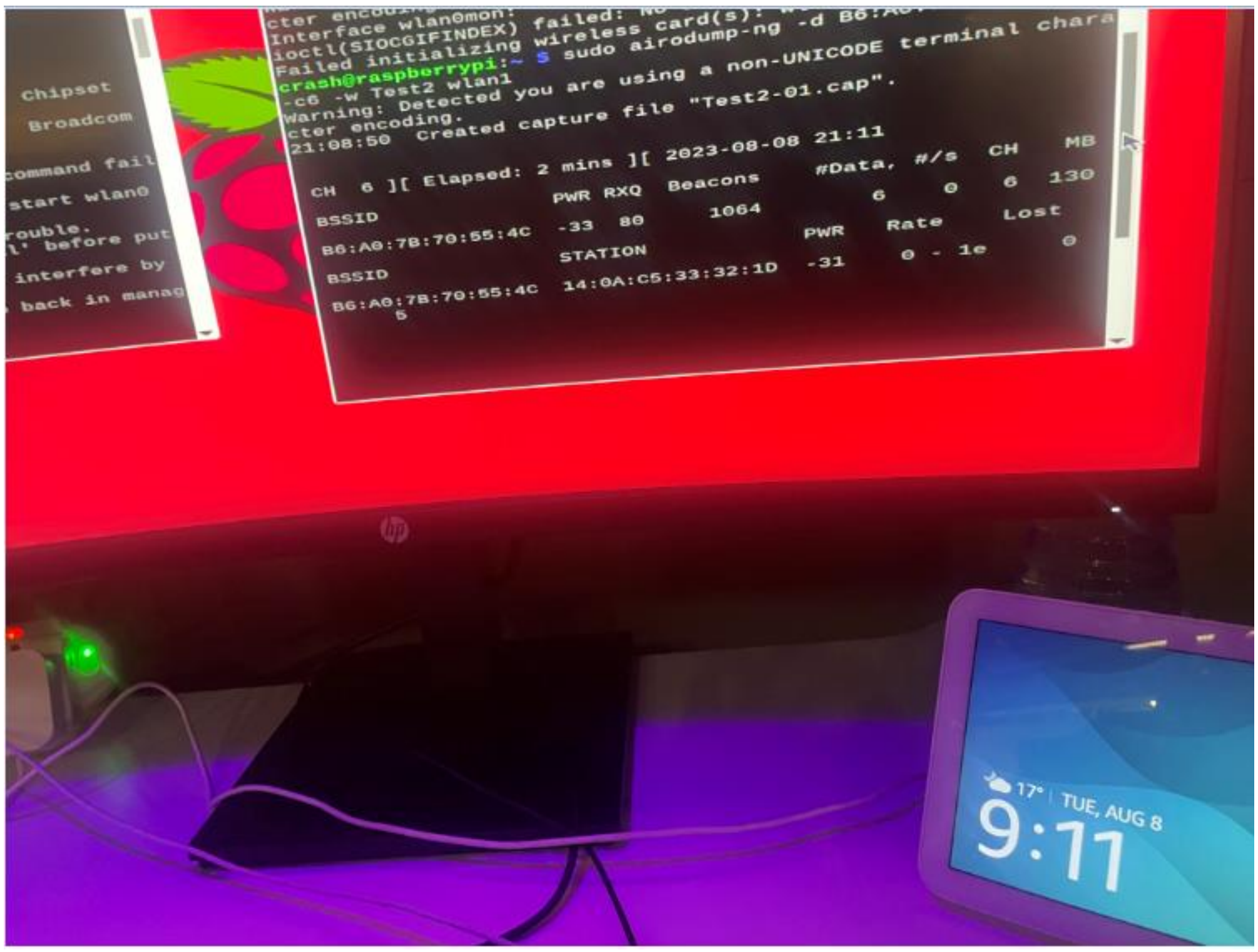

**Figure 24: Amazon Alexa Wi-Fi connection (Primary Source).**

The procedure of locating the MAC address of the Amazon Alexa device involved utilising the Arp protocol. Subsequently, the assault was initiated, with a specific focus on the gadget.

The wireless connectivity (WiFi) connection between the WiFi network and the Alexia device has been disconnected. From the perspective of the Alexia device, the software displays an error message stating "Link not available." This communication serves as a cautionary notice pertaining to the potential disruption or discontinuation of wireless network connectivity.

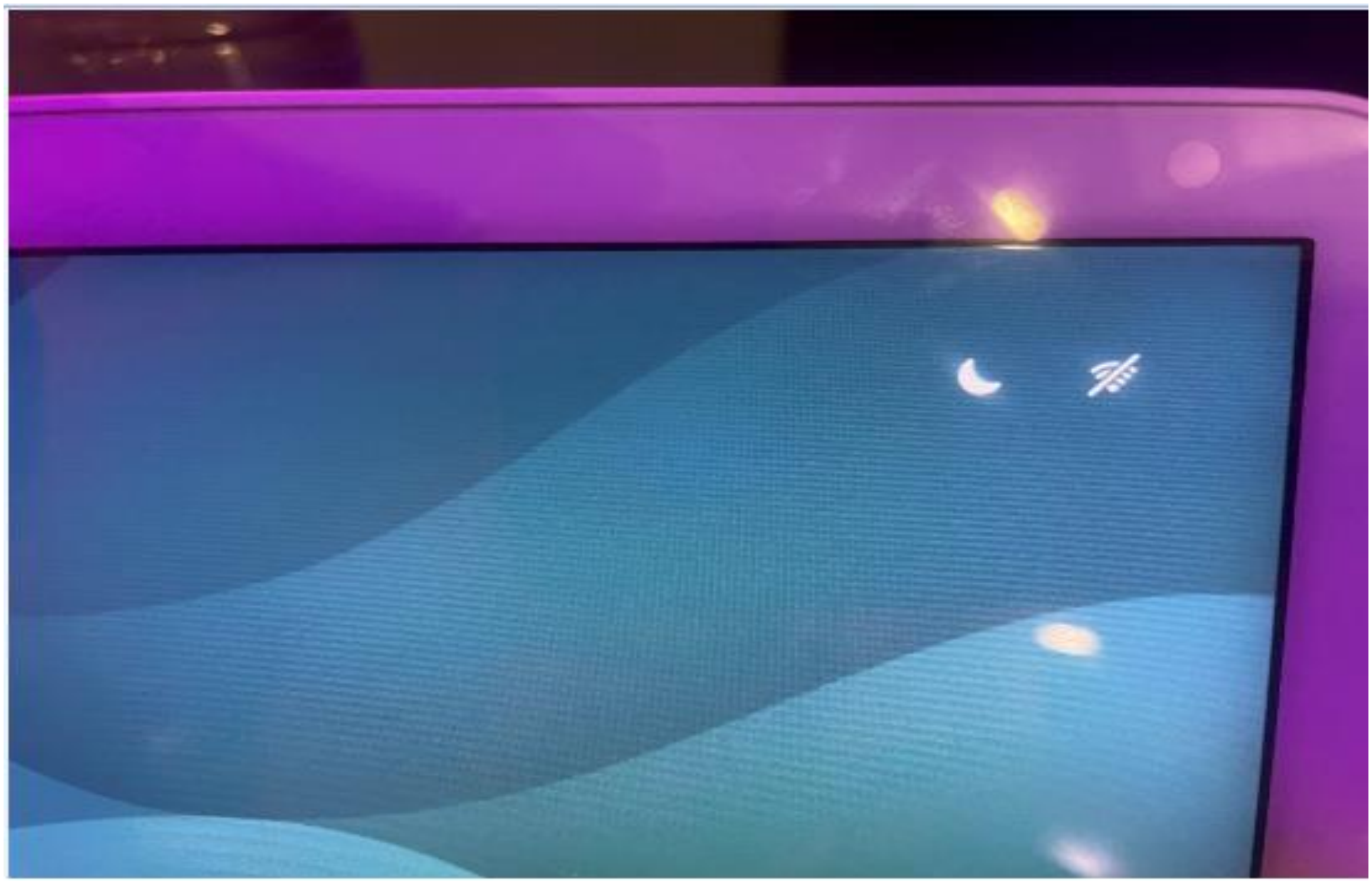

**Figure 25: Amazon Alexa Display no Wi-Fi connection (Primary Source).**

The DE authentication attacks proved effective on both devices. The same tools and methods were used. But this time I have more evidence that the network was attacked by deauth and it is clear that I attacked the Wi-Fi connection because I got a 4 way '**handshake**' which is what I was trying to achieve with my Tello drone wish did not have a WPA2 encryption.

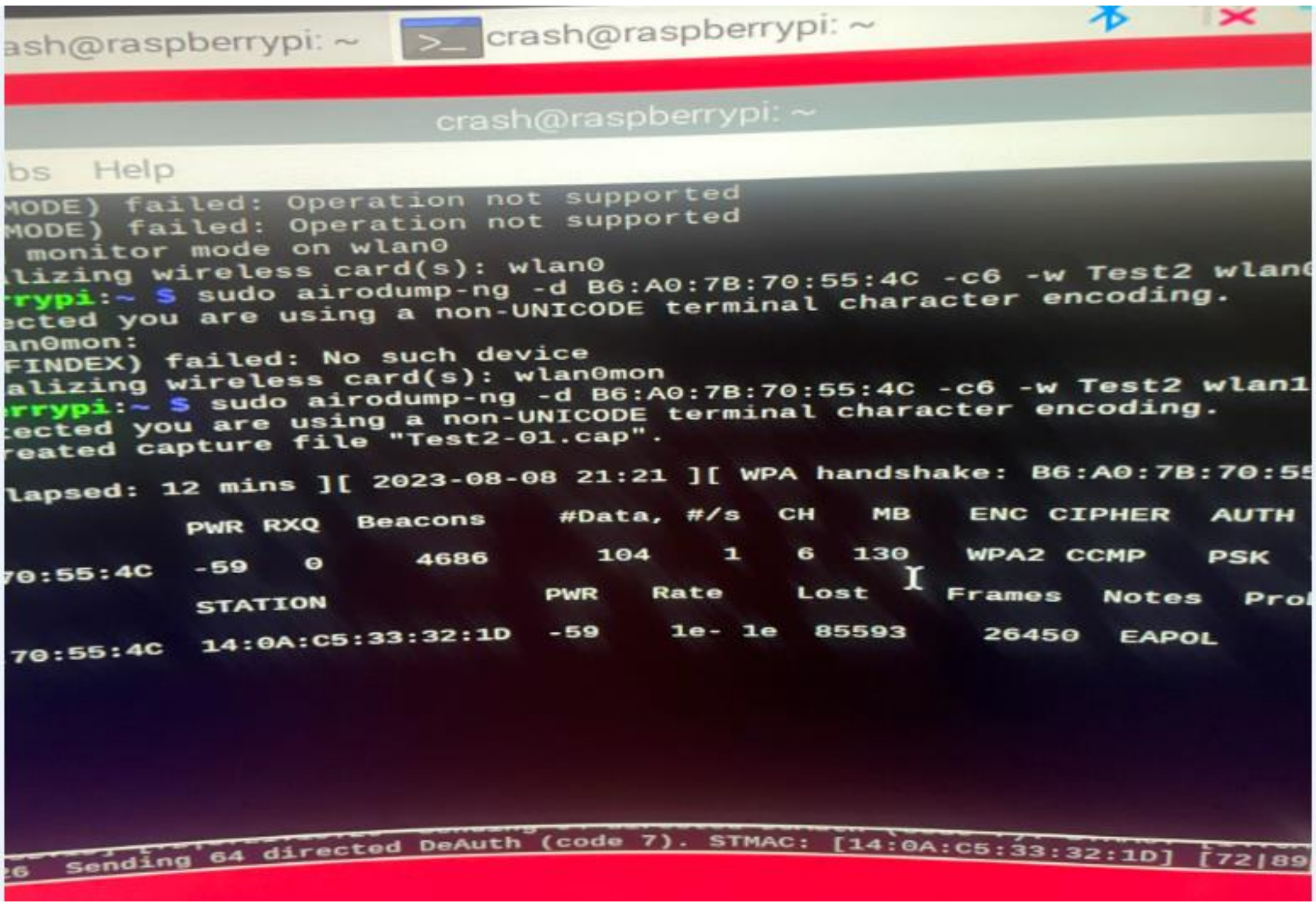

**Figure 26: The WPA handshake of the Amazon Alexa (Primary Source).**

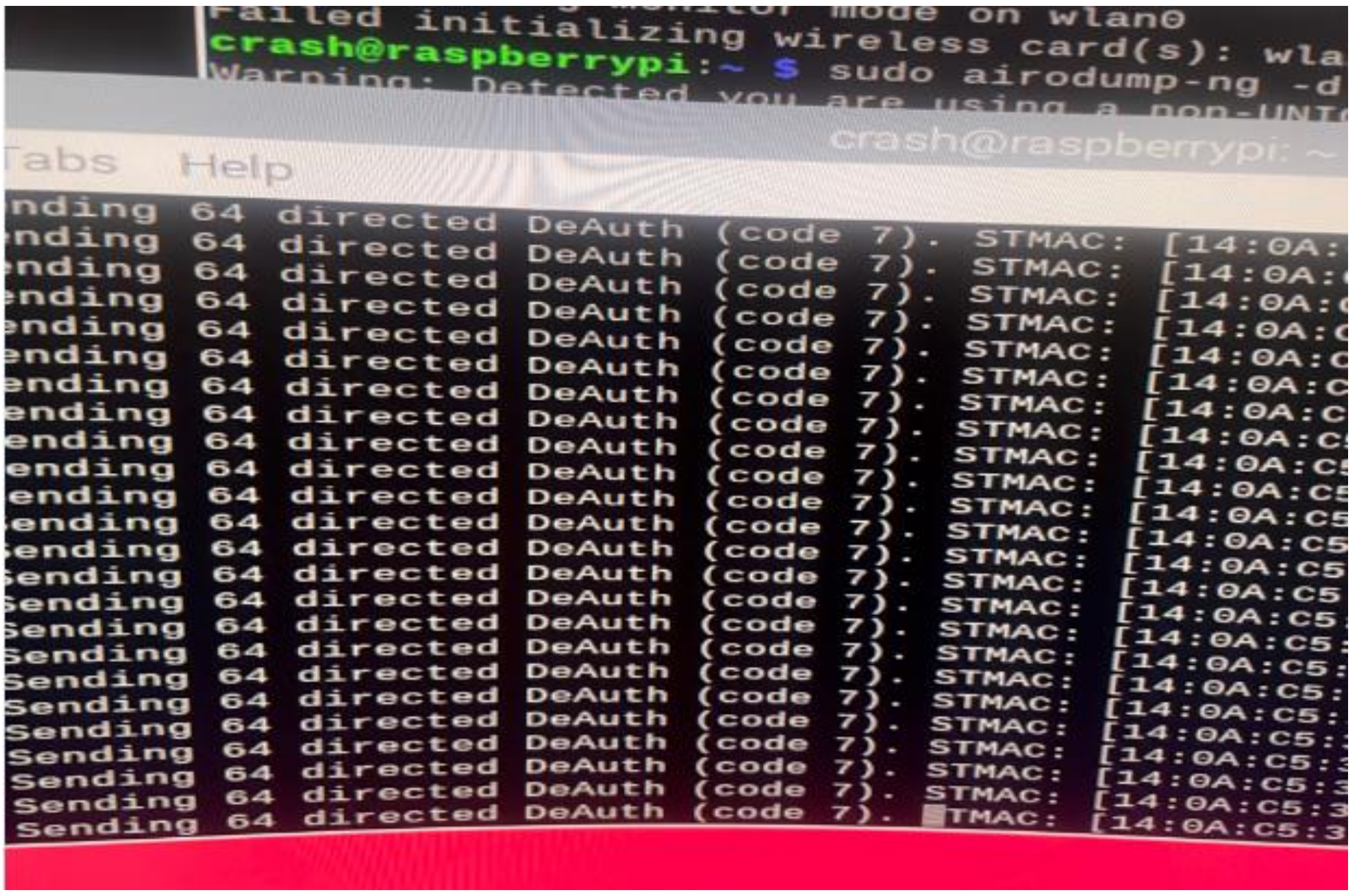

**Figure 27: shows the packets being sent and the Direct DeAuth (Primary Source).**

The jamming system used for both attacks was the Raspberry Pi 3. However, the tools used now are different but in World War 2 there was a similar method used for jamming radio frequencies.

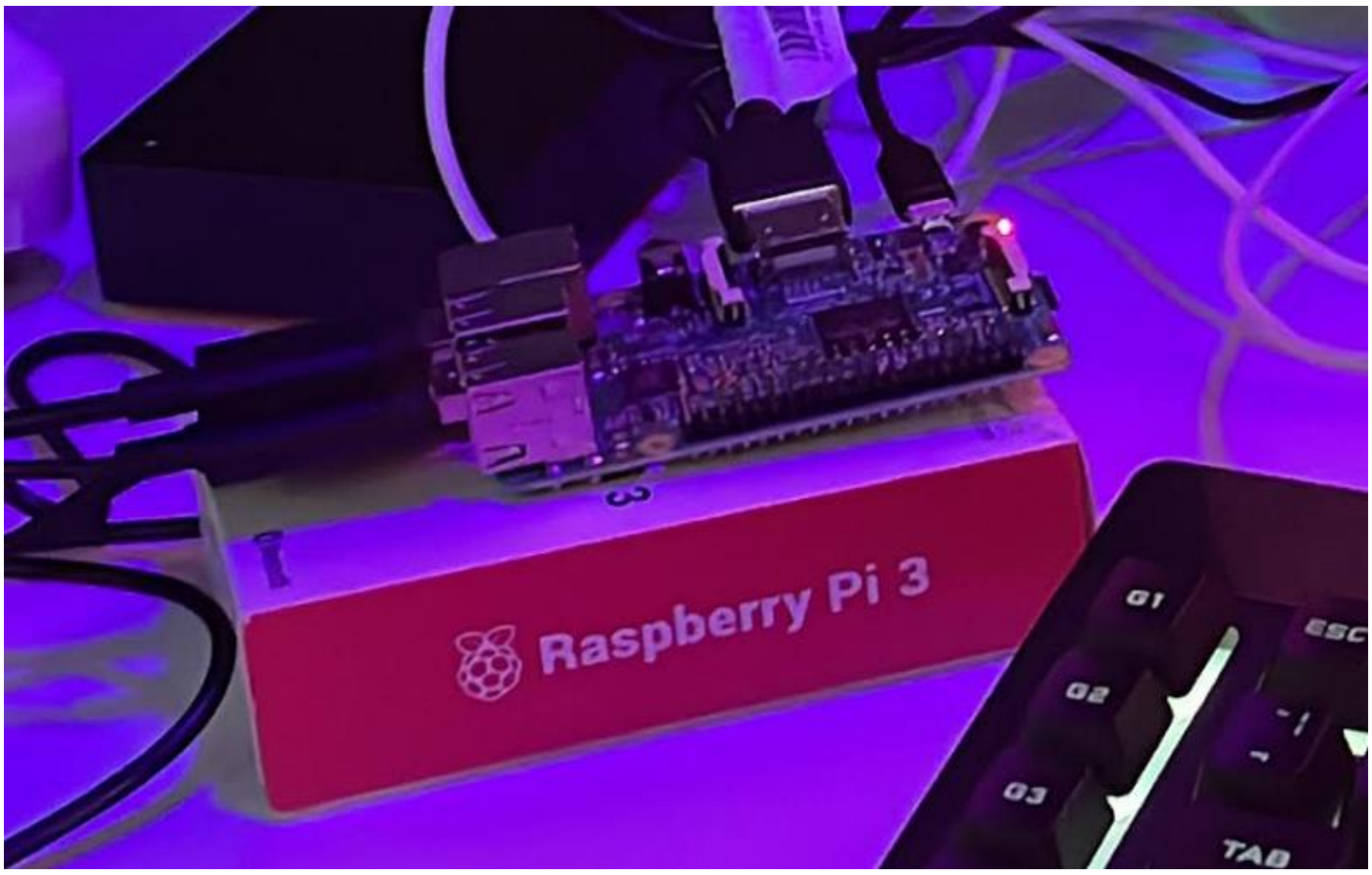

**Figure 28: The Raspberry Pi 3 was use for the attack (Primary Source).**

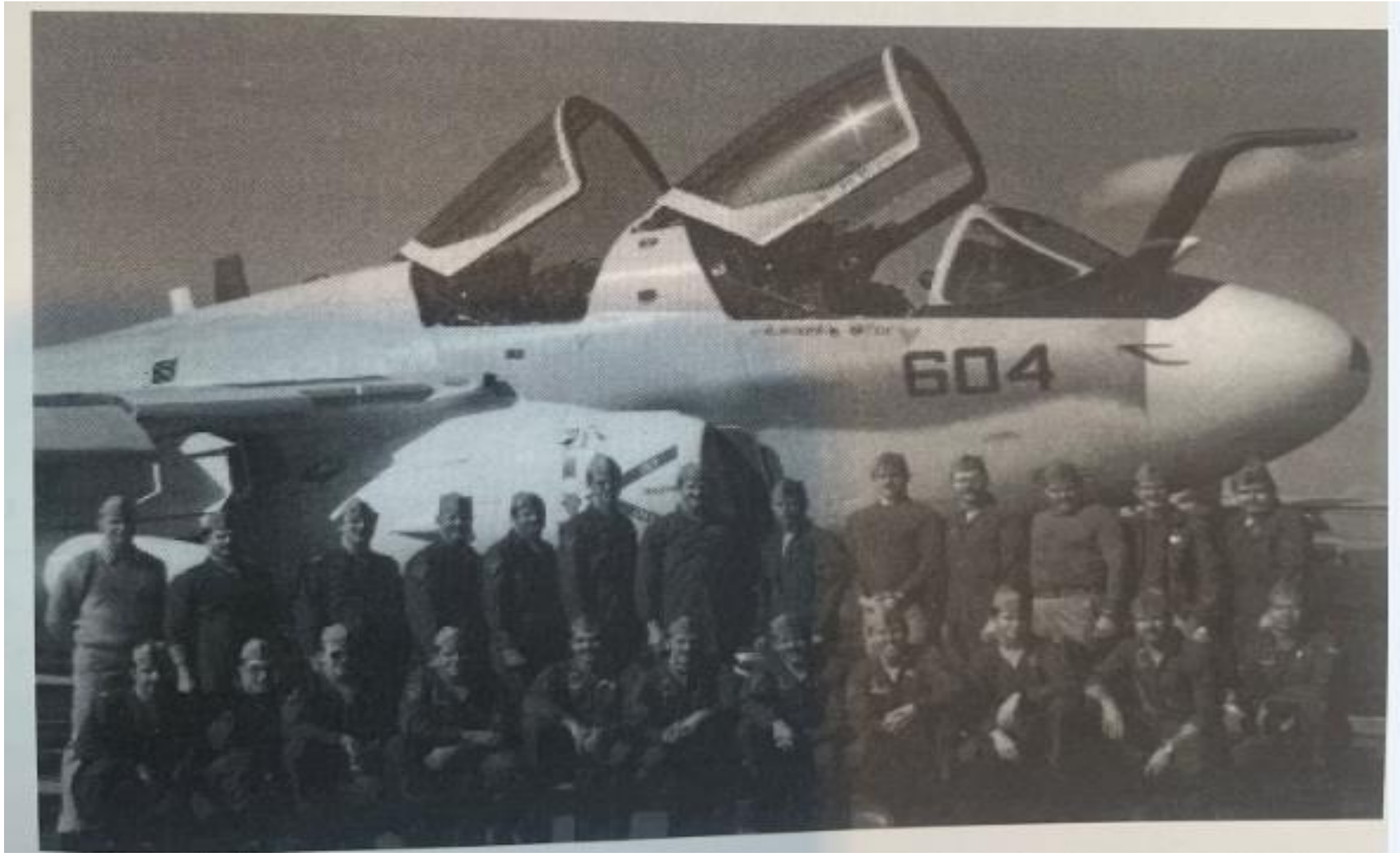

**Figure 29: Black Ravens, EA-6B tactical jamming aircraft (Frank O. Wi-fi Handbook 2018).**

The unit operated EA-6B tactical jammer aircraft. The aircraft is outfitted with an AlQ-99 jamming system, which is employed for the purpose of disrupting adversary radar and radio communications. Given the potential for interference in tactical applications, it is possible to undermine the advantages of wireless networks. (Frank Ohrtman., 2018)

## 4.11 Conclusion

In conclusion, the experiment provides clear evidence that vulnerabilities continue to exist despite the progress made in encryption and security measures. Devices such as the DJI drone, which includes inadequate encryption measures, may be viewed as easily exploitable targets for someone with malicious intent. Nevertheless, even devices equipped with encryption methods considered to be robust, such as the Amazon Alexa with WPA2 Wi-Fi encryption, can still be vulnerable to attacks, as evidenced by the outcomes of our experiments in capturing a 4 ways Handshake.

The instances of de-authentication assaults on both the drone and Amazon Alexa serve as a reminder of the vital significance of continuous and robust cybersecurity endeavours. This assertion holds particular validity when considering that the tools and methodology utilized in these attacks lack novelty and complexity, exhibiting similarities to jamming techniques deployed during World War II. The conducted experiments serve as an essential reminder that maintaining security is not a static endeavour but rather an ongoing and dynamic challenge.

The ability of the Raspberry Pi 3 to execute these attacks reinforces the notion that sophisticated attacks can be carried out not only with advanced technology but also with easily accessible and cost-effective means. This highlights the necessity of ensuring the security of wireless networks.

# Chapter 5: Data Analysis:

## 5.1 Introduction:

The data analysis introduction covers the technical analytics in proper connection for the previous chapters under the main topic. Here, the analysis and the findings are in the normal format without any primary software-based analysis. The entire data analysis and discussion chapter will be phased in a secondary format. The main basis of the dissertation is to analyse the cybersecurity threats and attack simulation process for unmanned aerial vehicles (UAV) like drones. The secondary analysis format ensures the usage of technical phases and systematic cases of the UAV threat analysis software that is to be installed on the vehicle network cases.

Here, the analysis findings and the discussions carry some major possibilities to track down the attack and threat with a specific simulation process. The chapter will deal with the types of threats or attacks that the cybersecurity process for unmanned aerial vehicles suffers and a basic threat network analysis. The cybersecurity threat analysis faces DDoS attacks or any other types of similar attack cases. The secondary mixed approach of the methodology cases connects with the main analysis terminals for the main topic. The correct attack simulation process and the detection of “**cybersecurity control**” phases under the UAV-CS systems are to be properly analysed.

The introduction in this chapter finds the types of threats and attacks carried out by the external forces without any prior cover and security. The analytics can cover up for the setting of the “**UAS type-1 INTERFACE**” based network tracking to prevent the drones' negative and unwanted attack simulation processes of the drones. The sections in this chapter properly complies with the main topic cases in the correct manner.

## 5.2 Finding and Analysis:

This chapter's findings and analysis section correctly complies with the threat analysis and attack simulation in the UAV (unmanned aerial vehicles) systems like the drones. The UAVs must be bound and processed with tough cybersecurity policies and systems to prevent unnecessary drone attacks on people or elsewhere (Hafeez et al. 2023). One possibility is that setting up a proper Blockchain assisted UAV network can phase down the major possibility of a potential cyber security threat or attack to the UAV systems.

The analysis process starts with building a strong blockchain interface process that can separate or divert the ongoing or incoming processes in the UAV vehicles and give alerts to the UAV system via blockchain-connected UAV laboratory software and hardware. The logical process follows mainly on the cryptography communication scale that legalises a proper communication channel for the UAV vehicle networks. The findings after the secondary based analysis are mainly assessed for the overall pointers that are very much essential to track down the working and the functionality of the UAV drone networks (Khan et al. 2022). Here, the tampering or any kind of processes are mainly compliant with the unmanned aerial vehicle movement or attack process. The "**Blockchain**" illustration shown in Figure is the central assessment of the overall UAV systems is very necessary to track down the positioning and human-led movement of the UAV systems without any prior communication network.

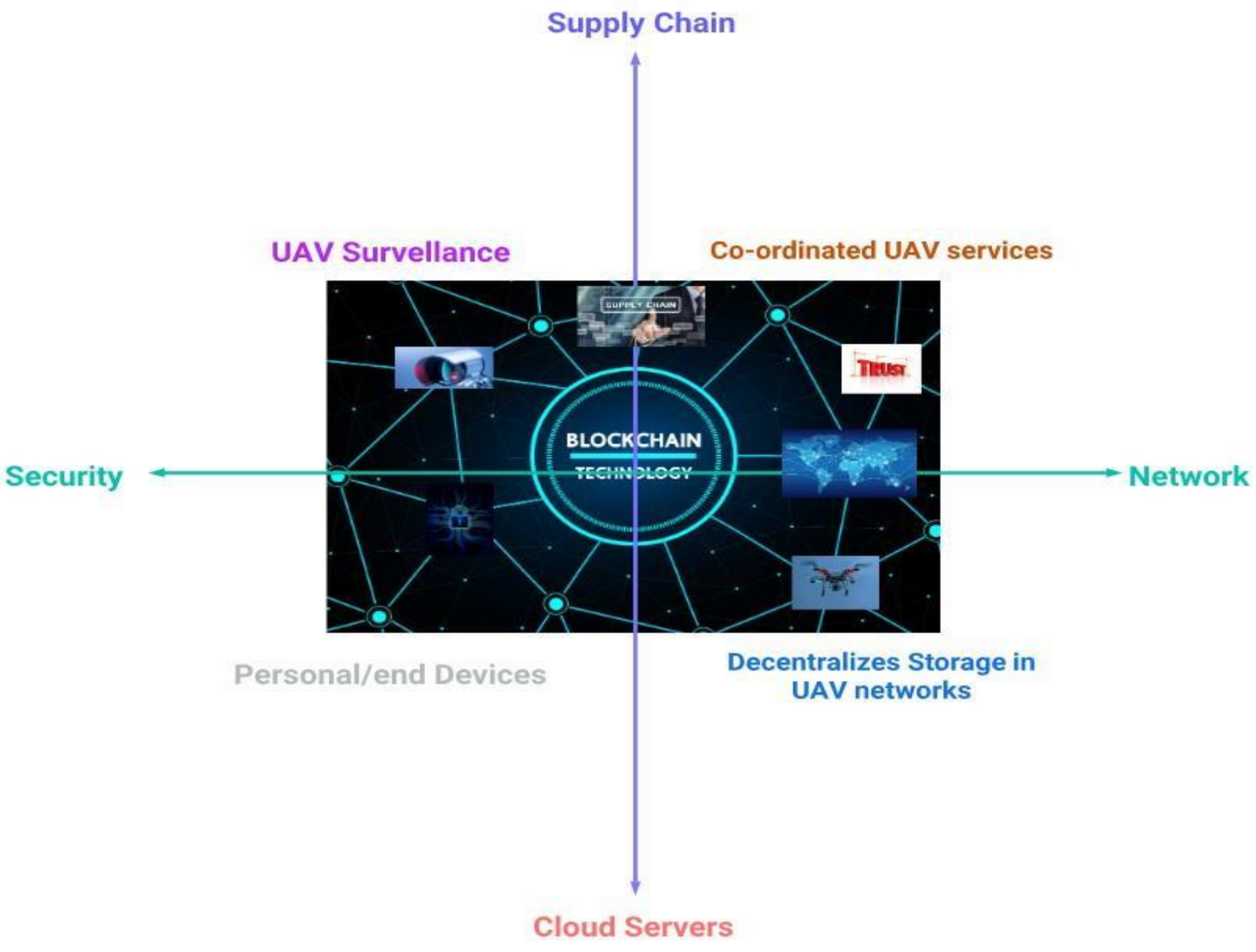

**Figure 30: Application of block chain in UAV (Primary Source).**

Here, the final blockchain-assisted analysis viewpoints on the proper cases of reporting any unwanted movement of the UAV drone systems (Al-Bkree, 2023). By ensuring safe and open data management, blockchain technology can improve UAV operations. It provides immutable records for tracking the supply chain, maintenance history, and flight data. Concerning transportation, monitoring, and autonomous missions, this improve UAV safety, transparency, and effectiveness. A tracking system is to be connected inside the UAV network system so that the server to the main peripheral line connects and stops the attacker from utilising and creating drone attack simulation. The data sensing layer combines with the blockchain processes and generates a locked track to separate and trace the attack simulation process via IPFS encoding (Benet Juan. 2023) The location and the data privacy also matter the most.

### 5.2.1 Results/Findings:

The findings from a secondary study indicate the possibility of combining blockchain technology with unmanned aerial vehicle (UAV) systems to augment the communication interface signals utilised by pilots (Ly & Ly, 2021). The Microsoft STRIDE framework was utilised in this scenario to thoroughly analyse UAV networks to proactively mitigate potential vulnerabilities to various types of attacks, including DDoS, malicious hacking, buffer overflow, and AR. Drone packet spoofing. The results of the study suggest that a considerable proportion of unmanned aerial vehicle (UAV) drones, when linked through blockchain technology, exhibit waypoints that are susceptible to manipulation by unauthorised entities. A thorough examination of these unmanned aerial vehicles (UAVs) has shown distinct cyber-attack patterns, encompassing activities such as password cracking, identification of vulnerabilities in DJI, Parrot, and Bebop 2 drones, IP-34-Phantom assaults, and Telnet-associated Man-in-the-Middle (MITM) attacks. In order to address the reoccurrence of these cyber dangers, several preventive measures suggested.

- Creating a passage connection in directive order with the UAV architecture.
- Using “Wi-Fi-Frequency Identification” channels to automatically analyse the threat checkpoints.

These solutions can mainly stop the in-armature attacks on the UAV systems by directly establishing connective links with the interference-maker/hacker. The patent organisation table data is a significant retrieval in the secondary data sources and analysis functions. Here, the table shows the possible patent organisations of the **UAS** (unmanned aerial system), which is a great point for threat analysis.

| Organization | UAS Patents | Country of Origin | Organization Type |
|---|---|---|---|
| DJI | 685 | China | Firm |
| State Grid Corporation China | 359 | China | Firm (state-owned) |
| Boeing | 345 | U.S. | Firm |
| Ewatt Technology | 193 | China | Firm |
| Amazon | 167 | U.S. | Firm |
| Raytheon | 140 | U.S. | Firm |
| Honeywell | 133 | U.S. | Firm |
| University of Beihang | 131 | China | University |
| Guangdong Rongqi Intelligent Technology | 121 | China | Firm |
| Lockheed Martin Corporation | 119 | US | Firm |
| Avic Xian Aircraft Design and Research Institute | 105 | China | Firm |
| Haoxiang Electrical Energy | 101 | China | Firm |
| Southern Power Grid Company Limited | 100 | China | Firm (state-owned) |
| Shenzhen Autel Intelligent Aviation Tech. | 98 | China | Firm |
| IBM | 96 | US | Firm |
| University of Nanjing Aeronautics & Astronautics | 91 | China | University |
| Korea Aerospace Research Institute | 85 | South Korea | Government |
| Shenzhen AEE Aviation Technology Co. Ltd. | 79 | China | Firm |
| Northwestern Polytechnical University | 79 | China | University |
| Guangzhou Xaircraft Technology Co. Ltd. | 77 | China | Firm |
| Qualcomm | 75 | US | Firm |
| Wuhu Yuanyi Aviation Technology Co Ltd | 73 | China | Firm |
| United States Navy | 70 | US | Government |
| Zero UAV Beijing Intelligence Technology | 62 | China | Firm |
| BAE Systems | 61 | UK | Firm |
| Geer Technology Co. Ltd. | 61 | China | Firm |
| Prodrone Craft Technology Shenzhen Co. | 61 | China | Firm |
| Samsung Electro-Mechanics Co. | 58 | South Korea | Firm |
| Aerovironment, Inc. | 56 | Canada | Firm |
| China Academy of Aerospace Aerodynamics | 54 | China | Government |

**Figure 31: List of UAS (unmanned aerial system) patent organisations for threat analysis (Source: apps.dtic.mil).**

Here, the attacks that are mainly analysed and reported based on the secondary data collected like the UAS system patents table can be recollected with the possible threat simulation to determine loopholes in the UAV network cases (Abualsauod, 2022).

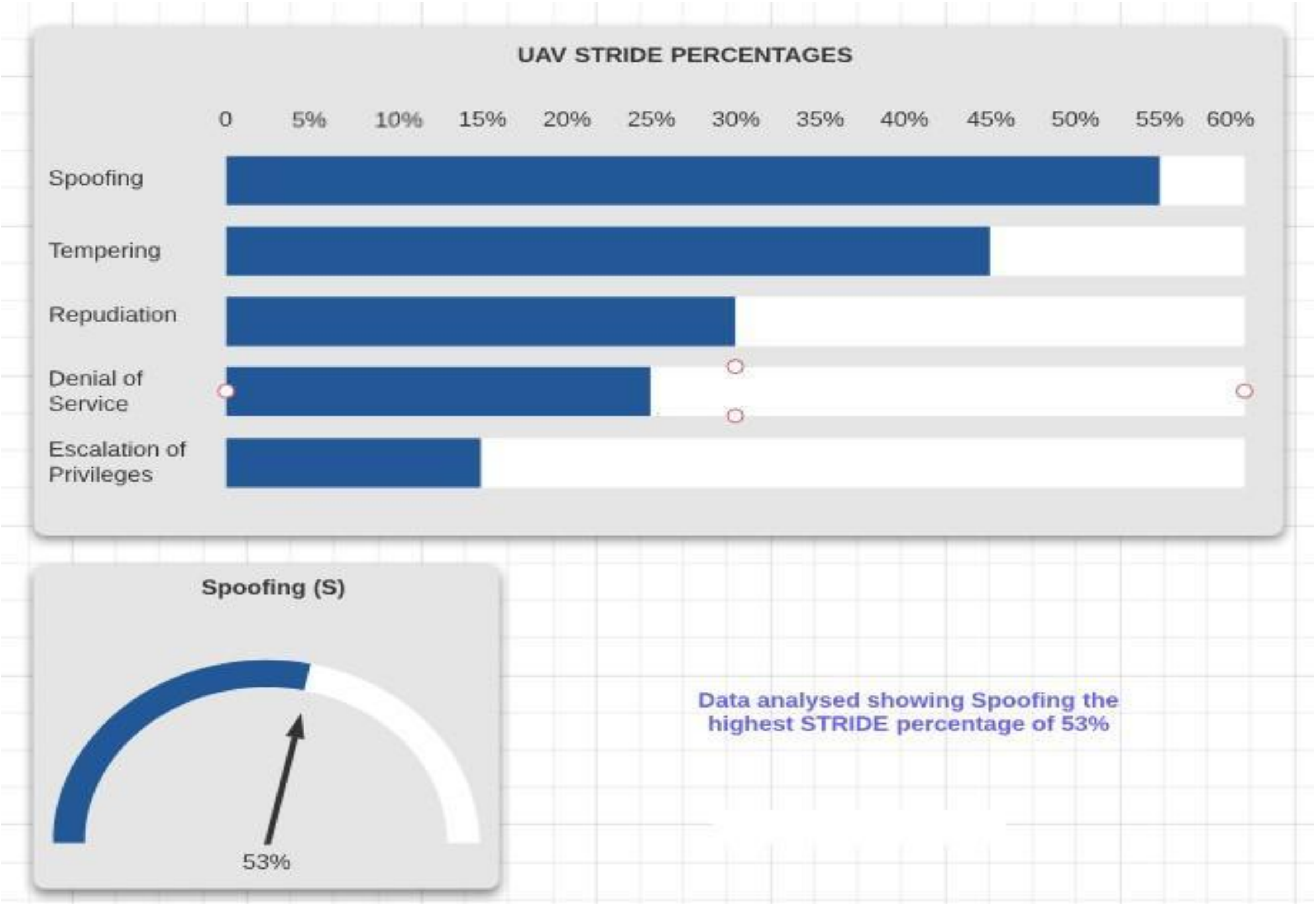

**Figure 32: Current cyber-attack overview in the STRIDE categories for the UAV (Primary Source graph but data obtained from source: researchgate.net).**

The above image shows the cyber-attack overview in the **STRIDE** categories where the attack simulation architecture without blockchain system targets and attacks the UAV device with a direct deviator pointer (Jacobsen, & Marandi, 2021). And with the blockchain system, the hacking process gets stopped midway to the UAV device with the help of jammers.

Here, the prominent cases follow the exact analysis per the UAV loader setup cases in a perfect alliance. The overall phases for the secondary analytical overview are to be correctly encased as per the UAV data readings uploaded on the software or hardware link phase (Tran et al. 2022). There are a lot of complications regarding the attack simulation phase, where the attacker/hacker can connect the drone processes via serial uploading of the malware and threats that can cause damage (JIANG et al. 2021). A successful STRIDE framework can create a defence of the attack simulation process. (Quentin R., BH., & JJ. 2021)

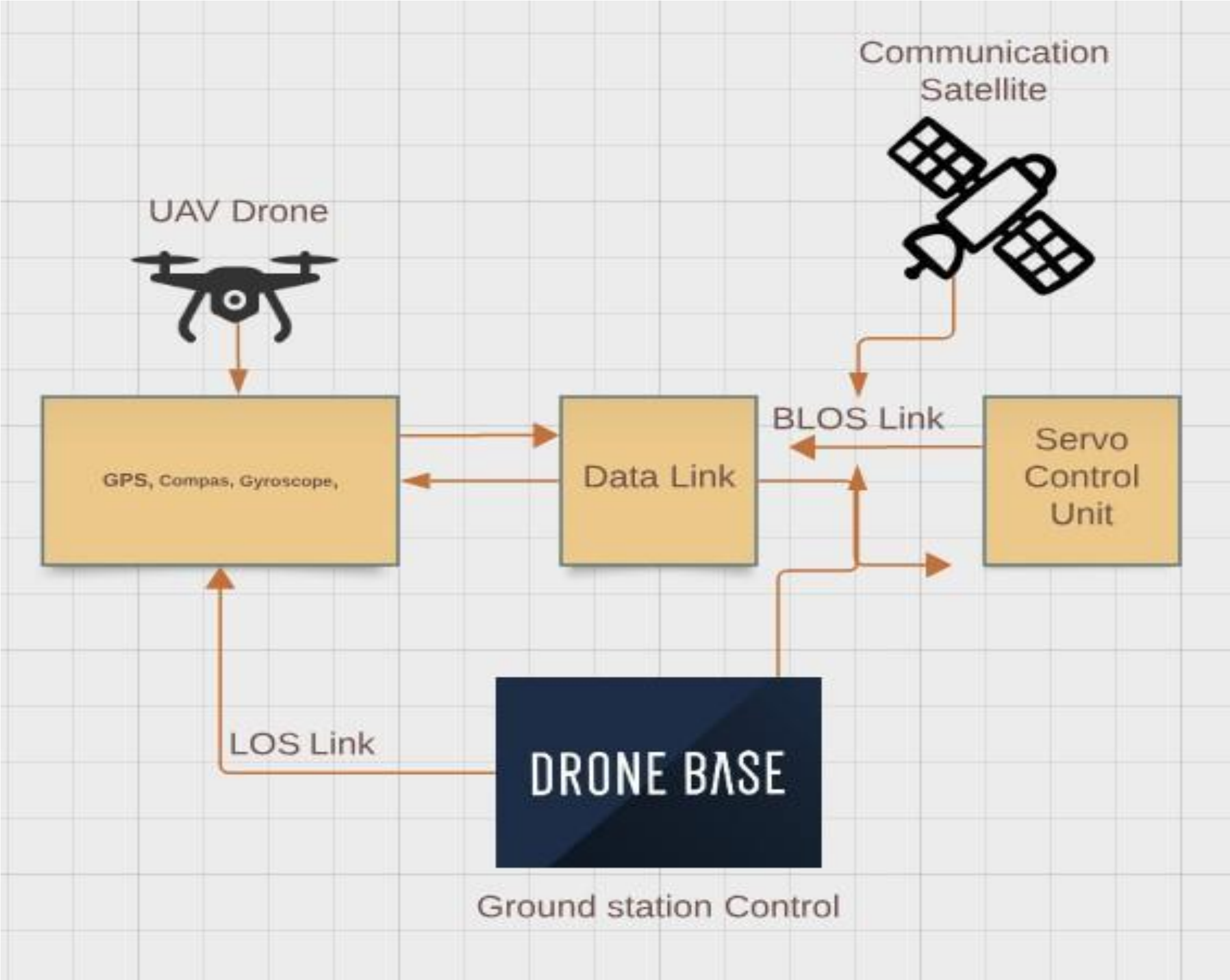

**Figure 33: High level architecture of UAV drone system. (Primary Source).**

The above image shows the patterned structure of the UAV communication network in the drone systems. A UAV drone system comprises three main parts: the UAV, which is fitted with sensors and systems for communication. The ground control station controls the operator, and the data processing centre analyses the data. Together, these elements make it possible for real-time decision-making, data collecting, and remote piloting for various applications, including mapping and surveillance. The analysis cases and collects the most recent type of drone attack system caused by hackers or terrorists who can initiate the misconfiguration of the stored credentials in the UAV (drones) (Manesh, & Kaabouch, 2019). A similar type of high-level setup of the communication network causes a diminution in the overall flight movement and the direction of the UAV drones.

As some of the attack analysis is preventable due to the implementation of the blockchain-enabled UAV structure cases, there is an indefinite possibility that the **DATC** station (Drone Air Traffic Control) might not detect (Tedeschi et al. 2023). In specific instances, the hackers use specific tools in their Kali Linux or Ubuntu Linux terminal interfaces like the "Dronesploit" that is an opensource pen-testing tool.

### 5.2.2 Discussion:

Here, in the discussion section, the attack simulation example that is shown by a proposed approach for the UAV system networks correctly complies with the real-time process. The strong encryption schemes such as the "**DES, AES**" are taken for the proper prevention of the cyber threats with proper analysis pointers in the UAV systems (McCoy, & Rawat, 2019). As the entire data analysis is based on the secondary data analysis like tables, surveys and report cases, an example attacks simulation result table data that is also a great form of the secondary data is found after the analysis via "**extra key generation**" technique of encryption process.

| Encryption Scheme | Attacks Prevented | Attacks Successful | Key Size (Bits) | Message Size (KB) | Memory Usage (KB) | Computational Time (MS) |
|---|---|---|---|---|---|---|
| DES | 48 | 6 | 142 | 1,164 | 7346 | 195 |
| AES | 51 | 3 | 132 | 2396 | 12439 | 253 |
| Proposed Approach | 54 | 0 | 448 | 2236 | 27567 | 1567 |

**Figure 34: Example of an Attack Simulation Results of the UAV threat analysis (Source: researchportal.hw.ac.uk).**

The above table data example shows the analysed findings of the cybersecurity attack simulation results with the prevention data that has been done with the help of the "DES and AES" extra key generation encryption processes. The attack results discuss that the attack's prevented percentage is bigger than the attack's successful case (Mairaj, & Javaid, 2022). The ratio is distributed in 9:2 where 9 is for the "Attacks prevented" and 2 for the "Attacks successful" in the UAV network systems. Implementing the extra key generation process in the UAV laboratory is highly successful for the blockchain-enabled UAV devices that mainly use the "**Pallier Homomorphic encryption**" standards. The ciphered data is provided in the UAV internal circuit cascades where the connection point of the UAV drone movement is almost controlled safely from the clutches of the hackers.

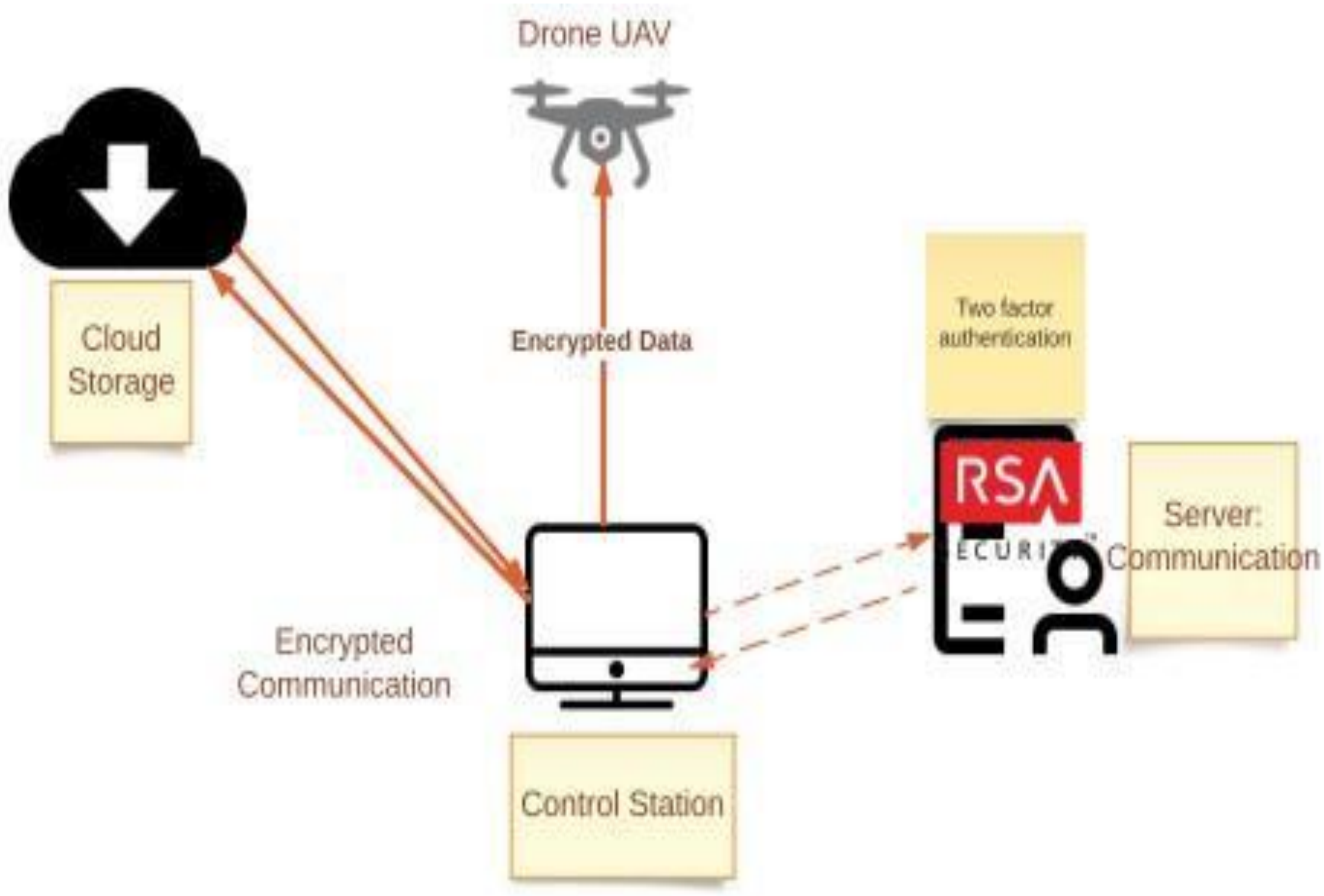

**Figure 35: Proposed approach system for the attack simulation prevention using encryption in the UAV network (Primary Source).**

The cybersecurity control approaches taken in proper feasibility for the UAV system tracking cases. Here, the ground control encrypts some data that is mainly regenerated using the “Pallier Homomorphic” blockchain cryptic system processes in the UAV drone (Son, & Kim, 2023). As shown above in Fig 19. The cloud sends encrypted communication signals to target the pinpoints of the hacked UAV movements that can cause either a crash landing or a rogue attack anywhere. The strong encrypted data patterns prevent the hacker's attack, and a perfect simulation result is achieved, showing the successful attack numbers, and contained attack numbers. This also is built in with a 2FA (two factor authentication) RSA. NASA also adopted the technique for implementing two-factor authentication (2FA) on servers and other computing resources that are not physically accessible by the user. (NASA. 2023) The drone attacks are caused by hackers who insert malicious spyware or ransomware (cyber virus threats) to misconfigure the UAV network movement.

Here, the analysis for the proposed approach system collects the specific instances after the encryption process is utilised in correct cases. The 'UAV-Cy' - a special thematic cyber security module is used in the UAV steering and correct manoeuvring process without misconfiguration or hacking (Haque, 2020). The analysis findings/ results retrieved after the successful, strong encryption of the "Pallier Homomorphic" process, the attack simulation example results show favourable data of the threat attack prevention and a small simulated process (Mohd, & Tesfa, 2023). A UAV kill chain process is initiated after the specific threat analysis of the consecutive attacks. Some analytics collect the under-retained data after sudden "brute force attack" systems.

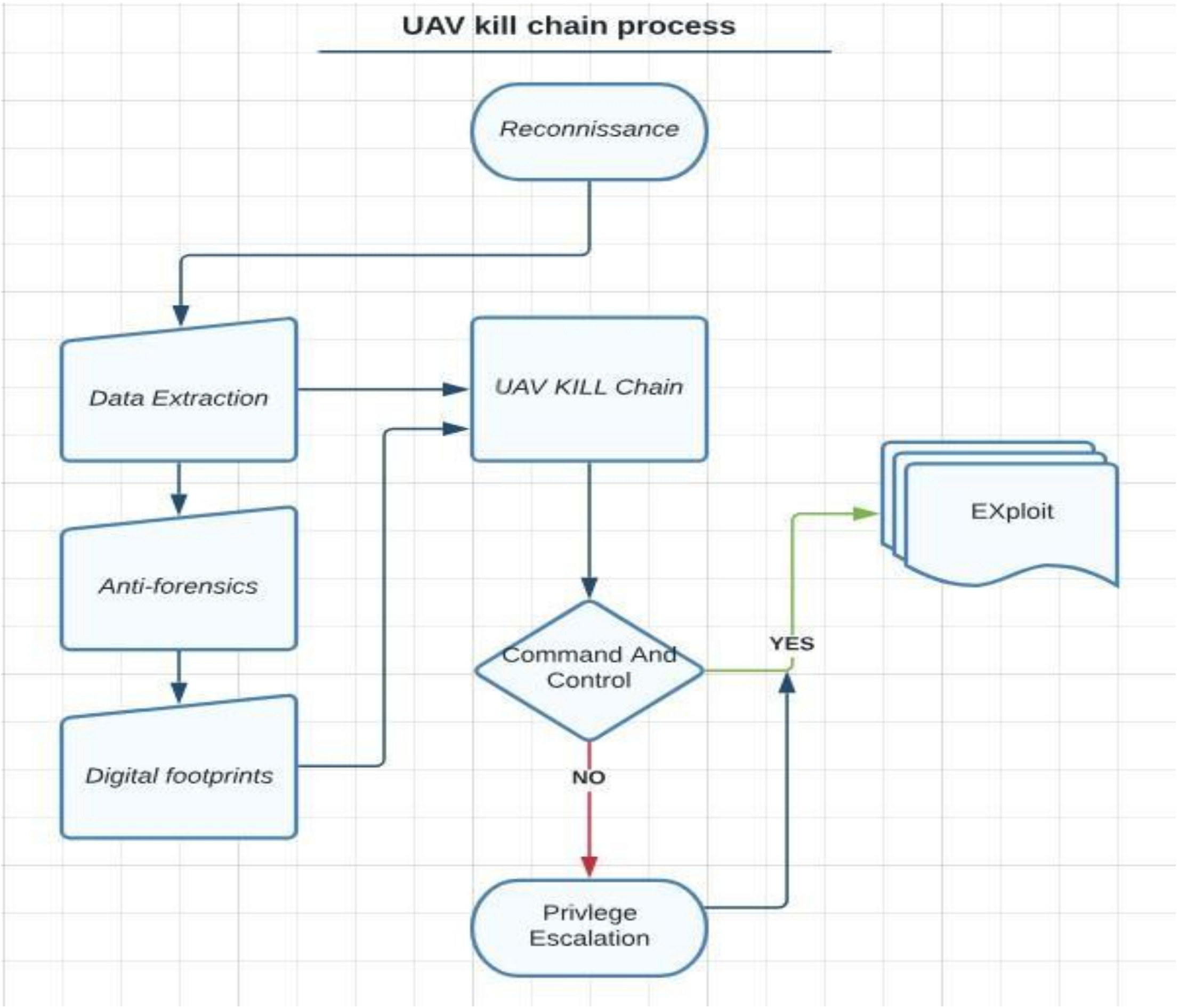

**Figure 36: UAV kill chain process for the threat attack analysis and simulation (Primary Source).**

The UAV network process analyses the entire kill chain system that is mainly accessed for the demonstrative purposes of attack detection and other techniques. Certain cases have been found in the case for the UAV patterns where the kill chain process is used to attack simulation styles (Alkadi et al. 2022). Some research phrases state that the kill chain states can direct the overall UAV network channel. As the data reaches the drone, the drone can detect the movement and then proceed with a favourable limitation.

Here, the UAV kill chain states some of the main processes for the UAV to attack processes where the first process, that is the hacked reconnaissance process, occurs. After this, the UAV is controlled by the main vulnerable target point controls for weaponization (Jahan, Sun, & Niyaz, 2020).

## 5.3 Limitations:

The overall limitations that are processed on the basis of the main topic in the research dissertation encase some provisions that cybersecurity compliance is possible. The varied application of the UAS system cases for the primary UAV devices can be void of the main cybersecurity frameworks and other stages (Wang et al. 2023). A normal limitation case regarding the overall UAV network architecture block process is mainly the normalised development in UAS data embedding and threat analysis. Some limitations can be considered while the proper cyber analysis and threat detection of the UAV network for drones. The limitations are as follows:

- In secondary analysis, there are bounded definitions for the UAV threat processing in linear cases.
- Different UAV drone sequences lead to the probably repeated crashing and rogue attacks as seen from the experiment 3 drones were tested.
- Military classification, where onsite research for drone variability can be studied.
- Source of funds to buy more high-tech drones and equipment for experimental data.

These limitations need to be analysed in correct proportions for the overall solutions to extract the corrupted data. Here the UAV network architecture block transforms the data cases from the UAV and SDN controller as the parameterised requirements in the UAV drone, such as the location, flight consolation and battery process, are utilised (Alzahrani et al. 2023). The limitations are an example for the overall UAV stem processes that can be mainly analysed in different encrypted states. The predefined connection processes the constraints and tries to point out in the context of the UAV network system hacking.

There is a strong case for the imitational changes that can be managed for the overall UAV network architectural process where the UAV and the SDN controllers collect the data from the final UAV network connection (Patel, Salot, & Parikh, 2022). The topic faces all the probabilities such that the UAV data can be changed after successfully implementing the limitation alterations. A small compliance factor regarding the overall limitations sections connects with the analysis sections where the UAV drones can be made clear of the limits and boundaries.

Here, all the limitations that can be managed in the term of the positive UAV stated armature positions, the possible analysis pointers can be coordinated keeping in mind the three IT security principles that is the “Availability, Integrity and Confidentiality” (Salamh et al. 2021). The possible conditions under which the entire analytical conditions can be possible is that the UAV workability boundary should be tightly compacted. The UAV model contacts the main operational limits like the connection of the ad hoc componential networks like “**FANET, MANET, and UANET**” (Khan, 2021). The model can also be initialised in correct order for the probable rechecking or final verification where the UAV drone systems operate a small correct simulation phase process.

Possible solutions can be implemented in the wake of the possible UAV threat attack (Guerber, Royer, & Larrieu, 2021). Some of the possible solutions that are needed in order to prevent the repeated disastrous threat entry and attack phase are:

- GPS data cross-checking.
- Hoping for the UAV frequency.
- Secured protocol bounding from large distances.
- Creating a strong UAS interference firewall for the drones.

The possible solutions are properly feasible to maintain the provided limitations in this dissertation.

## 5.4 Conclusion:

The analysis section in the dissertation work concludes the fact that the topic correctly interprets with the UAV systems and analyses in connection with the previous chapter phases. As the overall threat detection and the attack simulation of the UAV network for the drones collect proper information, there is a conduction of the final research process. The concluding factors demarcate the secondary analyses and the procedure of the threat analysis and attack simulation process for the UAV and UAS systems in the drones.

The concept of unmanned aerial processes separates the differentiation pattern that makes the hackers create a decryption process and lower the UAV aircraft via a “data-threat injection” scheme. The matters regarding the overall analysis and findings and the discussion cases cover all the phases in terms of the secondary data formats. The analysis chapter connects with the proper solutions to prevent unwanted access to the UAV drone consolation with the help of strong encryption standards.

# Chapter 6: Conclusion:

## 6.1 Introduction:

The introduction in the conclusion chapter summarises all the previous chapters under the provided dissertation topic case. The conclusion chapter creates the filters under the essential portions of the UAV analysis in secondary formats. The overall section connects with the possibilities or likeness of creating or drawing all kinds of summaries. The literature review cases and the correct methodology connect the data analysis and the concluding sections as the central part of the secondary data sources and the secondary form of analysis. to maintain and maintain the UAV navigation and proper management processes is to be analysed based on the threats and attack simulation process.

The probable cases for the overall analysis factors regarding UAV drone management and securing the network channels are correctly processed, as stated in this dissertation research case. There are other sections like the recommendations, future work where the topic analysis findings can be recommended to have proper analytical terms and the future additions or improvements for the UAV threat analysis and attack simulation. Some neutral changes are likely to be effectuated in the light of the proper UAV drone scaling for any malware or projected hacking for the additional implementations in the internal system phases.

## 6.2 Recommendations:

The recommendations section in this chapter faces some solution-based cases that are mainly to be followed in the present and future cases. The changes that are needed to make the UAV network architecture more portable and secure from external hacking or malware threat attacks are the only preliminary techniques. Enforce-Air is a company currently manufacturing a similar unite described, portable and secure. (Martin Broomhead., 2003) The recommendations are mainly for the significant analysis of future research work:

The recommendations are stated as follows:

- UAV drones must be checked thoroughly both internally as well as externally with the help of the UAV laboratory hardware/software tools.
- Creating a well-maintained and working strong anti-hacking and anti-malware drone systems.
- Processing the functionality process of the entire UAV network access through a secured window framework.
- Analysis of the main ad hoc network channels for resetting down all the vital UAV network lines in remote.
- Adding a strong blockchain encrypted framework in the drones for direct communication activity between the main user and the UAV system panel.

These recommendations are useful and are to be used in proper aspects for the overall UAV drone operations.

The recommendations serve as a correct activity case for the main term channelling of the UAV process and the accurate threat analysis and attack removal processes. All the above recommendation cases are to be utilised in the future work operations so that the future researchers can examine more from this dissertation and add improvised solutions and analysis cases properly.

## 6.3 Future Work:

The future work section entrails most of the possibilities of the analysis cases as well as the main topic justification changes. Here, in this section, the researchers can probably term down for the main analytical processes of having a correct secondary form of described threat analysis and attack simulation process. The future work cases that will be mainly added in the future studies for further examination are stated as follows:

- Adding a secure encrypted process to the threat attack analysis of the UAV systems. access control corresponding to a cable LAN is the goal that the authentication service for WLANs aims to achieve.

- Providing critical insights for the attacked system processes of the drones via "**backdoor-encryption**."
- Reading a full manually encrypted data encoder solution to phase and carve out a simple attack simulation structure of the UAV systems.

The future work of UAV threat detection analysis methods fall under the overall dissertation research. The recommendations develop important key structures that need to br observed in the UAV design. Also, the design and the structure of the UAV (drones) must be compact and error free so as to remove any possible reductions or threats inside the internal circuitry. All future work cases are to be maintained under the positive conditions.

## 6.4 Conclusion:

The topic of the dissertation research concludes that proper analysis of the threats faced by the UAV networks for the drones are all compiled and discussed in correct terms in all the chapters. The threat analysis process as well as the entire attack simulation cases are correctly informed and discussed with proper secondary data to connect with the chapter operations in proper terms. The introduction, literature review methodology, data analysis chapters are all properly discussed as per the topic cases. The recommendations as well as the future work shows the additions and improvements that are needed for the future researchers to examine and conduct more detailed research work derived from this dissertation research work.

The data analysis sections are mainly assessed for the proper data cases in the overall analytical records in the mainframe UAV data cases. There is a big compliance factor for the whole secondary type of data analysis and its collected data sources like in the table records, interviews, and other cases. As the entire research strongly emphasises the process of threat analysis, attack simulation and prevention, there is a basic need for future study.

# Appendices:

## Appendix 1: Parrot Drone

The Parrot Drone 2.0 which cause the battery power to smoke after several deauth attacks.

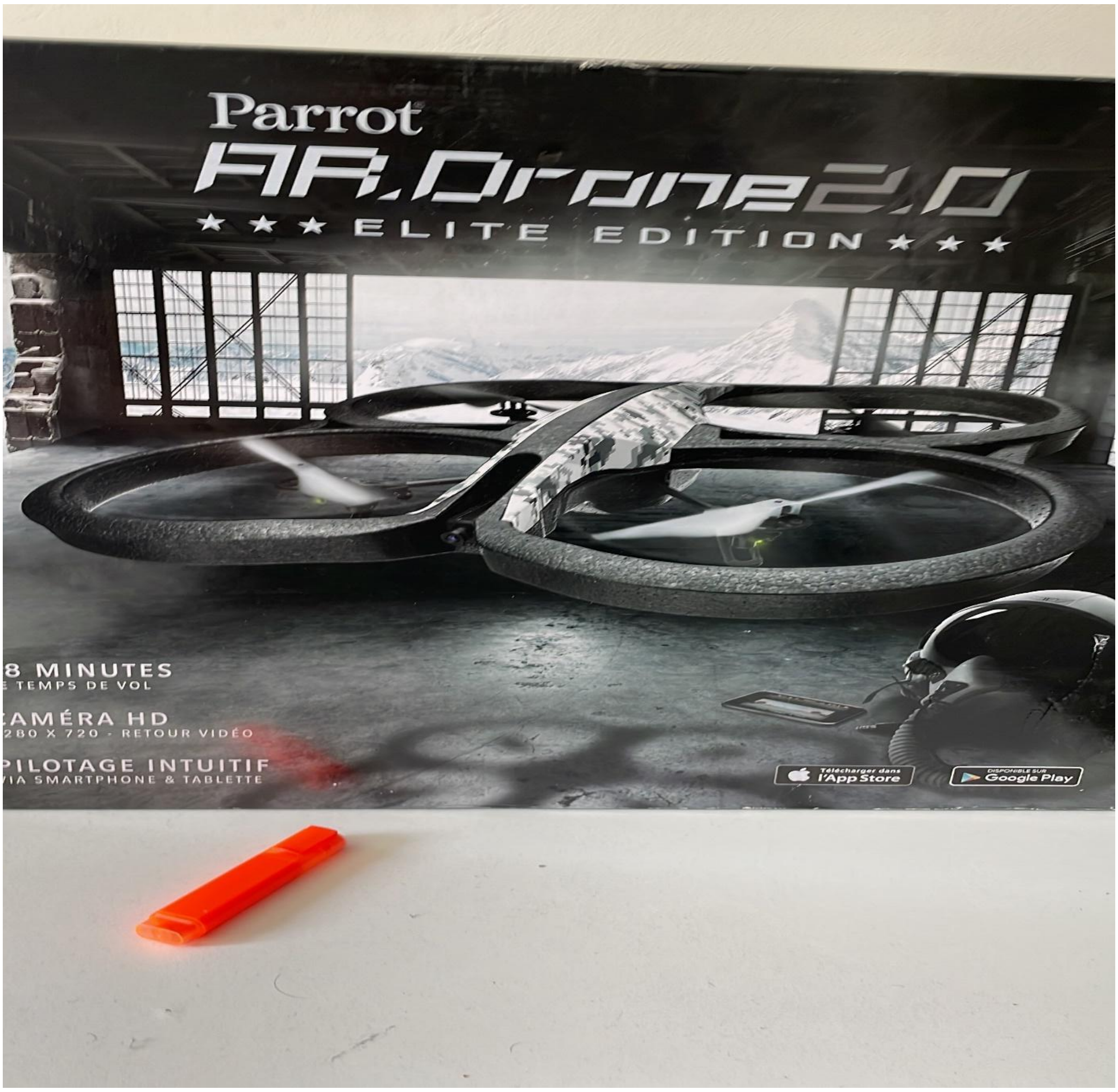

Box of Parrot Drone from e bay **(Primary Source).**

# Appendix: 2, Battery Power

The Battery power for the Parrot drone 1500mah, which got defective after the experiment.

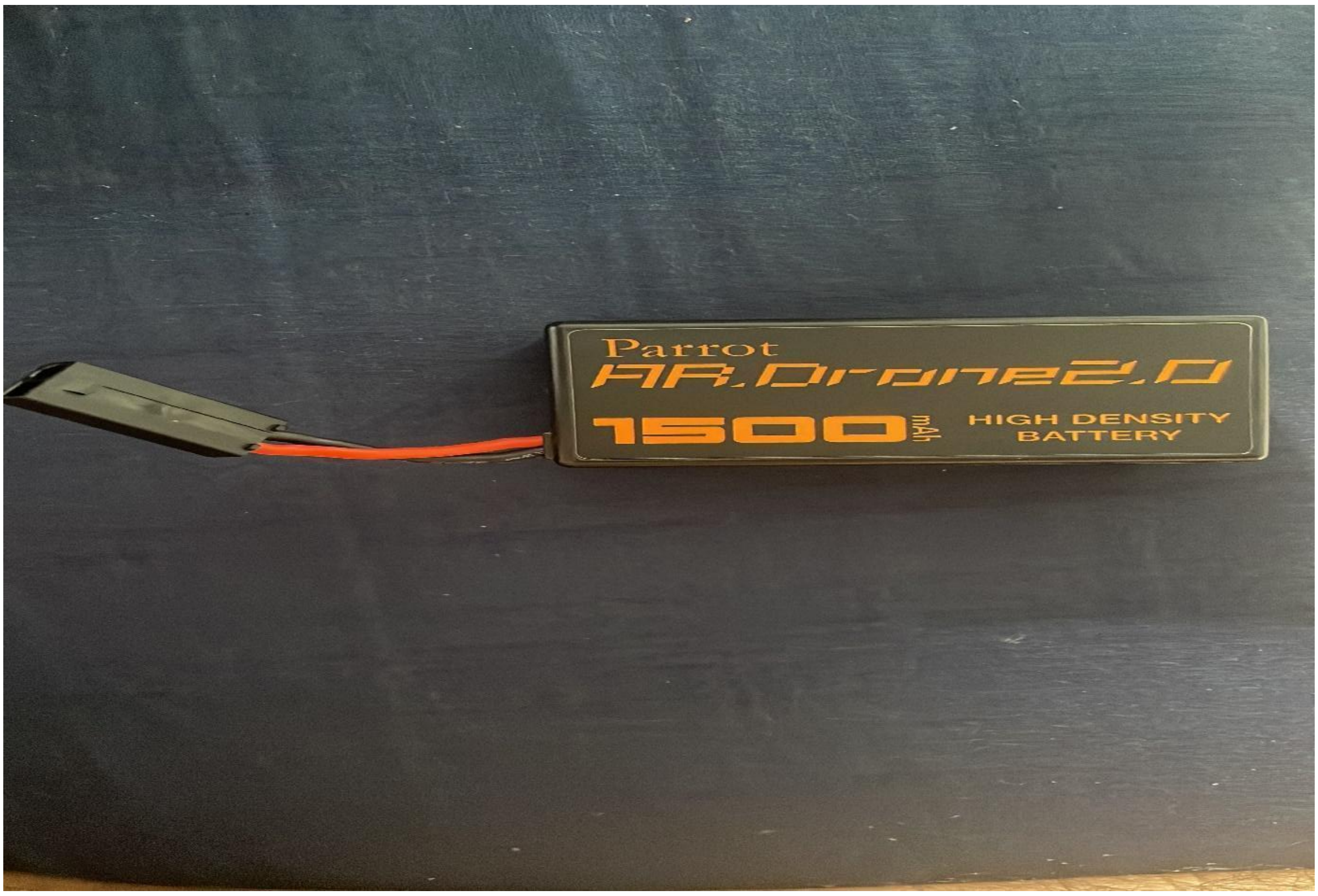

Battery Power Overheating after several pen-testing attacks **(Primary Source).**

## Appendix: 3. Emirates Drone Strike

When this report was written, Emirates Airline had their A380 Aircraft had a drone strike.

**‘Emirates confirms flight grounded after drone strike, safety ‘highest priority**
Dubai-based carrier Emirates reportedly cancelled the scheduled return flight to the emirate’
Staff Writer Arabian Business Tue 22 Aug 2023

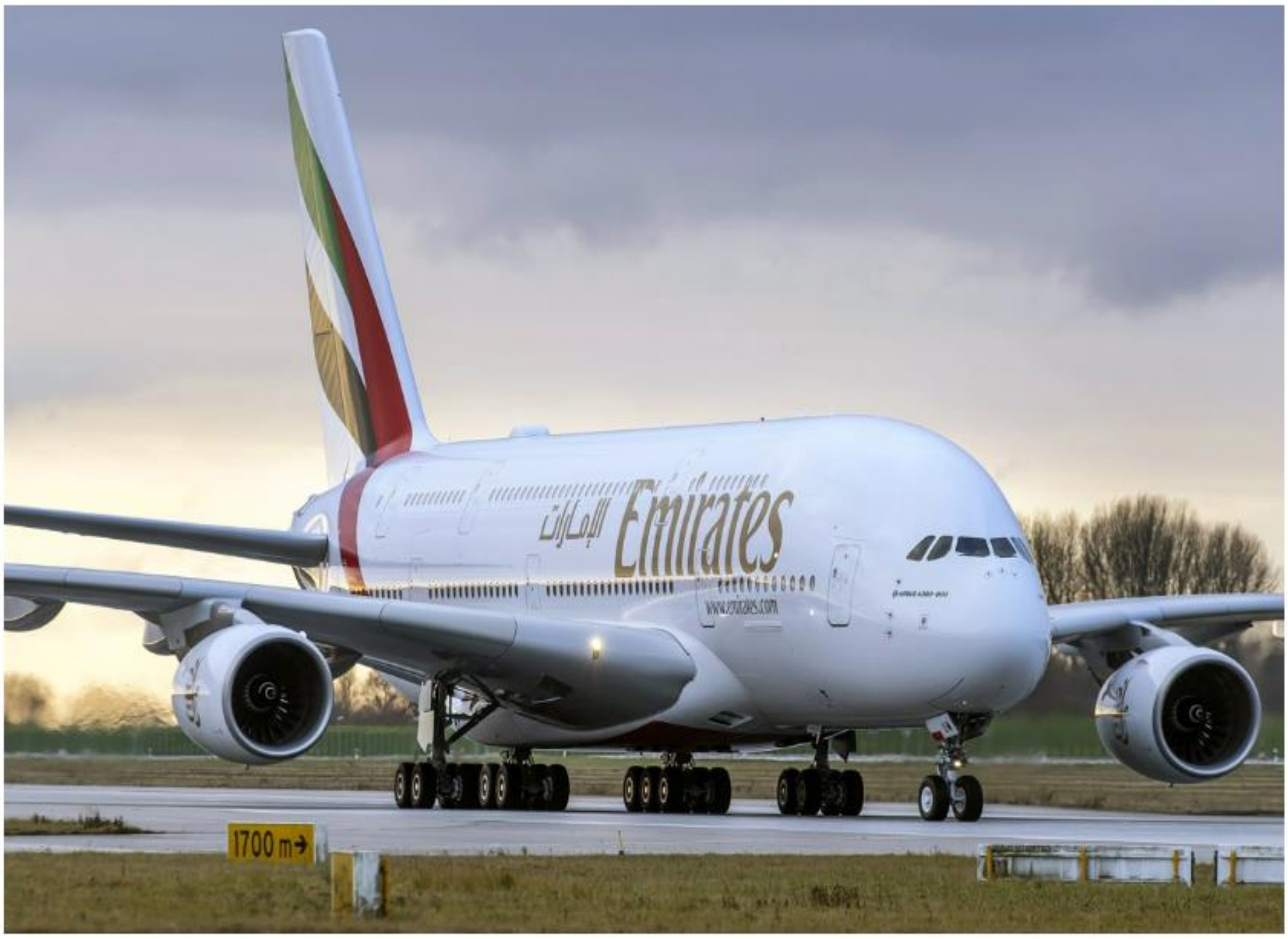

The damage was located in a section of the leading edge of the plane

Picture taken from Arabian Business 22/08/2023

## Appendix: 4, Simrex X500 Drone

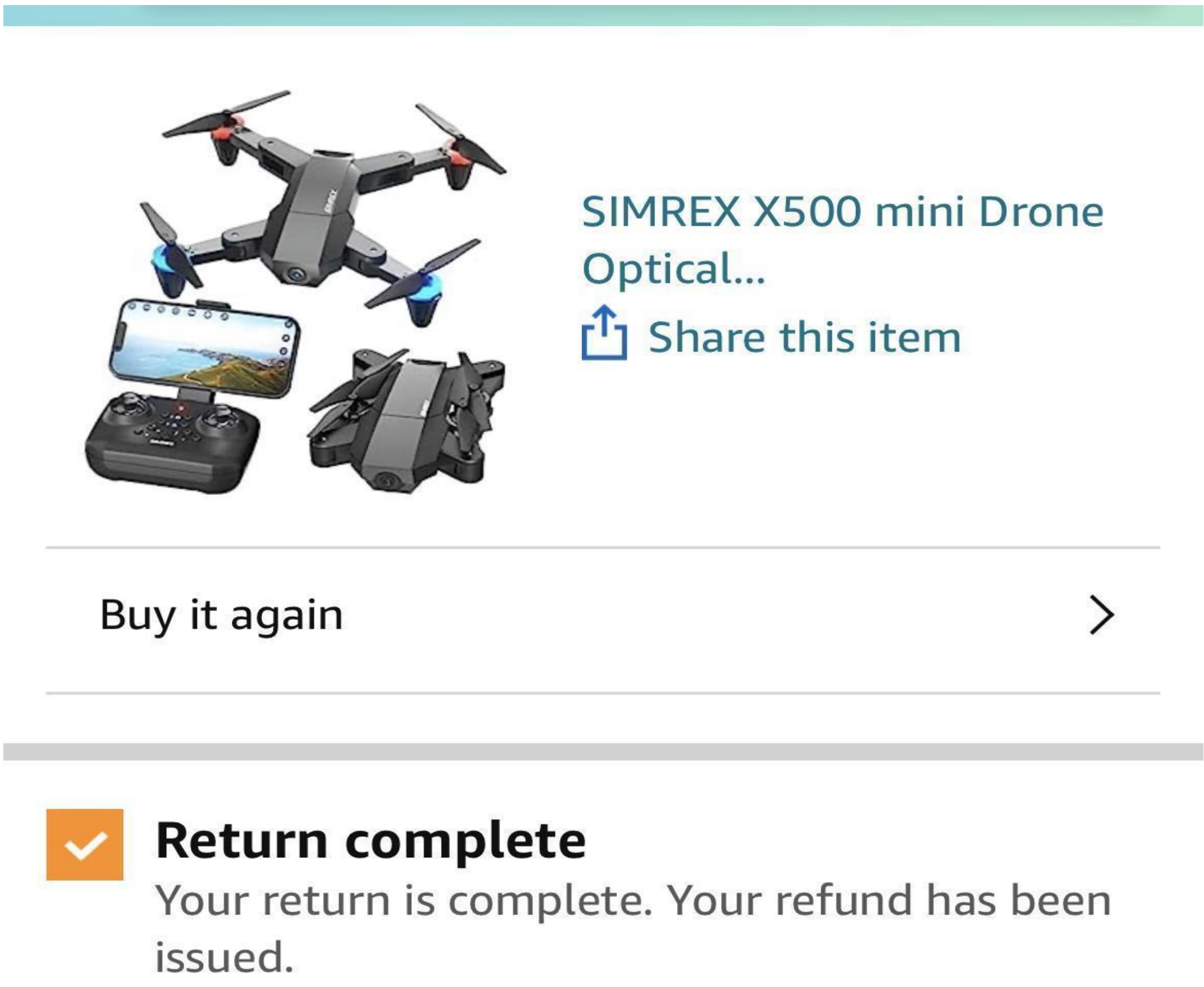

Amazon

The returned cheap toy drone

# Appendix 5: The drivers for Alpha network

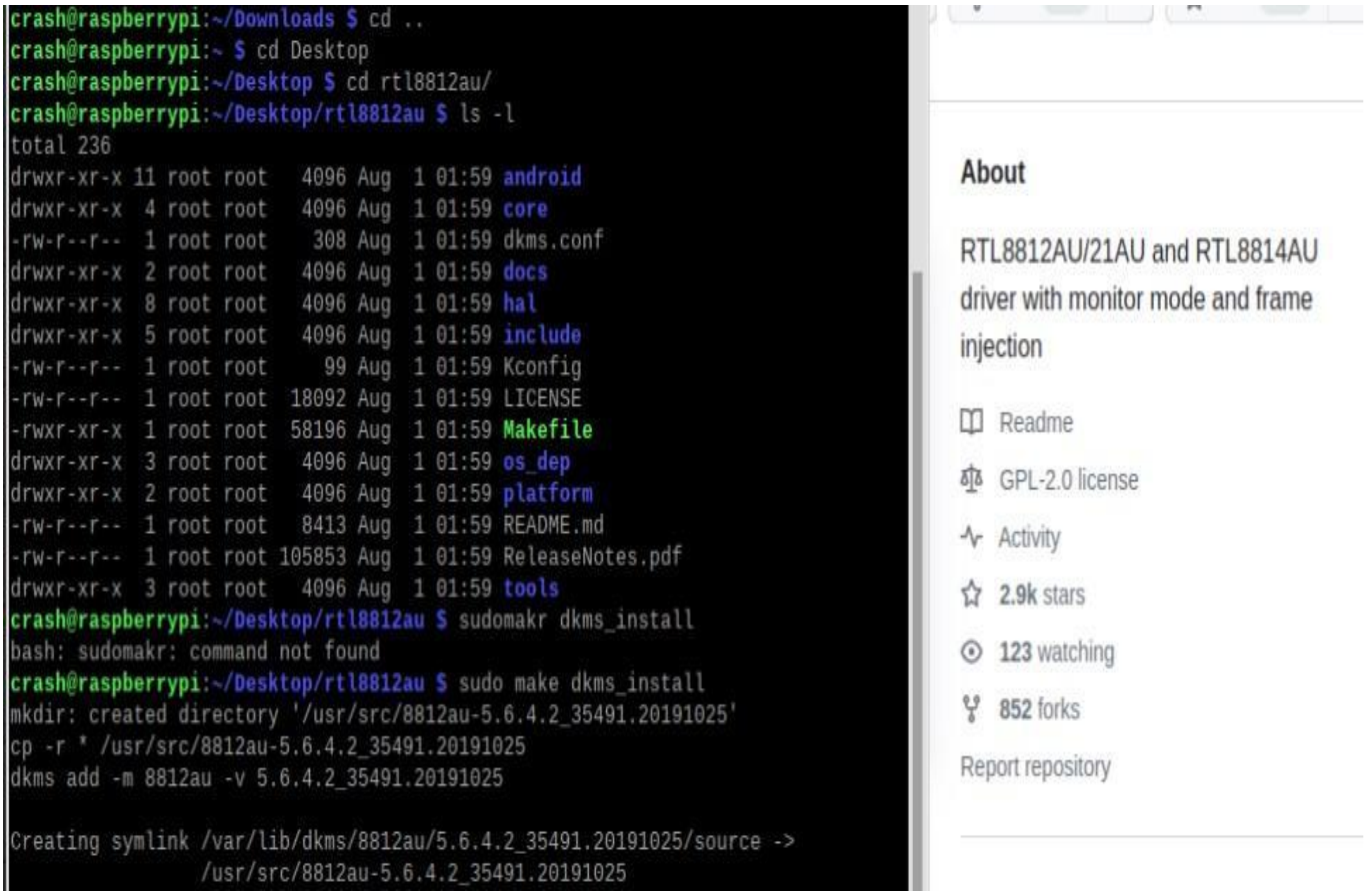

<u>Installation of Alpha network drivers **(Primary Source).**</u>

## Appendix 6: Monitor mode Alpha- Network

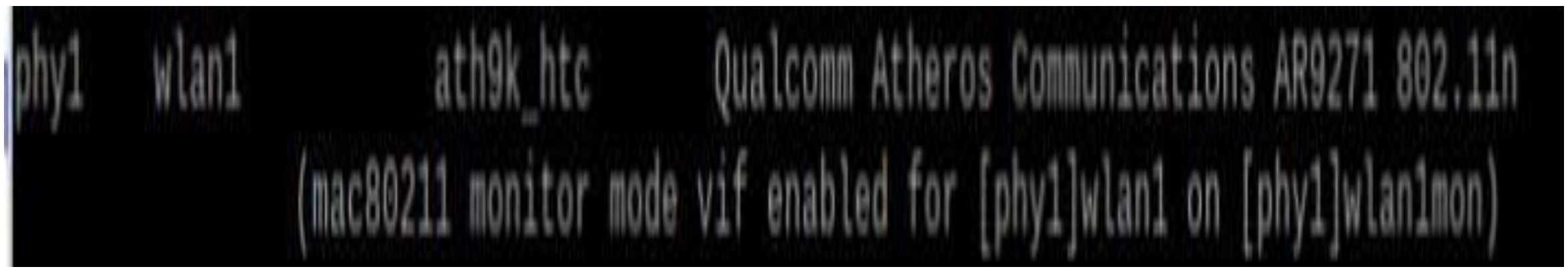

Alpha Driver Installed with Monitor Mode and Frame Injections **(Primary Source).**

## Appendix 7: Help options aircrack-ng

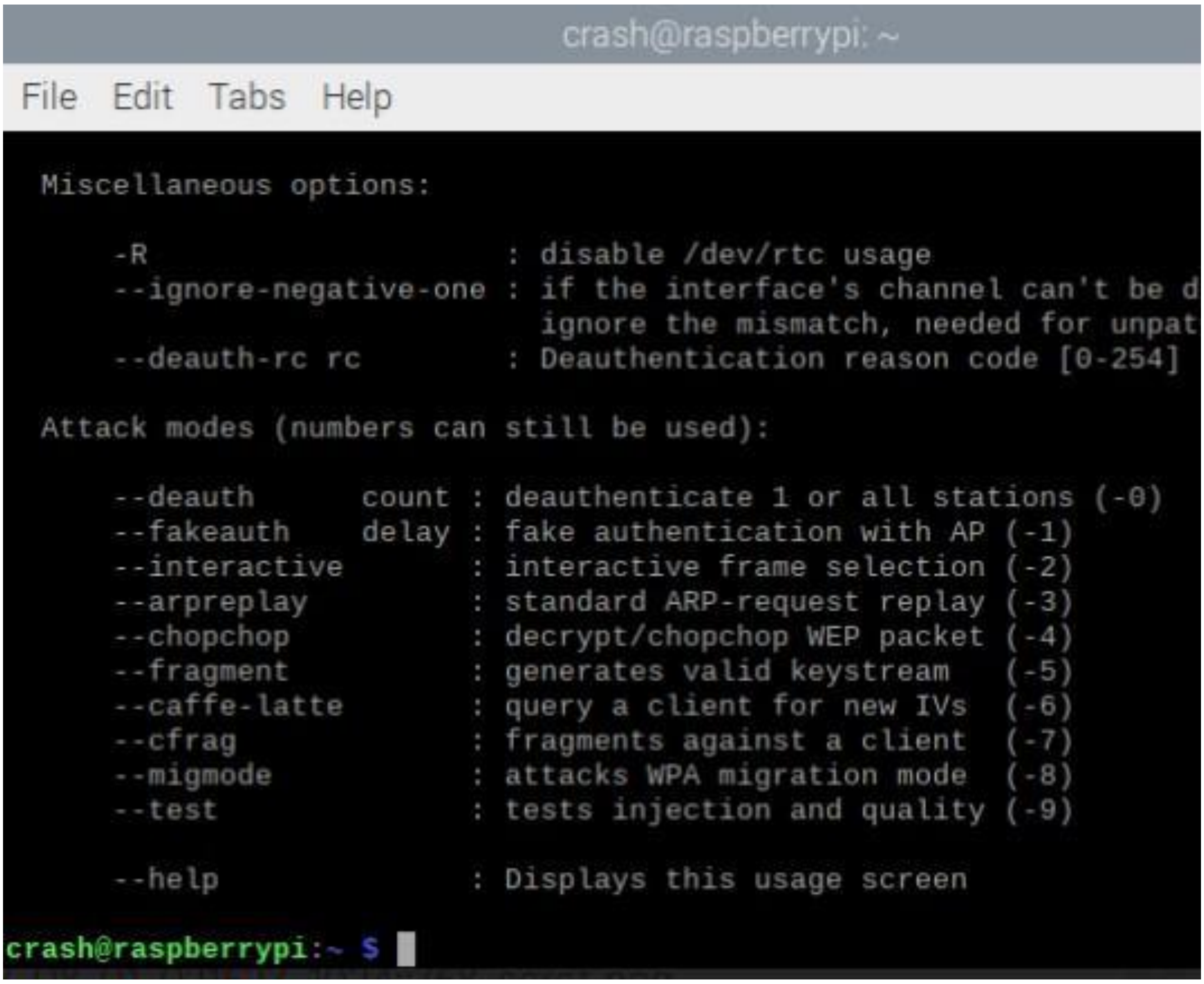

Other Attack Modes available on Aircrack-ng **(Primary Source).**

# Appendix 8: Project Planning

## Project Planning

The Gantt chart demonstrates the preparation of the overall project plan with time scales, resources and schedule:

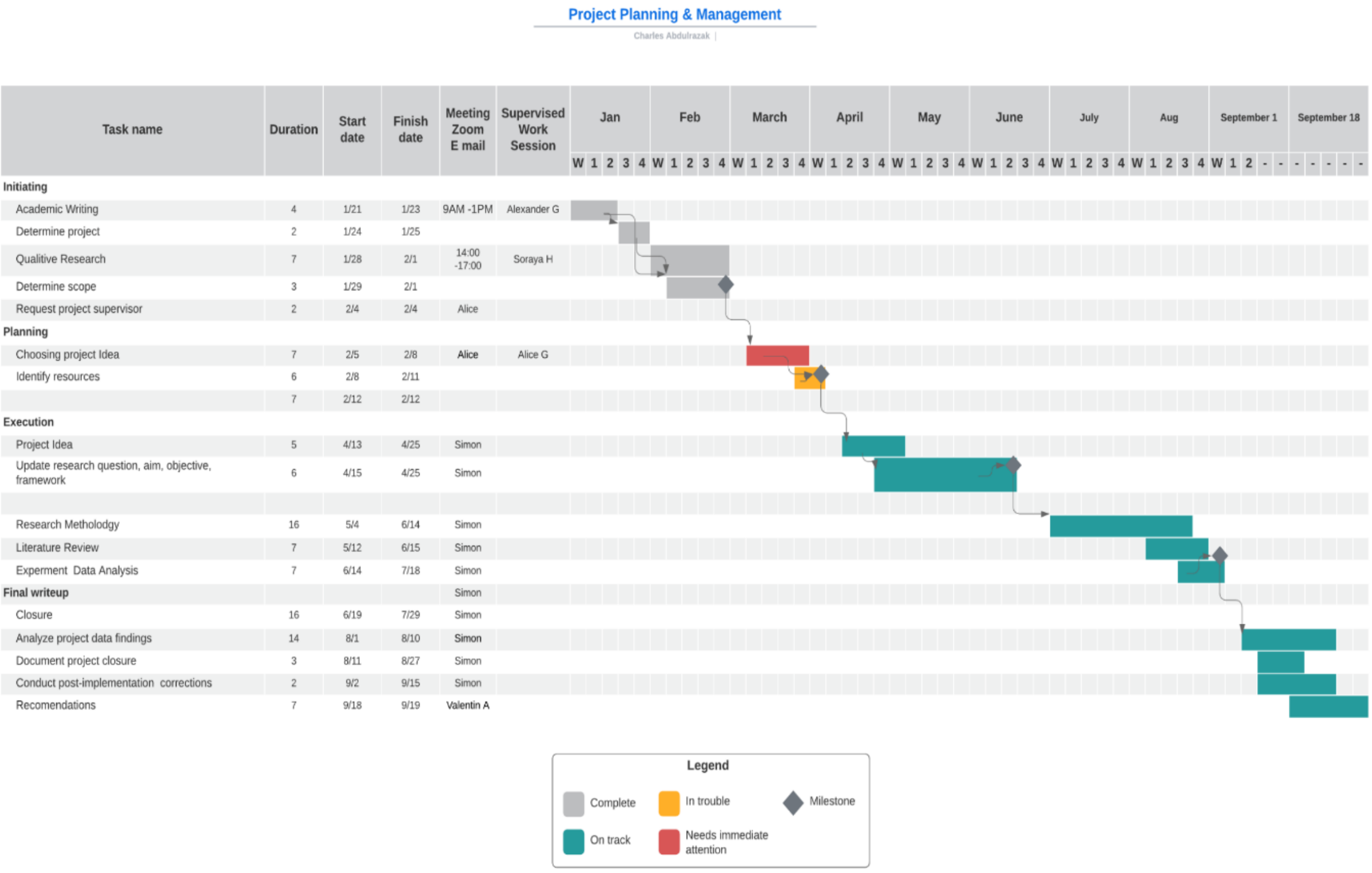